\documentstyle[preprint,prc,aps,eqsecnum,epsf,amssymb]{revtex}
\tighten
\begin{document}
\draft

\title{\hspace*{10cm}{\rm{JLAB-THY-00-15}} \\
Weak proton capture on $^{3}$He }
\author{L.E.\ Marcucci}
\address{Department of Physics, Old Dominion University,
         Norfolk, Virginia 23529}
\author{R.\ Schiavilla}
\address{Jefferson Lab, Newport News, Virginia 23606 \\
         and \\
         Department of Physics, Old Dominion University,
         Norfolk, Virginia 23529}
\author{M.\ Viviani}
\address{INFN, Sezione di Pisa, I-56100 Pisa, Italy}
\author{A.\ Kievsky}
\address{INFN, Sezione di Pisa, I-56100 Pisa, Italy}
\author{S.\ Rosati}
\address{Department of Physics, University of Pisa, I-56100 Pisa, Italy\\
and\\
INFN, Sezione di Pisa, I-56100 Pisa, Italy}
\author{J.F.\ Beacom}
\address{Physics Department 161-33, California Institute of 
Technology, Pasadena, California 91125}
\date{\today}

\maketitle

\begin{abstract}
The astrophysical $S$-factor for the proton weak capture on $^3$He
is calculated with correlated-hyperspherical-harmonics bound and
continuum wave functions corresponding to realistic Hamiltonians
consisting of the Argonne $v_{14}$ or Argonne $v_{18}$ two-nucleon and
Urbana-VIII or Urbana-IX three-nucleon
interactions.  The nuclear weak charge and current operators
have vector and axial-vector components, that include one- and many-body terms.
All possible multipole transitions connecting any of the $p\,^3$He S- and
P-wave channels to the $^4$He bound state are
considered.  The $S$-factor at a $p\,^3$He center-of-mass
energy of $10$ keV, close to the Gamow-peak energy, is predicted to be
$10.1 \times 10^{-20}$ keV~b with the AV18/UIX Hamiltonian, a factor of 
$\simeq$ 4.5 larger than the value adopted in the standard solar model.  
The P-wave transitions
are found to be important, contributing about 40 \% of the
calculated $S$-factor.  
The energy dependence is rather weak: the AV18/UIX zero-energy $S$-factor is 
$9.64 \times 10^{-20}$ keV~b, only 5 \% smaller than the 10 keV result 
quoted above.  The model dependence is also found to be weak: 
the zero-energy $S$-factor is calculated to be $ 10.2 \times 10^{-20}$
keV~b with the older AV14/UVIII model, only 6 \% larger than the AV18/UIX 
result. Our best estimate for the $S$-factor at 10 keV is 
therefore $(10.1\pm0.6) \times 10^{-20}$ keV~b, when the 
theoretical uncertainty due to the model dependence is included. 
This value for the
calculated $S$-factor is not as large as determined in fits to the
Super-Kamiokande data in which the $hep$ flux normalization is free.
However, the precise calculation of the $S$-factor and the consequent
absolute prediction for the $hep$ neutrino flux will allow much
greater discrimination among proposed solar neutrino oscillation
solutions.
\end{abstract}

\section{Introduction and Conclusions}
\label{sec:intro}

\subsection{Motivation}
\label{subsec:mot}

Recently, there has been a revival of interest in the reaction
$^3$He($p$,$e^+ \nu_e$)$^4$He~\cite{BK98,Fio98,Esc98,Hor99,Alb00,Alb002}.
This interest has been spurred by the Super-Kamiokande collaboration
measurements of the energy spectrum of electrons recoiling
from scattering with solar neutrinos~\cite{Fuk99,Smy99,Suz00}.  
Over most of the spectrum, a suppression $\simeq 0.5$ is observed
relative to the Standard Solar Model (SSM) predictions~\cite{BBP98}.  Above
12.5 MeV, however, there is an apparent excess of events.
The $hep$ process, as the proton weak capture on $^3$He
is known, is the only source of solar neutrinos
with energies larger than about 14 MeV--their end-point energy is
about 19 MeV.  This fact has naturally led
to questions about the reliability of calculations of the
$hep$ weak capture cross section, upon which is based 
the currently accepted SSM value for the astrophysical
$S$-factor at zero energy, $2.3 \times 10^{-20}$ keV~b~\cite{Sch92}.
In particular, Bahcall and Krastev have shown~\cite{BK98} that
a large enhancement, by a factor in the range 25--30, of
the SSM $S$-factor value given above would essentially fit the observed
excess~\cite{Fuk99} of recoiling electrons, in any of three different
neutrino scenarios--uniform suppression of the 
$^8$B flux, vacuum oscillations,
and matter-enhanced oscillations~\cite{MS86}.

The theoretical description of the $hep$ process, as well as that
of the neutron and proton radiative captures on $^2$H, $^3$H, and $^3$He,
constitute a challenging problem from the standpoint of
nuclear few-body theory.  Its difficulty can be appreciated
by comparing the measured values for the cross section
of thermal neutron radiative capture on $^1$H, $^2$H, and $^3$He.  Their
respective values are: $334.2 \pm 0.5$ mb~\cite{CWC65},
$0.508 \pm 0.015$ mb~\cite{JBB82},
and $0.055 \pm 0.003$ mb~\cite{Wol89,Wer91}.  Thus, in going from $A$=2 to 4
the cross section has dropped by almost four orders of magnitude.
These processes are induced by magnetic-dipole transitions between 
the initial two-cluster state in relative S-wave and the final bound state.
In fact, the inhibition of the $A$=3 and 4 captures has been understood
for a long time~\cite{Sch37}.  The $^3$H and $^4$He
wave functions, denoted, respectively, with $\Psi_3$ and $\Psi_4$ are, to
a good approximation, eigenfunctions of the magnetic dipole
operator $\bbox{\mu}$, namely $\mu_z \,\Psi_3 \simeq \mu_p \Psi_3$
and $\mu_z \, \Psi_4 \simeq 0$, where $\mu_p$=2.793 n.m. is the proton
magnetic moment (note that the experimental value of the
$^3$H magnetic moment is 2.979 n.m., while $^4$He has no magnetic
moment).  These relations would be exact, if the $^3$H and
$^4$He wave functions were to consist of a symmetric S-wave term only, for
example, 
$\Psi_4 = \phi_4({\rm S}) \, {\rm det}[ p\!\uparrow_1, p\!\downarrow_2,
n\!\uparrow_3, n\!\downarrow_4]$.  Of course,
tensor components in the nuclear interactions generate significant D-state
admixtures, that partially spoil this eigenstate property.
To the extent that it is approximately satisfied, though,
the matrix elements $\langle \Psi_3|\mu_z|\Psi_{1+2} \rangle$ and
$\langle \Psi_4|\mu_z|\Psi_{1+3}\rangle$ vanish due to
orthogonality between the initial and final states.  This orthogonality
argument fails in the case of the deuteron, since then
\begin{equation}
\mu_z \Psi_2 \simeq (\mu_p-\mu_n) \phi_2({\rm S}) \chi^0_0 \, \eta^1_0 \ ,
\end{equation}
where $\chi^S_{M_S}$ and $\eta^T_{M_T}$ are two-nucleon spin and isospin
states, respectively.  The magnetic dipole operator can therefore connect
the large S-wave component $\phi_2({\rm S})$ of the deuteron
to a $T$=1 $^1$S$_0$ $np$ state (note that the orthogonality
between the latter and the deuteron follows from the orthogonality
between their respective spin-isospin states). 

This quasi-orthogonality, while again
invalid in the case of the proton weak capture on protons,
is also responsible for inhibiting the $hep$ process.  Both these
reactions are induced by the Gamow-Teller operator, which differs
from the (leading) isovector spin part of the
magnetic dipole operator essentially by an isospin
rotation.  As a result, the $hep$ weak capture and $nd$, $pd$, 
$n\,^3$He, and $p\,^3$H radiative captures
are extremely sensitive to: (i) small
components in the wave functions, particularly the D-state admixtures 
generated by tensor interactions, and (ii) many-body terms in the
electro-weak current operator.  For example, two-body
current contributions provide, respectively, 
50 \% and over 90 \% of the calculated $pd$~\cite{Viv00} and 
$n\,^3$He~\cite{Sch92,Car90} 
cross sections at very low energies.

In this respect, the $hep$ weak capture is a particularly delicate
reaction, for two additional reasons: 
firstly and most importantly, the one- and two-body current contributions
are comparable in magnitude, but of opposite
sign~\cite{Sch92,Car91}; secondly, two-body axial currents, specifically
those arising from excitation of $\Delta$ isobars which have
been shown to give the dominant contribution, are
model dependent~\cite{Car91,CR71,Tow87}.

This destructive
interference between one- and two-body currents also occurs
in the $n\,^3$He (\lq\lq $hen$\rq\rq)
radiative capture~\cite{Sch92,Car90}, with the difference
that there the leading components of the two-body currents
are model independent, and give a much larger contribution than
that associated with the one-body current.

The cancellation in the $hep$ process between the one-
and two-body matrix elements has the effect of enhancing
the importance of P-wave capture channels, which would
ordinarily be suppressed.  Indeed, one of the results of the present
work is that these channels give about 40 \% of the
$S$-factor calculated value.  That the $hep$ process could proceed as easily
through P- as S-wave capture was not realized--or, at least,
not sufficiently appreciated~\cite{WB73}--in all
earlier studies of this reaction we are aware of, with the
exception of Ref.~\cite{Hor99}, where it was suggested, on the basis
of a very simple one-body reaction model, that the $^3$P$_0$ channel may
be important.

\subsection{Previous Studies of the $hep$ Capture}

The history of $hep$ cross section calculations has been
most recently reviewed by Bahcall and Krastev~\cite{BK98}.  The first estimate
of the cross section~\cite{Sal57} was based on the calculation of the overlap
of an s-wave proton continuum wave function and a 1s neutron 
wave function
in $^4$He.  It produced a large value for the $S$-factor, $630 \times 10^{-20}$
keV~b, and led to the suggestion by Kuzmin~\cite{Kuz65} that between 20 \% and
50 \% of the neutrinos in the high-energy end of the flux spectrum
could originate from the $hep$ reaction.  Of course, as already discussed
above and originally pointed out by Werntz and Brennan~\cite{WB67}, if the
$^4$He and $p\,^3$He states are approximated, respectively,
by (1s)$^4$ and (1s)$^3$2s$_c$ configurations (2s$_c$ is the continuum 
wave function), and antisymmetrized
in space, spin, and isospin, then the capture rate
vanishes identically.  Werntz and Brennan~\cite{WB67} attempted to relate
the matrix element of the axial current occurring in the
$hep$ capture to that of the electromagnetic current occurring
in the thermal neutron radiative capture on $^3$He, and
provided an upper limit for the $hep$ $S$-factor, $3.7 \times 10^{-20}$
keV~b, based on an experimental upper limit of 100 $\mu$b for
the $^3$He($n$,$\gamma$)$^4$He cross section known at the time.

Werntz and Brennan assumed: (i) the validity
of isospin symmetry, apart from differences in the neutron (in $hen$ capture)
and proton (in $hep$ capture) continuum wave functions, which they related
to each other via $|\psi_p(r)/\psi_n(r)| \simeq C_0$ ($C_0$ is
the usual Gamow penetration factor); (ii) that two-body currents
dominated both the weak and radiative captures, and that their matrix
elements could be put in relation to each other through
an isospin rotation.  These authors refined their earlier estimate
for the $hep$ $S$-factor in a later
publication~\cite{WB73}, by using hard-sphere
phase shifts to obtain a more realistic value for the ratio of the
neutron to proton continuum wave functions, and by including
the contributions due to P-wave capture channels.  These refinements led
to an $S$-factor value, $8.1 \times 10^{-20}$ keV~b, considerably
larger than they had obtained previously.  They found, though, that
the P-waves only contribute at the 10 \% level. 

Subsequent studies of the $hep$ process also attempted
to relate it to the $hen$ radiative capture, but recognized
the importance of D-state components in the $^3$He and $^4$He wave
functions--these had been ignored in
Refs.~\cite{WB73,WB67}--, and used the Chemtob-Rho
prescription~\cite{CR71} (with some short-range modification) for the two-body
terms in the electroweak current operator.  Tegn\'er and Bargholtz~\cite{TEG83}
and Wervelman {\it et al.}~\cite{Wer91} found, using a shell-model description
of the initial and final states, that two-body
current contributions do not dominate the capture processes, in
sharp contrast with the assumptions of Refs.~\cite{WB73,WB67}
and the later conclusions of
Refs.~\cite{Sch92,Car90,Car91}.  These two groups as well as
Wolfs {\it et al.}~\cite{Wol89}
arrived, nevertheless, to contradictory results, due
to the different values calculated for the ratio of weak to electromagnetic
matrix elements.  Tegn\'er and Bargholtz~\cite{TEG83}
obtained an $S$-factor value of
$(17 \pm 8)\times 10^{-20}$ keV~b, the spread being due to the uncertain
experimental value of the thermal neutron capture cross section
before 1983.  This prediction was sharpened by Wolfs {\it et al.}~\cite{Wol89},
who measured the $hen$ cross section precisely.  They quoted
an $hep$ $S$-factor value of $(15.3 \pm 4.7) \times 10^{-20}$ keV~b.
Wervelman {\it et al.}~\cite{Wer91} also measured the $hen$
cross section, reporting a value of $(55\pm 3)$ $\mu$b
in excellent agreement with the Wolfs {\it et al.} measurement of $(54 \pm 6)$
$\mu$b, but estimated an $hep$ $S$-factor in the range
$(57 \pm 8)\times 10^{-20}$ keV~b.  These discrepancies are 
presumably due to the schematic wave functions used in the
calculations.

In an attempt to reduce the uncertainties in the predicted
values for both the radiative and weak capture rates, fully
microscopic calculations of these reactions were performed 
in the early nineties~\cite{Car90,Car91}, based on ground- and scattering-state
wave functions obtained variationally from a realistic Hamiltonian
with two- and three-nucleon interactions.  The main part of
the electromagnetic current operator
(denoted as \lq\lq model independent\rq\rq) was constructed
consistently from the two-nucleon interaction model.  The less
well known (\lq\lq model dependent\rq\rq) electroweak currents associated
with the excitation of intermediate $\Delta$ isobars
and with transition couplings, such as the electromagnetic
or axial $\rho \pi$ current, were also included.  However, it was
emphasized that their contribution was to be viewed as numerically
uncertain, as very little empirical information is available
on their coupling constants and short-range behavior.  These studies
showed that both the $hen$ and $hep$ reactions
have large (in the case of the radiative capture,
dominant) contributions from two-body currents.
Indeed, the values obtained with one-body only and full currents
for the $hep$ $S$-factor (radiative capture cross section)
were, respectively, $5.8 \times 10^{-20}$ and $1.3 \times 10^{-20}$ keV~b
(6 and 112 $\mu$b).  These results indicated
that the common practice of inferring the $hep$ $S$-factor
from the measured radiative capture cross section
is bound to be misleading, because of different
initial-state interactions in the $n\,^3$He and
$p\,^3$He channels, and because of the large contributions
associated with the two-body components of the electroweak
current operator, and their destructive interference with
the one-body current contributions.  Yet, the substantial
overprediction of the $hen$ cross section, 112 $\mu$b
versus an experimental value of 55 $\mu$b,
was unsatisfactory.  It became clear that the contributions
of the \lq\lq model dependent\rq\rq currents, particularly those due to
the $\Delta$ isobar, were unreasonably large (about
40 $\mu$b out of the total 112 $\mu$b).  It was therefore
deemed necessary to include the $\Delta$ degrees of freedom
explicitly in the nuclear wave functions, rather
than eliminate them in favor of effective two-body
operators acting on nucleon coordinates, as it had
been done in earlier studies.  This led to the development
of the transition-correlation operator (TCO)
method~\cite{Sch92}--a scaled-down approach
to a full $N$+$\Delta$ coupled-channel
treatment.  The radiative capture cross section was
now calculated to be between 75 and 80 $\mu$b~\cite{Sch92}
(excluding the small contribution of the
\lq\lq uncertain\rq\rq  $\omega \pi \gamma$
current), the spread depending on whether the $\pi$$N$$\Delta$
coupling constant in the transition interactions is taken
either from experiment or from the quark model.  In this approach,
the $hep$ $S$-factor was calculated to be in the range
between $1.4 \times 10^{-20}$ and $3.1 \times 10^{-20}$ keV~b~\cite{Sch92},
the spread due to whether the axial $N$$\Delta$
coupling was determined by fitting the Gamow-Teller
matrix element in tritium $\beta$-decay or, again, taken from
the quark model (uncertainties in the values of the
$\pi$$N$$\Delta$ coupling had a much smaller impact).
In fact, the SSM value for the $hep$ $S$-factor
now quoted in the literature~\cite{BK98,Fio98} is the average
of these last two results.

\subsection{Overview of Present Calculations}

Improvements in the modeling of two- and three-nucleon interactions and
the nuclear weak current, and the significant
progress made in the last few years in the description of the bound and
continuum four-nucleon wave functions, have prompted
us to re-examine the $hep$ reaction.  The nuclear Hamiltonian
has been taken to consist of the Argonne $v_{18}$
two-nucleon~\cite{WSS95} and Urbana-IX
three-nucleon~\cite{Pud95} interactions.  To make contact with the earlier
studies~\cite{Sch92,Car91}, however, and to
have some estimate of the model dependence
of the results, the older Argonne $v_{14}$ two-nucleon~\cite{WSA84}
and Urbana-VIII three-nucleon~\cite{Wir91}
interaction models have also been used. 
Both these Hamiltonians, the AV18/UIX and AV14/UVIII, reproduce
the experimental binding energies and charge
radii of the trinucleons and $^4$He in exact Green's function
Monte Carlo (GFMC) calculations~\cite{Pud97,Car90a}. 

The correlated-hyperspherical-harmonics (CHH) method is used here to solve
variationally the bound- and scattering-state
four-nucleon problem~\cite{VKR95,VRK98}.  The
binding energy of $^4$He calculated with the
CHH method~\cite{VKR95,Viv99b} is within 1--2 \%,
depending on the Hamiltonian model, of that obtained with the GFMC
method.  The accuracy of the CHH method to calculate
scattering states has been successfully
verified in the case of the trinucleon systems, by comparing results for
a variety of $Nd$ scattering observables obtained by a number
of groups using different
techniques~\cite{Kie98}.  Indeed, the numerical uncertainties
in the calculation of the trinucleon continuum have been so drastically
reduced that $Nd$ scattering observables can now be used to directly study
the sensitivity to two- and three-nucleon interaction models--the $A_y$
\lq\lq puzzle\rq\rq constitutes an excellent example of
this type of studies~\cite{Glo96}.  

Studies along similar lines show~\cite{Viv98} that the
CHH solutions for the four-nucleon continuum are also highly accurate.
The CHH predictions~\cite{VRK98} for the
$n\,^3$H total elastic cross section,
$\sigma_T= \pi\, (\, |a_{\rm s}|^2+3\,|a_{\rm t}|^2\, )$,
and coherent scattering length, 
$a_{\rm c}=a_{\rm s}/4 +3\, a_{\rm t}/4$, measured
by neutron interferometry techniques--$a_{\rm s}$ and $a_{\rm t}$ are
the singlet and triplet scattering lengths--have
been found to be in excellent agreement with the corresponding
experimental values.  The $n\,^3$H cross section is known over
a rather wide energy range, and its extrapolation to zero energy
is not problematic~\cite{Phi80}.  The situation is
different for the $p\,^3$He
channel, for which the scattering lengths have been determined
from effective range extrapolations of data taken above 1 MeV, and
are therefore somewhat uncertain, $a_{\rm s}=(10.8 \pm 2.6)$ fm~\cite{AK93} and
$a_{\rm t}=(8.1 \pm 0.5)$ fm~\cite{AK93}
or $(10.2 \pm 1.5)$ fm~\cite{TEG83}.  Nevertheless, 
the CHH results are close to the experimental values above.  For example,
the AV18/UIX Hamiltonian predicts~\cite{VRK98}
$a_{\rm s}=10.1$ fm and $a_{\rm t}=9.13$ fm.

In Refs.~\cite{Sch92,Car91} variational Monte Carlo
(VMC) wave functions had been used
to describe both bound and scattering states.  The triplet scattering
length was found to be 10.1 fm with the AV14/UVIII Hamiltonian
model, in satisfactory agreement with the experimental determination and the
value obtained with the more accurate
CHH wave functions.  However, the present work includes all S- and
P-wave channels, namely $^1$S$_0$, $^3$S$_1$, 
$^3$P$_0$, $^1$P$_1$, $^3$P$_1$, and $^3$P$_2$, while
all previous works only retained the $^3$S$_1$ channel, which
was thought, erroneously, to be the dominant one.

The nuclear weak current consists of vector and axial-vector parts, with
corresponding one-, two-, and many-body components.  The weak vector current
is constructed from the isovector part of the
electromagnetic current, in accordance with the conserved-vector-current
(CVC) hypothesis.  
Two-body weak vector currents have \lq\lq model-independent\rq\rq
and \lq\lq model-dependent\rq\rq components. The model-independent
terms are obtained from the nucleon-nucleon interaction, and by
construction satisfy current conservation with it.
The leading two-body weak vector current
is the \lq\lq$\pi$-like\rq\rq operator, obtained from the isospin-dependent
spin-spin and tensor nucleon-nucleon interactions.
The latter also generate an isovector \lq\lq$\rho$-like\rq\rq current,
while additional isovector two-body currents arise from the isospin-independent
and isospin-dependent central and momentum-dependent interactions.  These
currents are short-ranged, and numerically far less important
than the $\pi$-like current.  With the exception of the $\rho$-like current,
they have been neglected in the present work.  The model-dependent
currents are purely transverse, and therefore cannot be directly
linked to the underlying two-nucleon interaction.  The present
calculation includes the isovector currents associated with excitation
of $\Delta$ isobars which, however, are found to give a rather small 
contribution in weak-vector transitions, as compared to that due to
the $\pi$-like current.  The $\pi$-like and $\rho$-like
weak vector charge operators have also been retained in the
present study.

The leading two- and many-body terms in the
axial current, in contrast to the case of the
weak vector (or electromagnetic) current, are those due to $\Delta$-isobar
excitation, which are treated within the TCO scheme.  This scheme 
has in fact been extended~\cite{Mar98} to include
three-body connected terms which were
neglected in the earlier work~\cite{Sch92}.  The
axial charge operator includes
the long-range pion-exchange term~\cite{Kub78}, required by low-energy
theorems and the partially-conserved-axial-current relation,
as well as the (expected) 
leading short-range terms constructed from the central
and spin-orbit components of the nucleon-nucleon interaction, following
a prescription due to Riska and collaborators~\cite{Kir92}.  

The largest model dependence is in the weak axial current.  To minimize it,
the poorly known $N \rightarrow \Delta$ transition
axial coupling constant has been adjusted to reproduce the experimental value
of the Gamow-Teller matrix element in tritium $\beta$-decay.  While this
procedure is inherently model dependent, its actual model dependence is in fact
very weak, as has been shown in Ref.~\cite{Sch98}.  The analysis 
carried out there could be extended to the present case.

\subsection{Conclusions}

We present here a discussion of the results for the astrophysical 
$S$-factor and their implications for the Super-Kamiokande (SK) solar 
neutrino spectrum.

\subsubsection{Results for the $S$-factor}
Our results for the astrophysical $S$-factor, defined as 
\begin{equation}
S(E) = E\, \sigma(E)\,
{\rm exp}( 4\, \pi \, \alpha/v_{\rm rel}) \ ,
\end{equation}
where $\sigma(E)$ is the $hep$ cross
section at center-of-mass energy $E$, $v_{\rm rel}$ is the $p\,^3$He relative
velocity, and $\alpha$ is the fine
structure constant, are reported in Table~\ref{tb:sfact}.  By 
inspection of the table, we note that: (i) 
the energy dependence is
rather weak: the value at $10$ keV is only about 4 \% larger
than that at $0$ keV; (ii) the P-wave capture states are found to
be important, contributing about 40 \% of the calculated
$S$-factor.  However, the contributions from D-wave channels
are expected to be very small.  We have verified explicitly
that they are indeed small in $^3$D$_1$ capture. 
(iii) The many-body axial currents
associated with $\Delta$ excitation play a crucial
role in the (dominant) $^3$S$_1$ capture, where they reduce
the $S$-factor by more than a factor of four; thus
the destructive interference between the one- and many-body
current contributions, first obtained in Ref.~\cite{Car91}, is
confirmed in the present study, based on more accurate
wave functions.  The (suppressed) one-body contribution
comes mostly from transitions involving the D-state
components of the $^3$He and $^4$He wave functions, while
the many-body contributions are predominantly due to transitions
connecting the S-state in $^3$He to the D-state in $^4$He, or viceversa.

It is important to stress the differences between the present
and all previous studies.  Apart from
ignoring, or at least underestimating, the contribution
due to P-waves, the latter only considered the long-wavelength
form of the weak multipole operators, namely, their $q$=$0$ limit, 
where $q$ is the magnitude of the momentum transfer.
In $^3$P$_0$ capture, for example, only the $C_0$-multipole,
associated with the weak axial charge, survives in this limit,
and the corresponding $S$-factor is calculated to be $2.2 \times 10^{-20}$
keV~b, including two-body contributions.  However,
when the transition induced by the longitudinal component of
the axial current (via the $L_0$-multipole, which
vanishes at $q$=$0$) is also taken into account, the $S$-factor
becomes $0.82 \times 10^{-20}$ keV~b, because of destructive
interference between the $C_0$ and $L_0$ matrix
elements (see discussion in Sec.~\ref{sec:xs}).  Thus use of the
long-wavelength approximation in the calculation of the
$hep$ cross section leads to inaccurate results.

Finally, besides the differences listed above, the present
calculation also improves that of Ref.~\cite{Sch92} in a number of
other important respects: firstly, it uses
CHH wave functions, corresponding to the latest generation
of realistic interactions; secondly, the model for the
nuclear weak current has been extended to include
the axial charge as well as the vector charge and current operators.
Thirdly, the one-body operators now take
into account the $1/m^2$ relativistic
corrections, which had previously been neglected.  In $^3$S$_1$
capture, for example, these terms increase by 25 \% the
dominant (but suppressed) $L_1$ and $E_1$ matrix elements calculated with the
(lowest order) Gamow-Teller operator.  These improvements in the
treatment of the one-body 
axial current indirectly affect also the contributions
of the $\Delta$-excitation currents, since the $N\Delta$ transition
axial coupling constant is determined by reproducing the Gamow-Teller
matrix element in tritium $\beta$-decay, as discussed in Sec.~\ref{sec:vadlt}
below.

The chief conclusion of the present work is  
that the $hep$ $S$-factor is
predicted to be $\simeq$ 4.5 times larger than the value
adopted in the SSM.  This enhancement,
while very significant, is smaller than that first suggested 
in Refs.~\cite{BK98,Esc98}, and then reconsidered by the SK 
collaboration in Ref.~\cite{Suz00}.  A discussion of the 
implications of our results for the 
SK solar neutrino spectrum is given below.  

Even though our result is
inherently model dependent, it is unlikely that the model dependence
is large enough to accommodate a drastic increase in the value obtained here.  
Indeed, calculations using Hamiltonians based on the AV18 two-nucleon 
interaction only and the older AV14/UVIII two- and
three-nucleon interactions predict zero energy $S$-factor values of 
$12.1 \times 10^{-20}$ keV~b and $10.2 \times 10^{-20}$ keV~b, respectively.
It should be stressed, however, that the AV18 model, in contrast to the
AV14/UVIII, does not reproduce the experimental binding energies and low-energy
scattering parameters of the three- and four-nucleon systems.
The AV14/UVIII prediction is only 6 \% larger than the AV18/UIX 
zero-energy result.  This 6 \% variation should provide
a fairly realistic estimate of the theoretical uncertainty due to
the model dependence.  It would be very
valuable, though, to repeat the present study with a Hamiltonian consisting of 
the CD-Bonn interaction~\cite{Mac96} which, in contrast
to the AV14 and AV18 models, has strongly non-local central
and tensor components.  We would expect the CD-Bonn calculation
to predict an $S$-factor value close to that
reported here, provided the axial current in that calculation were again
constrained to reproduce the known Gamow-Teller matrix
element in tritium $\beta$-decay~\cite{Sch98}.

To conclude, our best estimate for the 
$S$-factor at 10 keV c.m.\ energy is therefore 
$(10.1\pm 0.6) \times 10^{-20}$ keV~b.

\subsubsection{Effect on the Super-Kamiokande Solar Neutrino Spectrum}

Super-Kamiokande (SK) detects solar neutrinos by neutrino-electron
scattering.  The energy is shared between the outgoing neutrino and
scattered electron, leading to a very weak correlation
between the incoming neutrino energy and the measured electron energy.
The electron angle relative to the solar direction is also measured,
which would in principle allow reconstruction of the incoming neutrino
energy.  However, the kinematic range of the angle is very forward,
and is comparable to the angular resolution of the detector.
Furthermore, event-by-event reconstruction of the neutrino energy
would be prevented by the detector background.  Above its threshold of
several MeV, SK is sensitive to the $^8$B electron neutrinos.  These
have a total flux of $5.15 \times 10^6$ cm$^{-2}$ s$^{-1}$ in the
SSM~\cite{BBP98}.  While the flux is uncertain
to about 15 \%, primarily due to the nuclear-physics uncertainties in
the $^7$Be($p$,$\gamma$)$^8$B cross section, the
spectral shape is more precisely known~\cite{8Bshape}.

\begin{figure}[bth]
\let\picnaturalsize=N
\def\picsize{5in}
\def\picfilenamea{ratio_col.eps}
\ifx\nopictures Y\else{\ifx\epsfloaded Y\else\input epsf \fi
\let\epsfloaded=Y
\centerline{
\ifx\picnaturalsize N\epsfxsize \picsize\fi \epsfbox{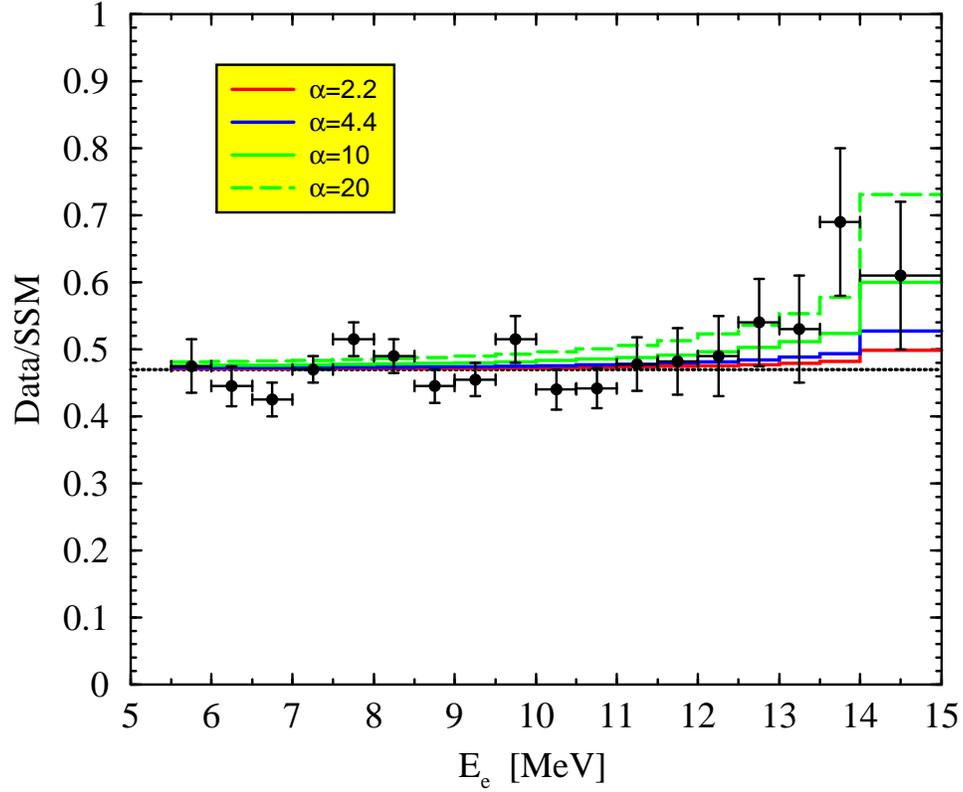}
 }}\fi
\caption{Electron energy spectrum for the ratio 
between the Super-Kamiokande 825-days data and the expectation 
based on unoscillated $^8$B neutrinos~\protect\cite{BBP98}. The data 
were extracted graphically from Fig. 8 of Ref.~\protect\cite{Suz00}. 
The 5 curves correspond respectively to no $hep$ contribution (dotted line), 
and an enhancement $\alpha$ of 2.2 (solid red line), 4.4 (solid blue line), 
10 (solid green line) and 20 (long-dashed green line).}
\label{fig:ratio}
\end{figure}

The SK results are presented as the ratio of the measured electron
spectrum to that expected in the SSM with no neutrino oscillations.
Over most of the spectrum, this ratio is constant at $\simeq 0.5$.  At
the highest energies, however, an excess relative to $0.5 \times$SSM
is seen (though it has diminished in successive data sets).  The SK
825-day data, determined graphically from Fig.~8 of Ref.~\cite{Suz00},
are shown by the points in Fig.~\ref{fig:ratio} (the error bars denote the
combined statistical and systematic error).  The excess above 12.5
MeV may be interpreted as neutrino-energy dependence in the neutrino
oscillation probability that is not completely washed out in the
electron spectrum.
This excess has also been interpreted as possible evidence for a large
$hep$ flux~\cite{BK98,Esc98,Suz00} (though note that the data never
exceeds the {\it full} SSM expectation from $^8$B neutrinos).  In the
SSM, the total $hep$ flux is very small, $2.10 \times 10^3$ cm$^{-2}$
s$^{-1}$.  However, its endpoint energy is higher than for the $^8$B
neutrinos, 19 MeV instead of about 14 MeV, so that the $hep$ neutrinos
may be seen at the highest energies.  This is somewhat complicated by
the energy resolution of SK, which allows $^8$B events beyond their
nominal endpoint.  The ratio of the $hep$ flux to its value in the SSM
(based on the $hep$ S-factor prediction of Ref.~\cite{Sch92}) will be
denoted by $\alpha$, defined as

\begin{equation}
\alpha \equiv \frac{S_{\rm new}}{S_{{\rm SSM}}} \times P_{\rm osc}\ ,
\label{eq:alphadef}
\end{equation}
where $P_{\rm osc}$ is the $hep$-neutrino suppression constant. 
In the present work, $\alpha = (10.1\times 10^{-20}{\rm\ keV~b})
/(2.3\times 10^{-20}{\rm\ keV~b}) = 4.4$, if 
$hep$ neutrino oscillations are ignored.
The solid lines in Fig.~\ref{fig:ratio} 
indicate the effect of
various values of $\alpha$ on the ratio of the electron spectrum with
both $^8$B and $hep$ to that with only $^8$B (the SSM).  Though some
differences are expected in the $hep$ spectral shape due to P-wave
contributions, here we simply use the standard $hep$ spectrum
shape~\cite{hepshape}.  In calculating this ratio, the $^8$B flux in
the numerator has been suppressed by 0.47, the best-fit constant value
for the observed suppression.  If the $hep$ neutrinos are suppressed
by $\simeq 0.5$, then $\alpha = 2.2$.  Two other arbitrary values
of $\alpha$ (10 and 20) are shown for comparison.  As for the SK data,
the results are shown as a function of the total electron energy in
0.5 MeV bins.  The last bin, shown covering 14 -- 15 MeV, actually
extends to 20 MeV.  The SK energy resolution was approximated by
convolution with a Gaussian of energy-dependent width, chosen to match
the SK LINAC calibration data~\cite{Nak99}.

The effects of a larger $hep$ flux should be compared to other
possible distortions of the ratio.  The data show no excess at low
energies, thus limiting the size of a neutrino magnetic moment
contribution to the scattering~\cite{Bea99}.  The $^8$B neutrino
energy spectrum has recently been remeasured by Ortiz {\it et
al.}~\cite{Ort00} and their spectrum is significantly larger at high
energies than that of Ref.~\cite{8Bshape}.  Relative to the standard
spectrum, this would cause an increase in the ratio at high energies
comparable to the $\alpha = 4.4$ case.  The measured electron spectrum
is very steep, and the fraction of events above 12.5 MeV is only $\sim
1\%$ of the total above threshold.  Thus, an error in either the
energy scale or resolution could cause an apparent excess of events at
high energy.  However, these are known precisely from the SK
LINAC~\cite{Nak99} calibration; an error in either could explain the
data only if it were at about the 3- or 4-sigma level~\cite{Suz00}.

The various neutrino oscillation solutions can be distinguished by
their neutrino-energy dependence, though the effects on the electron
spectrum are small.  Generally, the ratio is expected to be rising at
high energies, much like the effect of an increased $hep$ flux.  The
present work predicts $\alpha = 4.4$ (and $\alpha = 2.2$ if the $hep$
neutrinos oscillate).  From Fig.~\ref{fig:ratio}, 
this effect is smaller than the
distortion seen in the data or found in Refs.~\cite{BK98,Esc98,Suz00},
where the $hep$ flux was fitted as a free parameter.  However, the
much more important point is that this is an {\it absolute}
prediction.  Fixing the value of $\alpha$ will significantly improve
the ability of SK to identify the correct oscillation solution.

In the remainder of the paper we provide details of the calculation
leading to these conclusions.  In Sec.~II we derive the $hep$
cross section in terms of reduced matrix elements of the
weak current multipole operators.  In Sec.~III
we discuss the calculation of the bound- and scattering-state wave functions
with the CHH method, and summarize a number of results obtained for
the $^4$He binding energy and
$p\,^3$He elastic scattering observables, comparing them to experimental data.
In Sec.~IV we review the model for the nuclear weak current 
and charge operators, while
in Sec.~V we provide details about the calculation of
the matrix elements and resulting cross section.  
Finally, in Sec.~VI we summarize
and discuss our results.  
\section{Cross Section}
\label{sec:xsec}

In this section we sketch the derivation of the cross section
for the $p$$\,^3$He weak capture process.  The center-of-mass
(c.m.) energies of interest are of the order of 10 keV--the Gamow-peak
energy is 10.7 keV--and it is therefore convenient to expand the
$p$$\,^3$He scattering state into partial waves, and perform
a multipole decomposition of the nuclear weak charge and current
operators.  The present study includes S- and P-wave capture channels,
i.e. the $^1$S$_0$, $^3$S$_1$, $^3$P$_0$, $^1$P$_1$, $^3$P$_1$, and $^3$P$_2$
states in the notation 
$^{2S+1}$L$_J$ with $S=0,1$, and retains all contributing
multipoles connecting these states to the $J^\pi$=0$^+$ $^4$He ground
state.  The relevant formulas are given in the next three subsections.
Note that the $^1$P$_1$ and $^3$P$_1$, and $^3$S$_1$ and $^3$D$_1$ channels are
coupled.  For example, a pure $^1$P$_1$ incoming wave will produce
both $^1$P$_1$ and $^3$P$_1$ outgoing waves.  The degree of
mixing is significant, particularly for the P-waves, as discussed
in Sec.~\ref{sec:chhs}.
\subsection{The Transition Amplitude}
\label{sec:transampl}

The capture process $^{3}$He($p$,$e^{+}\nu_{e}$)$^{4}$He is induced by 
the weak interaction Hamiltonian~\cite{WAL95}
\begin{equation}
H_{W}=\frac{G_{V}}{\sqrt{2}}\int d{\bf{x}}\,{\rm{e}}^{-{\rm{i}}
({\bf{p}}_{e}+{\bf{p}}_{\nu})\cdot{\bf{x}}}\,l_{\sigma}\,j^{\sigma}({\bf{x}}) 
\ , \label{eq:hw}
\end{equation}
where $G_{V}$ is the Fermi coupling constant 
($G_{V}$=1.14939 10$^{-5}$ GeV$^{-2}$~\cite{Har90}), $l_{\sigma}$ is the
leptonic weak current 
\begin{equation}
l_{\sigma} = \overline{u}_{\nu}\gamma_{\sigma}(1-\gamma_5)v_{e}
\equiv (\>\overline{l}_0,-{\bf l}) \ ,
\label{eq:lmu}
\end{equation}
and $j^{\sigma}({\bf{x}})$ is the hadronic weak current density. 
The positron and (electron) neutrino momenta and spinors are 
denoted, respectively, by ${\bf{p}}_{e}$ and ${\bf{p}}_{\nu}$, and 
$v_{e}$ and ${u}_{\nu}$. The Bjorken and Drell~\cite{BJO64} conventions 
are used for the metric tensor
$g^{\sigma\tau}$ and $\gamma$-matrices. However, 
the spinors are normalized as
$v_{e}^{\dag}v_{e}={u}_{\nu}^{\dag}{u}_{\nu}=1$.

The transition amplitude in the c.m.\ frame is then 
given by
\begin{eqnarray}
\langle f|H_{W}|i\rangle&=&
\frac{G_{V}}{\sqrt{2}}\,l^{\sigma} \langle -{\bf{q}}; ^{4}\!{\rm{He}}|
j_{\sigma}^{\dag}({\bf{q}})|{\bf{p}}; p\,^{3}{\rm{He}}\rangle \ ,
\label{eq:tra}
\end{eqnarray}
where ${\bf{q}}={\bf{p}}_{e}+{\bf{p}}_{\nu}$,
$|{\bf{p}}; p\,^{3}{\rm{He}}\rangle$ and 
$|-{\bf{q}}; ^{4}\!{\rm{He}} \rangle$ represent the $p\,^{3}$He scattering 
state with relative momentum ${\bf{p}}$ and $^{4}$He bound state 
recoiling with momentum $-{\bf{q}}$, respectively, and 
\begin{equation}
j^{\sigma}({\bf{q}})=\int d{\bf{x}}\, 
{\rm{e}}^{{\rm{i}}{\bf{q}}\cdot{\bf{x}}} \,j^{\sigma}({\bf{x}})\equiv
(\rho({\bf{q}}),{\bf{j}}({\bf{q}})) \ .
\label{eq:jmuq}
\end{equation}
The dependence of the amplitude upon the spin-projections of the 
proton and $^{3}$He is understood. It is useful to 
perform a partial-wave expansion of the $p\,^{3}$He scattering wave function
\begin{equation}
\Psi^{(+)}_{{\bf{p}}, s_{1} s_{3}} =
\sqrt{4\pi}\sum_{LSJJ_z}\sqrt{2L+1}\,
{\rm{i}}^{L} \langle \frac{1}{2} s_1 , \frac{1}{2} s_3|S J_z\rangle 
\langle S J_z , L 0|J J_z\rangle 
\overline{\Psi}_{1+3}^{LSJJ_z} \ , 
\label{eq:psi+}
\end{equation}
with 
\begin{equation}
\overline{\Psi}_{1+3}^{LSJJ_z} = {\rm{e}}^{{\rm{i}}\sigma_{L}}
\sum_{L^{\prime}S^{\prime}} [1-{\rm{i}}\,R^J]^{-1}_{LS,
L^{\prime}S^{\prime}}
\Psi_{1+3}^{L^{\prime}S^{\prime}JJ_z} \ ,
\label{eq:psi13}
\end{equation}
where $s_1$ and $s_3$ are the proton and $^3$He spin projections, 
$L$, $S$, and $J$ are the relative orbital angular 
momentum, channel spin ($S$=0,1), and total angular 
momentum (${\bf{J}}={\bf{L}}+{\bf{S}}$), respectively, 
$R^J$ is the $R$-matrix in channel $J$, and $\sigma_{L}$ is 
the Coulomb phase shift, 
\begin{eqnarray}
\sigma_{L}&=&{\rm{arg}}[\Gamma(L+1+{\rm{i}}\eta)] \ , \label{eq:sigml} \\
\eta&=&\frac{2\alpha}{v_{\rm rel}} \ .  \label{eq:eta}
\end{eqnarray}
Here $\alpha$ is the fine-structure constant and $v_{\rm rel}$ is
the $p\,^{3}$He relative velocity, $v_{\rm rel}=p/\mu$, $\mu$
being the reduced mass, $\mu=m m_3/(m+m_3)$ ($m$ and $m_3$ are
the proton and $^3$He rest masses, respectively).
Note that $\Psi^{(+)}$ has been constructed 
to satisfy outgoing wave boundary conditions, and that the 
spin quantization axis has been chosen
to lie along $\hat{\bf{p}}$, which defines the $z$-axis.
Finally, the scattering wave function
$\Psi_{1+3}^{LSJJ_z}$ as well as the $^{4}$He
wave function $\Psi_{4}$ are obtained variationally with the 
correlated-hyperspherical-harmonics (CHH) method, as described in 
Sec.~\ref{sec:chh}.

The transition amplitude is then written as
\begin{eqnarray}
\langle f|H_{W}|i \rangle&=&\frac{G_{V}}{\sqrt{2}}
\sqrt{4\pi}\sum_{LSJJ_z}\sqrt{2L+1}\,{\rm{i}}^{L} 
\langle \frac{1}{2} s_1 , \frac{1}{2} s_3|S J_z\rangle 
\langle S J_z , L 0|J J_z\rangle \nonumber \\
&\times &\Bigg[\overline{l}_{0}\langle\Psi_{4}|\rho^{\dag}({\bf{q}})|
\overline{\Psi}_{1+3}^{LSJJ_z}\rangle-\sum_{\lambda=0,\pm 1} l_{\lambda}
\langle\Psi_{4}|{\hat{\bf{e}}}_{q\lambda}^{*}\cdot
{\bf{j}}^{\dag}({\bf{q}})|
\overline{\Psi}_{1+3}^{LSJJ_z}\rangle \Bigg] \ , \label{eq:fhwi2}
\end{eqnarray}
where, with the future aim of a multipole decomposition of the 
weak transition operators, the lepton vector ${\bf l}$ has been
expanded as
\begin{equation}
{\bf{l}}=\sum_{\lambda=0,\pm 1} l_{\lambda}{\hat{\bf{e}}}_{q\lambda}^{*} \ ,
\label{eq:ll}
\end{equation}
with $l_{\lambda}={\hat{\bf{e}}}_{q\lambda}\cdot{\bf{l}}\ ,$ and
\begin{eqnarray}
\hat{{\bf{e}}}_{q0}&\equiv&
{\hat{\bf e}}_{q3} \ , \label{eq:eq0} \\
\hat{{\bf{e}}}_{q\pm{1}}&\equiv&
\mp\frac{1}{\sqrt{2}}(\hat{{\bf{e}}}_{q1}
\pm\,{\rm{i}}\,\hat{{\bf{e}}}_{q2}) \ . \label{eq:eqpm1}
\end{eqnarray}
The orthonormal basis $\hat{\bf{e}}_{q1}$, $\hat{\bf{e}}_{q2}$, 
$\hat{\bf{e}}_{q3}$ is defined by 
$\hat{\bf{e}}_{q3}=\hat{\bf q}$,
$\hat{\bf{e}}_{q2}={\bf p} \times {\bf q}/|{\bf p} \times {\bf q}|$,
$\hat{\bf{e}}_{q1}= \hat{\bf{e}}_{q2} \times \hat{\bf{e}}_{q3}$.
\subsection{The Multipole Expansion}
\label{sec:multex}

Standard techniques~\cite{WAL95} can now be used to perform the 
multipole expansion of the weak charge and current matrix 
elements occurring in Eq.~(\ref{eq:fhwi2}).  The spin quantization axis
is along ${\hat{\bf p}}$ rather than along ${\hat{\bf q}}$.  
Thus, we first express
the states quantized along ${\hat{\bf{p}}}$
as linear combinations of those quantized along ${\hat{\bf{q}}}$:
\begin{equation}
|J\,J_{z}\rangle_{\hat{\bf{p}}}=\sum_{J_{z}^{\prime}}\,
{D}_{J_{z}^{\prime}J_{z}}^{J}(-\phi,\theta,\phi)
\,|J\,J_{z}^{\prime}\rangle_{\hat{\bf{q}}} \ , 
\label{eq:jjz}
\end{equation}
where ${D}_{J_{z}^{\prime}J_{z}}^{J}$
are standard rotation matrices~\cite{WAL95,EDM57} 
and the angles $\theta$ and $\phi$ specify the
direction ${\hat{\bf{q}}}$.  We then 
make use of the transformation properties under rotations 
of irreducible tensor operators to arrive at
the following expressions:
\begin{equation}
\langle\Psi_{4}\,|\,\rho^{\dag}({\bf{q}})\,|\,
{\overline{\Psi}}_{1+3}^{LSJJ_{z}}\rangle 
= \sqrt{4\pi} (-{\rm{i}})^{J} 
(-)^{J-J_z} D_{-J_{z},0}^{J}(-\phi,-\theta,\phi)\> 
C_{J}^{LSJ}(q) \ , \label{eq:c}
\end{equation}
\begin{equation}
\langle\Psi_{4}\,|\,{\hat{\bf{e}}}^{*}_{q0}\cdot
{\bf{j}}^{\dag}({\bf{q}})\,|\,
{\overline{\Psi}}_{1+3}^{LSJJ_{z}}\rangle 
=\sqrt{4\pi} (-{\rm{i}})^{J} 
(-)^{J-J_z} D_{-J_{z},0}^{J}(-\phi,-\theta,\phi) \>
L_{J}^{LSJ}(q) \ , \label{eq:l}
\end{equation}
\begin{eqnarray}
\langle\Psi_{4}\,|\,{\hat{\bf{e}}}^{*}_{q\lambda}\cdot
{\bf{j}}^{\dag}({\bf{q}})\,|\,
{\overline{\Psi}}_{1+3}^{LSJJ_{z}}\rangle
=&-&\sqrt{2\pi} (-{\rm{i}})^{J}
(-)^{J-J_z} D_{-J_{z},-\lambda}^{J}(-\phi,-\theta,\phi) \nonumber \\
&\times&\left[ \lambda M_{J}^{LSJ}(q) +E_{J}^{LSJ}(q)\right] \ .
\label{eq:me}
\end{eqnarray}
Here $\lambda = \pm 1$, and $C_{J}^{LSJ}$, $L_{J}^{LSJ}$, 
$E_{J}^{LSJ}$ and $M_{J}^{LSJ}$ denote the reduced matrix 
elements of the Coulomb $(C)$, longitudinal $(L)$, 
transverse electric $(E)$ and transverse magnetic $(M)$ 
multipole operators, explicitly given by~\cite{WAL95}
\begin{equation}
C_{l l_z}(q)=\int d{\bf{x}}\> \rho({\bf x})\>
j_{l}(qx)\,Y_{ll_z}({\hat{\bf{x}}}) \ , \label{eq:cdef} 
\end{equation}
\begin{equation}
L_{l l_z}(q)= \frac{{\rm{i}}}{q} \int d{\bf{x}}\>
{\bf j}({\bf x}) \cdot \nabla j_{l}(qx)\,
Y_{ll_z}({\hat{\bf{x}}}) \ , \label{eq:ldef}
\end{equation}
\begin{equation}
E_{l l_z}(q)= \frac{1}{q} \int d{\bf{x}}\>
{\bf j}({\bf x}) \cdot \nabla \times j_{l}(qx)\,{\bf Y}_{l l_z}^{l1}
\ , \label{eq:edef}
\end{equation}
\begin{equation}
M_{l l_z}(q)= \int d{\bf{x}}\> 
{\bf j}({\bf x}) \cdot j_{l}(qx)\, {\bf Y}_{l l_z}^{l1} \ , 
\label{eq:mdef}
\end{equation}
where ${\bf Y}_{l l_z}^{l1}$ are vector spherical harmonics.

Finally, it is useful to consider the transformation properties 
under parity of the multipole operators.  The weak
charge/current operators 
have components of both scalar/polar-vector (V) 
and pseudoscalar/axial-vector (A) character, and hence 
\begin{equation}
T_{l l_z}=T_{l l_z}({\rm V})+T_{l l_z}({\rm A}) \ ,
\label{eq:tav}
\end{equation}
where $T_{ll_z}$ is any of the multipole operators above.
Obviously, the parity of $l$th-pole V-operators is opposite of that of 
$l$th-pole A-operators.  The parity of Coulomb, longitudinal, and
electric $l$th-pole V-operators is $(-)^l$, while that of magnetic 
$l$th-pole V-operators is $(-)^{l+1}$.
\subsection{The Cross Section}
\label{sec:xs}

The cross section for the $^{3}$He($p$,$e^{+}\nu_{e}$)$^{4}$He reaction 
at a c.m.\ energy $E$ is given by
\begin{eqnarray}
\sigma(E)=\int&& 2\pi \, \delta\left (\Delta m  + E -
\frac{q^{2}}{2 m_{4}} - E_e
- E_\nu\right )\frac{1}{v_{\rm rel}} \nonumber \\
&&\times \frac{1}{4}\sum_{s_e s_\nu}\sum_{s_1 s_3} 
|\langle f\,|\,H_{W}\,|\,i\rangle|^{2} 
\frac{d{\bf{p}}_{e}}{(2\pi)^3} \frac{d{\bf{p}}_{\nu}}{(2\pi)^3} \ ,
\label{eq:xsc1}
\end{eqnarray}
where $\Delta m = m + m_3 - m_4 $ = 19.287 MeV
($m_4$ is the $^4$He rest mass), and $v_{\rm rel}$
is the $p\,^{3}$He relative velocity
defined above.  It is convenient to write:
\begin{equation}
\frac{1}{4}\sum_{s_e s_\nu}\sum_{s_1 s_3} 
|\langle f\,|\,H_{W}\,|\,i\rangle|^{2} = (2 \pi)^2\> G_V^2\> L_{\sigma \tau} \>
N^{\sigma\tau} ,
\label{eq:lwst}
\end{equation}
where the lepton tensor $L^{\sigma \tau}$ is defined as
\begin{eqnarray}
L^{\sigma \tau}\equiv \frac{1}{2} \sum_{s_e s_\nu}l^{\sigma}{l^{\tau}}^{*} 
&=&\frac{1}{2}{\rm tr}
\bigg[\gamma^{\sigma}(1-\gamma_5)\frac{({\not{p}}_{e}-m_{e})}{2E_{e}}
\gamma^{\tau}(1-\gamma_5)\frac{{\not{p}}_{\nu}}{2E_{\nu}}\bigg] \nonumber \\
&=&{\rm v}_{e}^{\sigma}{\rm v}_{\nu}^{\tau}+
{\rm v}_{\nu}^{\sigma}{\rm v}_{e}^{\tau}-g^{\sigma\tau}
{\rm v}_{e} \cdot {\rm v}_{\nu}+
{\rm{i}}\>\epsilon^{\sigma\alpha\tau\beta}
{\rm v}_{e,\alpha} {\rm v}_{\nu,\beta}\ , 
\label{eq:lalb}
\end{eqnarray}
with $\epsilon^{0123}=-1$,  ${\rm v}_e^\sigma=p_e^\sigma/E_e$ and
${\rm v}_\nu^\sigma=p_\nu^\sigma/E_\nu$. 
The nuclear tensor $N^{\sigma \tau}$ is defined as
\begin{equation}
N^{\sigma \tau} \equiv \sum_{s_1 s_3} W^\sigma({\bf q};s_1 s_3)
W^{\tau *}({\bf q}; s_1 s_3) \ ,
\label{nuclt}
\end{equation}
where
\begin{equation} 
W^{\sigma=0}({\bf q}; s_1 s_3)=\sum_{LSJ} X^{LSJ}_0(\hat{\bf q};s_1 s_3) 
C^{LSJ}_J(q) \ ,
\label{eq:w0}
\end{equation}
\begin{equation}
W^{\sigma=3}({\bf q}; s_1 s_3)=\sum_{LSJ} X^{LSJ}_0(\hat{\bf q};s_1 s_3) 
L^{LSJ}_J(q) \ ,
\label{eq:w3}
\end{equation}
\begin{equation}
W^{\sigma=\pm 1}({\bf q}; s_1 s_3)=
-\frac{1}{\sqrt{2}} \sum_{LSJ} X^{LSJ}_{\mp 1}(\hat{\bf q};s_1 s_3)
\left [\pm M^{LSJ}_J(q) + E^{LSJ}_J(q) \right ] \ .
\label{eq:w1}
\end{equation}
The dependence upon the direction $\hat{\bf q}$ and proton and 
$^3$He spin projections $s_1$ and
$s_3$ is contained in the functions $X_\lambda^{LSJ}$ given by
\begin{eqnarray}
X_\lambda^{LSJ}(\hat{\bf q};s_1 s_3) = \sum_{J_z}
\sqrt{2L+1}\,{\rm{i}}^{L} (-{\rm i})^J (-)^{J-J_z}
&&\langle \frac{1}{2} s_1 , \frac{1}{2} s_3|S J_z\rangle
\langle S J_z , L 0|J J_z\rangle \nonumber \\
&&\times D^J_{-J_z,\lambda}(-\phi,-\theta,\phi) \ ,
\label{nuclx}
\end{eqnarray}
with $\lambda=0, \pm 1$.  Note that the Cartesian components
of the lepton and nuclear tensors ($\sigma,\tau=1,2,3$) are
relative to the orthonormal basis $\hat{\bf e}_{q1}$,
$\hat{\bf e}_{q2}$, $\hat{\bf e}_{q3}$, defined at the end
of Sec.~\ref{sec:transampl}.

The expression for the nuclear tensor can be further simplified
by making use of the reduction formulas for the product
of rotation matrices~\cite{EDM57}.  In fact, it can easily
be shown that the dependence of $N^{\sigma \tau}$
upon the angle ${\rm cos}\, \theta =
\hat{\bf p} \cdot \hat{\bf q}$ can be expressed in terms
of Legendre polynomials $P_n({\rm cos}\, \theta)$ and
associated Legendre functions $P^m_n({\rm cos}\, \theta)$ with
$m=1,2$.  However, given the large number of channels
included in the present study (all S- and P-wave
capture states), the resulting equations for $N^{\sigma \tau}$
are not particularly illuminating, and will not be given here.
Indeed, the calculation of the cross section, Eq.~(\ref{eq:xsc1}),
is carried out numerically with the techniques discussed
in Sec.~\ref{sec:calxs}.

It is useful, though, to discuss the simple case in which 
only the contributions involving transitions
from the $^3$S$_1$ and $^3$P$_0$ capture states
are considered.  In the limit $q=0$, one then finds
\begin{equation}
\sigma(E) \simeq \frac{2}{\pi}\, \frac{G_V^2}{v_{\rm rel}}\,m_{e}^{5}
\, f_0(E)\, \left[ \left | L^{011}_1({\rm A})\right |^2
                  +\left | E^{011}_1({\rm A})\right |^2
                  +\left | C^{110}_0({\rm A})\right |^2 \right ] \ ,
\label{eq:simple}
\end{equation}
where $L^{011}_1({\rm A})$ and $E^{011}_1({\rm A})$ are
the longitudinal and transverse electric
axial current reduced matrix elements (from $^3$S$_1$ capture), and
$C^{110}_0({\rm A})$ is the Coulomb axial charge reduced matrix element
(from $^3$P$_0$ capture) at $q$=0.  Here the \lq\lq Fermi function\rq\rq 
$f_0(E)$ is defined as
\begin{equation}
f_0(E) = \int_1^{x_0} dx \, x \, \sqrt{ x^2-1} \, (x_0-x)^2 \ ,
\end{equation}
with $x_0=( \Delta m+E)/m_e$.  The expression in Eq.~(\ref{eq:simple})
can easily be related, {\it mutatis mutandis},
to that given in Ref.~\cite{Car91}.  

Although the $q$=0 approximation can appear to be adequate
for the $hep$ reaction, for which $q\leq 20$ MeV/c and 
$q R \simeq 0.14$ or less ($R$ being the $^4$He radius), the expression
for the cross section given in Eq.~(\ref{eq:simple}) is in fact
inaccurate.  To elaborate this point further, consider the
$^3$P$_0$ capture.  The long-wavelength forms of the
$C_0(q;{\rm A})$ and $L_0(q;{\rm A})$ multipoles, associated
with the axial charge and longitudinal component of the axial
current, are constant and linear in $q$, respectively, as can
be easily inferred from Eqs.~(\ref{eq:cdef})--(\ref{eq:ldef}). 
The corresponding reduced matrix elements are, to leading order
in $q$,

\begin{eqnarray}
C_0^{110}(q;{\rm A}) &\simeq& c_0 +\dots \ ,\\
L_0^{110}(q;{\rm A}) &\simeq& l_0\, q +\dots \ ,
\end{eqnarray}
where $c_0=C_0^{110}({\rm A})$ in the notation of
Eq.~(\ref{eq:simple}).  The $^3$P$_0$ capture cross section
can be written, in this limit, as

\begin{equation}
\sigma(E; ^3\!{\rm P}_0) \simeq \frac{2}{\pi}\, \frac{G_V^2}
{v_{\rm rel}}\,m_{e}^{5}
\, \left[ f_0(E)\,      \left | c_0 \right |^2 +
          f_1(E)\,m_e^2 \left | l_0 \right |^2 -
      2\, f_2(E)\,m_e \,  \Re(c_0^* l_0) \right ] \ .
\label{eq:simple3p0_2}
\end{equation}
When the full model for the nuclear axial
charge and current is considered,
the constants $c_0$ and $l_0$, at zero $p\,^3$He relative
energy, are calculated to be $c_0= {\rm i}\, 0.043$ fm$^{3/2}$
and $l_0= {\rm i}\, 0.197$ fm$^{5/2}$ (note that they are purely
imaginary at $E=0$).  The \lq\lq Fermi functions\rq\rq
$f_0(E)$, $f_1(E)$, and $f_2(E)$, that
arise after integration over the phase space, at $E=0$ have
the values $f_0(0)=2.54\times 10^{6}$, $f_1(0)=3.61\times 10^9$, 
and $f_2(0)=9.59\times 10^7$.  The zero energy $S$-factor obtained by
including only the term $c_0$ is 2.2$\times 10^{-20}$ keV~b.  However,
when both the $c_0$ and $l_0$ terms are retained, it becomes 
0.68$\times$10$^{-20}$ keV~b. 

In fact, this last value is still inaccurate:
when not only the leading, but also the 
next-to-leading order terms are considered
in the expansion of the multipoles in powers of $q$
(see Sec.~\ref{sec:calxs}), the $S$-factor
for $^3$P$_0$ capture increases 
to 0.82$\times$10$^{-20}$ keV~b, its fully converged
value.  The conclusion of this discussion is that use of the
long-wavelength approximation in the $hep$ reaction leads
to erroneous results. 

Similar considerations also apply to the case of $^3$S$_1$ capture:
at values of $q$ different from zero, the transition 
can be induced not only by the axial current via the 
$E_1({\rm A})$ and $L_1({\rm A})$ multipoles, but 
also by the axial charge and vector current via the $C_1({\rm A})$ and 
and $M_1({\rm V})$ multipoles.
While the contribution of $M_1({\rm V})$ is
much smaller than that of the leading 
$E_1({\rm A})$ and $L_1({\rm A})$, the
contribution of $C_1({\rm A})$ is relatively large, and its
interference with that of $L_1({\rm A})$ cannot be neglected.
This point is further discussed in Sec.~\ref{subsec:3s1}.

As a final remark, we note that the general expression
for the cross section in Eq.~(\ref{eq:xsc1}) as follows
from Eqs.~(\ref{eq:lwst})--(\ref{nuclx}) contains
interference terms among the reduced matrix elements
of multipole operators connecting different capture channels.
However, these interference contributions have been found to account for
less than 2 \% of the total $S$-factor at zero $p\,^3$He c.m.\ energy. 
\section{Bound- and Scattering-State Wave Functions}
\label{sec:chh}

The $^4$He bound-state and $p\,^{3}$He scattering-state
wave functions are obtained variationally with the
correlated-hyperspherical-harmonics (CHH) method from
realistic Hamiltonians consisting of the Argonne $v_{18}$ 
two-nucleon~\cite{WSS95}
and Urbana-IX three-nucleon~\cite{Pud95} interactions (the AV18/UIX model), or
the older Argonne $v_{14}$ two-nucleon~\cite{WSA84}
and Urbana-VIII three-nucleon~\cite{Wir91} interactions (the AV14/UVIII model).
The CHH method, as implemented in the calculations reported in the
present work, has been developed by Viviani, Kievsky, and Rosati in
Refs.~\cite{VKR95,VRK98,Kie93,Kie94}.  Here, it will be reviewed briefly
for completeness, and a summary of relevant results obtained for
the three- and four-nucleon bound-state properties,
and $p\,^{3}$He effective-range parameters will be presented.
\subsection{The CHH Method}
\label{sec:chh1}

In the CHH approach a four-nucleon wave function $\Psi$ is
expanded as
\begin{equation}
\Psi =\sum_{p} \Bigl[
          \psi_A({\bf x}_{Ap}, {\bf y}_{Ap},{\bf z}_{Ap})+
          \psi_B({\bf x}_{Bp}, {\bf y}_{Bp},{\bf z}_{Bp})\Bigr]\ ,
      \label{eq:4bwf}
\end{equation}
where the amplitudes $\psi_A$ and $\psi_B$ correspond, respectively,
to the partitions 3+1 and 2+2, and the index $p$ runs
over the even permutations of particles $ijkl$.  The dependence on the 
spin-isospin variables is understood. 
The overall antisymmetry of the wave function
$\Psi$ is ensured by requiring that both $\psi_A$ and $\psi_B$
change sign under the exchange $ i \rightleftharpoons j$.

The Jacobi variables corresponding to the partition 3+1
are defined as
\begin{eqnarray}
{\bf x}_{Ap}&=& {\bf r}_j-{\bf r}_i\ , \\
{\bf y}_{Ap}&=& \sqrt{4/3} ({\bf r}_k-{\bf R}_{ij} )\ ,  \\
{\bf z}_{Ap}&=& \sqrt{3/2} ({\bf r}_l- {\bf R}_{ijk}  ) \ ,
\label{eq:jaca}
\end{eqnarray}
while those corresponding to the partition 2+2 are defined
as
\begin{eqnarray}
{\bf x}_{Bp}&=& {\bf r}_j-{\bf r}_i \ , \\
{\bf y}_{Bp}&=& \sqrt{2} ( {\bf R}_{kl} - {\bf R}_{ij} ) \ ,\\
{\bf z}_{Bp}&=& {\bf r}_l- {\bf r}_k \ , 
\label{eq:jacb}
\end{eqnarray}
where ${\bf R}_{ij}$ (${\bf R}_{kl}$) and ${\bf R}_{ijk}$
denote the c.m.\ positions of particles $ij$
($kl$) and $ijk$, respectively.
In the $LS$-coupling scheme, the amplitudes $\psi_A$ and
$\psi_B$ are expanded as
\begin{equation}
 \psi_A({\bf x}_{Ap},{\bf y}_{Ap},{\bf z}_{Ap})=
 \sum_\alpha  F_{\alpha,p}\>
 \phi^A_\alpha(x_{Ap},y_{Ap},z_{Ap}) \>Y^A_{\alpha,p} \ ,
\label{eq:ampli1}
\end{equation}
\begin{equation}
 \psi_B({\bf x}_{Bp},{\bf y}_{Bp},{\bf z}_{Bp})=
 \sum_\alpha  F_{\alpha,p} \>
 \phi^B_\alpha(x_{Bp},y_{Bp},z_{Bp}) \> Y^B_{\alpha,p} \ ,
 \label{eq:ampli2} 
\end{equation}
where 
\begin{eqnarray}
Y^A_{\alpha,p}=
\biggl \{\Bigl [\bigl[ Y_{\ell_{1\alpha}}(\hat {\bf z}_{Ap})
Y_{\ell_{2\alpha}}(\hat {\bf y}_{Ap})\bigr]_{\ell_{12\alpha}}
Y_{\ell_{3\alpha}}(\hat {\bf x}_{Ap}) \Bigr ]_{L_\alpha}
&&\biggl [\Bigl[\bigl[ s_i s_j \bigr]_{S_{a\alpha}}
          s_k\Bigr]_{S_{b\alpha}}  s_l  \biggr]_{S_\alpha}
\biggr \}_{JJ_z} \nonumber \\
&&\times \biggl [\Bigl[\bigl[ t_i t_j \bigr]_{T_{a\alpha}}
  t_k\Bigr]_{T_{b\alpha}}
  t_l  \biggr]_{TT_z}\ ,    \label{eq:Y1}
\end{eqnarray}                                                           
\begin{eqnarray}
Y^B_{\alpha, p}=
\biggl \{\Bigl [\bigl[ Y_{\ell_{1\alpha}}(\hat {\bf z}_{Bp})
Y_{\ell_{2\alpha}}(\hat {\bf y}_{Bp})\bigr]_{\ell_{12\alpha}}
Y_{\ell_{3\alpha}}(\hat {\bf x}_{Bp}) \Bigr ]_{L_\alpha}
&&\biggl [\bigl[ s_i s_j \bigr]_{S_{a\alpha}}
\bigl[ s_k s_l \bigr]_{S_{b\alpha}}
\biggr]_{S_\alpha} \biggr \}_{JJ_z} \nonumber \\
&&\times \biggl [\bigl[ t_i t_j \bigr]_{T_{a\alpha}}
   \bigl[ t_k t_l \bigr]_{T_{b\alpha}}
   \biggr]_{TT_z} \ .   \label{eq:Y2}                                                
\end{eqnarray}
Here a channel $\alpha$ is specified by: orbital angular
momenta $\ell_{1\alpha}$, $\ell_{2\alpha}$, $\ell_{3\alpha}$,
$\ell_{12\alpha}$, and $L_\alpha$; spin angular momenta 
$S_{a\alpha}$, $S_{b\alpha}$, and $S_\alpha$; isospins
$T_{a\alpha}$ and $T_{b\alpha}$.  The total orbital and spin
angular momenta and cluster isospins are then coupled to the assigned
$JJ_z$ and $TT_z$.  

The correlation factors $F_{\alpha,p}$ consist of the
product of pair-correlation functions, that are obtained
from solutions of two-body Schr\"odinger-like equations,
as discussed in Ref.~\cite{VKR95}.  These correlation
factors take into account the strong state-dependent 
correlations induced by the nucleon-nucleon interaction, and
improve the behavior of the wave function at small interparticle
separations, thus accelerating the convergence of the calculated
quantities with respect to the number of required hyperspherical
harmonics basis functions, defined below.
      
The radial amplitudes $\phi^A_\alpha$ and $\phi^B_\alpha$
are further expanded as
\begin{eqnarray}
\phi^A_\alpha (x_{Ap},y_{Ap},z_{Ap})& = &
\sum_{n,m} { u^{\alpha}_{nm}(\rho) \over \rho^4}
z_{Ap}^{\ell_{1\alpha}} \>y_{Ap}^{\ell_{2\alpha}}
\>x_{Ap}^{\ell_{3\alpha}}
\>X_{nm}^{\alpha}(\phi^A_{2p},\phi_{3p}) \ , \label{eq:CHH1}  \\
 \noalign{\medskip}
\phi^B_\alpha (x_{Bp},y_{Bp},z_{Bp})& = &
\sum_{n,m} {w^{\alpha}_{nm}(\rho) \over \rho^4}
z_{Bp}^{\ell_{1\alpha}}\> y_{Bp}^{\ell_{2\alpha}}
\>x_{Bp}^{\ell_{3\alpha}}
\>X_{nm}^{\alpha}(\phi^B_{2p},\phi_{3p})\ , \label{eq:CHH2}
\end{eqnarray}
where the magnitudes of the Jacobi variables have been
replaced by the hyperspherical coordinates, i.e. the
hyperradius $\rho$
\begin{equation}
\rho=\sqrt{ x^2_{Ap}+ y^2_{Ap}+z^2_{Ap}}
    =\sqrt{ x^2_{Bp}+ y^2_{Bp}+z^2_{Bp}}\ ,
\label{eq:hypr}
\end{equation}
which is independent of the permutation $p$ considered, and
the hyperangles appropriate for partitions A and $B$.
The latter are given by 
\begin{eqnarray}
       \cos \phi_{3p}&=&  x_{Ap}/\rho =  x_{Bp}/\rho\ , \\
 \noalign{\medskip}
      \cos \phi^A_{2p}&=&  y_{Ap}/(\rho \sin\phi_{3p})  \ , \\
 \noalign{\medskip}
      \cos \phi^B_{2p}&=& y_{Bp}/(\rho \sin\phi_{3p} ) \ .
       \label{eq:hypzx}
\end{eqnarray}
Finally, the hyperangle functions $X^\alpha_{nm}$ consist
of the product of Jacobi polynomials
\begin{equation}
X_{nm}^{\alpha}(\beta,\gamma) = N_{nm}^{\alpha}\; (\sin\beta)^{2 m}
P_n^{K_{2\alpha},\ell_{3\alpha}+{1\over2}}(\cos2\beta)
P_m^{\ell_{1\alpha}+{1\over2},\ell_{2\alpha}
+{1\over2}}(\cos2\gamma)\ ,
\label{eq:Yfun}
\end{equation}
where the indices $m$ and $n$ run, in principle,
over all non-negative integers, $K_{2\alpha}=\ell_{1\alpha}+\ell_{2\alpha}
+2m+2$, and $N^\alpha_{nm}$ are normalization factors~\cite{VKR95}.

Once the expansions for the radial amplitudes $\phi^A$ and $\phi^B$
are inserted into Eqs.~(\ref{eq:ampli1})--(\ref{eq:ampli2}), the
wave function $\Psi$ can schematically be written as
\begin{equation}
\Psi = \sum_{\alpha n m} 
\Bigg [ {z^{\alpha ,A}_{nm}(\rho) \over \rho^4}
    Z^{\alpha ,A}_{nm}(\rho,\Omega) + 
        {z^{\alpha ,B}_{nm}(\rho) \over \rho^4}
    Z^{\alpha ,B}_{nm}(\rho,\Omega) \Bigg ] \ , \label{eq:4bwf2}
\end{equation}
where $z^{A}(\rho)\equiv u(\rho)$ and $z^{B}(\rho)\equiv w(\rho)$ are 
yet to be determined, 
and the factors $Z^{\alpha ,W}_{nm}$, with $W=A,B$, include the dependence
upon the hyperradius $\rho$ due to the correlation
functions, and the angles and hyperangles, denoted
collectively by $\Omega$, and are given by:
\begin{equation}
Z^{\alpha ,W}_{nm}(\rho,\Omega) = \sum_{p} F_{\alpha ,p}\, Y^{W}_{\alpha ,p}
\, z_{W,p}^{\ell_{1\alpha}}\, y_{W,p}^{\ell_{2\alpha}}\, x_{W,p}^{\ell_{3\alpha}}
\, X_{n,m}^{\alpha}(\phi_{2p}^{W},\phi_{3p}) \ .
\label{eq:zaw}
\end{equation}

The CHH method for three-nucleon systems has been
most recently reviewed in Ref.~\cite{Viv00}, and
will not be discussed here.  It leads, in
essence, to wave functions having the same structure as
in Eq.~(\ref{eq:4bwf2}) with suitably defined $Z(\rho,\Omega)$. 
\subsection{The $^3$He and $^4$He Wave Functions}
\label{sec:chhb}

The Rayleigh-Ritz variational principle
\begin{equation}
<\delta_z\Psi|H-E|\Psi>=0 \label{eq:rrvar}
\end{equation}
is used to determine the hyperradial functions $z^\alpha_{nm}(\rho)$ in
Eq.~(\ref{eq:4bwf2}) and bound state energy $E$.  Carrying out the
variations with respect to the functions $z^\alpha_{nm}$ leads to a set
of coupled second-order linear differential equations in the
variable $\rho$ which, after discretization, 
is converted into a generalized eigenvalue problem and solved 
by standard numerical techniques~\cite{VKR95}. 

The present status of $^3$He~\cite{KVR95} and $^4$He~\cite{VKR95,Viv99b}
binding energy calculations with the CHH method is summarized in
Tables~\ref{tb:be} and~\ref{tb:scl}.  The binding energies
calculated with the CHH method using the AV18 or AV18/UIX
Hamiltonian models are within 1.5 \% of
corresponding \lq\lq exact\rq\rq Green's function Monte Carlo
(GFMC) results~\cite{Pud97}, 
and of the experimental value (when the three-nucleon
interaction is included).  The agreement between
the CHH and GFMC results is less
satisfactory when the AV14 or AV14/UVIII models are considered, presumably
because of slower convergence of the CHH expansions for the
AV14 interaction.  This interaction has tensor components which
do not vanish at the origin.
\subsection{The $p\,^{3}$He Continuum Wave Functions}
\label{sec:chhs}

The $p\,^{3}$He cluster wave function $\Psi_{1+3}^{LSJJ_z}$, having
incoming orbital angular momentum $L$ and channel spin $S$ ($S=0, 1$)
coupled to total angular $JJ_z$, is expressed as 

\begin{equation}
\Psi_{1+3}^{LSJJ_z}=\Psi_C^{JJ_z}+\Psi_A^{LSJJ_z} \ ,
\label{eq:psica}
\end{equation}
where the term $\Psi_C$ vanishes in the
limit of large intercluster separations, and
hence describes the system in the region where the particles are close
to each other and their mutual interactions are strong.
The term $\Psi_A^{LSJJ_z}$ describes the system
in the asymptotic region, where intercluster interactions are negligible.
It is given explicitly as:

\begin{eqnarray}
\Psi_A^{LSJJ_z}=&&\frac{1}{\sqrt{4}}
\sum_i \sum_{L^\prime S^\prime} 
\left [ [ s_i \otimes \phi_3(jkl) ]_{S^\prime} \otimes
Y_{L^\prime}(\hat{\bf y}_i) \right ]_{JJ_z} \nonumber \\
&&\times \bigg[\delta_{L L^\prime} \delta_{S S^\prime}
\frac{F_{L^\prime}(py_i)}{py_i} + {R}^J_{LS,L^\prime S^\prime}(p)
\frac{G_{L^\prime}(py_i)}{py_i} g(y_i) \bigg] \ ,
\label{eq:psia}
\end{eqnarray}
where $y_i$ is the distance between the proton (particle $i$) and
$^3$He (particles $jkl$), $p$ is the magnitude of the relative momentum
between the two clusters, $\phi_3$ is the $^3$He wave function, and
$F_L$ and $G_L$ are the regular and irregular Coulomb functions, respectively.
The function $g(y_i)$ modifies the $G_L(py_i)$ at small $y_i$ by
regularizing it at the origin, and $g(y_i) \rightarrow 1$
as $y_i \gtrsim 10$ fm, thus
not affecting the asymptotic behavior of $\Psi_{1+3}^{LSJJ_z}$.
Finally, the real parameters ${R}^J_{LS,L^\prime S^\prime}(p)$ are the
$R$-matrix elements introduced in Eq.~(\ref{eq:psi13}),
which determine phase shifts and (for coupled
channels) mixing angles at the energy
$p^2/(2 \mu)$ ($\mu$ is $p\,^{3}$He reduced
mass).  Of course, the sum over $L^\prime$ and $S^\prime$ is over
all values compatible with a given $J$ and parity.

The \lq\lq core\rq\rq wave function $\Psi_C$ is expanded in the
same CHH basis as the bound-state wave function, and both
the matrix elements ${R}^J_{LS,L^\prime S^\prime}(p)$ and
functions $z^\alpha_{nm}(\rho)$ occurring in the expansion of
$\Psi_C$ are determined by making the functional
\begin{equation}
[{R}^J_{LS,L^\prime S^\prime}(p)]= {R}^J_{LS,L^\prime S^\prime}(p)
-\frac{m}{\sqrt{6}}
\langle\Psi^{L^\prime S^\prime JJ_z }_{1+3} |H-E_3-\frac{p^2}{2 \mu} |
\Psi^{LSJJ_z}_{1+3}\rangle \ , 
\label{eq:kohn}
\end{equation}
stationary with respect to variations in the ${R}^J_{LS,L^\prime S^\prime}$
and $z^\alpha_{nm}$ (Kohn variational principle).  Here $E_3=-7.72 $ MeV
is the $^3$He ground-state energy.  It is important to emphasize
that the CHH scheme, in contrast to Faddeev-Yakubovsky momentum space
methods, permits the straightforward inclusion of Coulomb distortion
effects in the $p\,^{3}$He channel.

The $p\,^3$He singlet and triplet scattering lengths predicted
by the Hamiltonian models considered in the present work
are listed in Table~\ref{tb:scl}, and are found in good agreement
with available experimental values, although these are rather
poorly known.  The experimental scattering lengths have been
obtained, in fact, from effective range parametrizations
of data taken above $1$ MeV, and therefore might have 
large systematic uncertainties. 

The most recent determination of phase-shift and mixing-angle parameters for
$p\,^3$He elastic scattering has been performed in Ref.~\cite{AK93}
by means of an energy-dependent phase-shift analysis (PSA), including
almost all data measured prior 1993 (for a listing of old PSAs, see
Ref.~\cite{AK93}).  New measurements are currently under way
at TUNL~\cite{K99} and Madison~\cite{GK99}.  At low
energies ($E < 4$ MeV) the process is dominated by scattering in
$L$=$0$ and $1$ waves, with a small contribution from $L$=$2$ waves.
Therefore, the important channels are:
$^1$S$_0$, $^3$P$_0$, $^3$S$_1$-$^3$D$_1$, $^1$P$_1$-$^3$P$_1$, $^3$P$_2$,
$^1$D$_2$-$^3$D$_2$ and $^3$D$_3$, ignoring channels with
$L>2$.  The general trend is the following:
(i) the energy dependence of the S-wave phase shifts
indicates that the $L$=0 channel interaction
between the $p$ and $^{3}$He is repulsive (mostly, due to
the Pauli principle), while that of the four
P-wave phase shifts ($^3$P$_0$, $^1$P$_1$, $^3$P$_1$, and $^3$P$_2$)
shows that in these channels there is a strong 
attraction.  Indeed, this fact has led to speculations
about the existence of four resonant states~\cite{A4}.
(ii) The D-wave phase shifts are rather tiny, even at $E > 2$ MeV.
(iii) The only mixing-angle 
parameter playing an important role at $E < 4$ MeV is
$\epsilon(J^\pi=1^-)$, in channel $^1$P$_1$-$^3$P$_1$.

\begin{figure}[bth]
\let\picnaturalsize=N
\def\picsize{5in}
\def\picfilenamea{xsu.eps}
\ifx\nopictures Y\else{\ifx\epsfloaded Y\else\input epsf \fi
\let\epsfloaded=Y
\centerline{
\ifx\picnaturalsize N\epsfxsize \picsize\fi \epsfbox{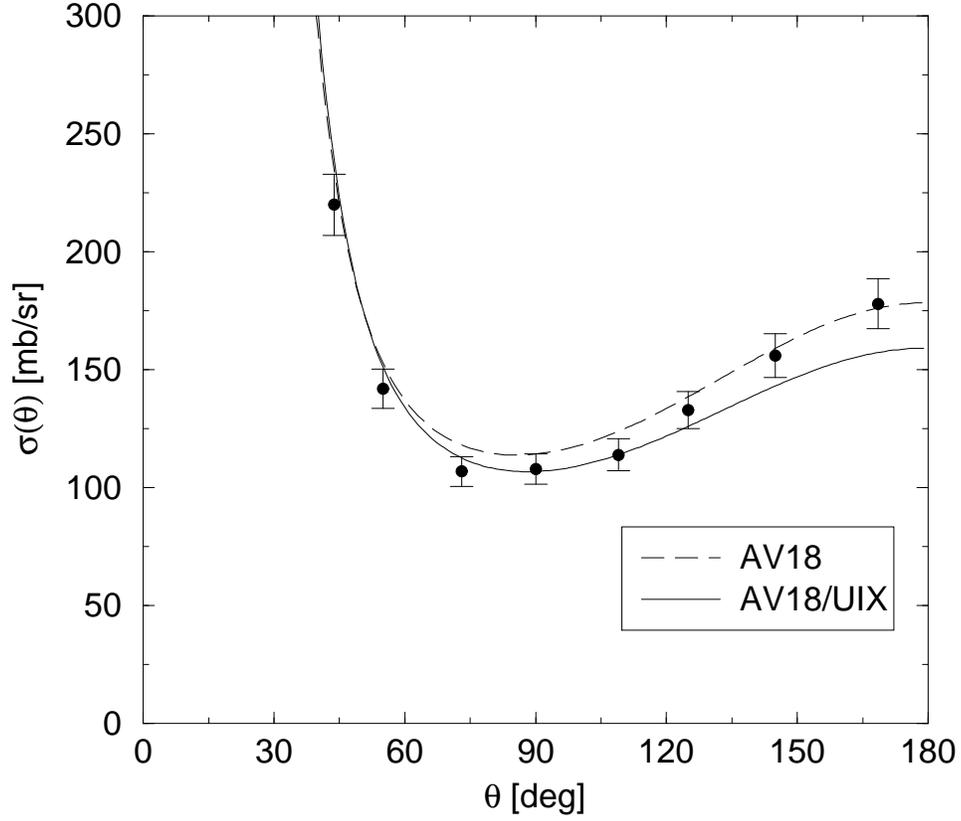}
 }}\fi
\caption{Differential cross section $\sigma(\theta)$ as function of the 
c.m.\ scattering angle $\theta$ at c.m.\ energy of $1.2$ MeV.
The experimental data are taken from Ref.~\protect\cite{Fea54}.  
The long-dashed and solid lines
correspond, respectively, to the CHH calculations with the AV18 and AV18/UIX
Hamiltonian models.}
\label{fig:012}
\end{figure}

Precise measurements have been taken at a c.m.\ energy of $1.2$ MeV, and
consist in differential cross section $\sigma(\theta)$~\cite{Fea54} and proton
analyzing power $A_y(\theta)$~\cite{GK99} data ($\theta$ is the
c.m.\ scattering angle).  The theoretical predictions for $\sigma(\theta)$, 
obtained from the AV18 and AV18/UIX interactions,
are compared with the corresponding experimental data in Fig.~\ref{fig:012}.
Inspection of the figure shows that the differential cross section calculated
with the AV18/UIX model is in excellent agreement with the data,
except at backward angles.  

By comparing, in Table~\ref{tb:ps}, the calculated phase-shift and
mixing-angle parameters with those extracted from the PSA~\cite{AK93} at
$E=1.2$ MeV, one observes a qualitative agreement, except for
the $^3$P$_1$ and $^3$P$_2$ phase shifts which are significantly
underestimated in the calculation.
The mixing-angle parameter $\epsilon(1^-)$ is found to be rather large,
$\simeq -14^\circ$, in qualitative agreement with
that obtained from the PSA (it is worth pointing out, however,
that in the PSA the mixing angle was
constrained to vanish at $E=0$, which may be unphysical).
The experimental error for each parameter quoted in Ref.~\cite{AK93}
is an average uncertainty over the whole
energy range considered, and it is therefore only indicative.
It would be very interesting to relate these discrepancies
to the $Nd$ $A_y$ puzzle and to specific deficiencies in the
nuclear interaction models.  A detailed study of $p\,^3$He elastic
scattering is currently underway and will published elsewhere~\cite{KRV00}.
\section{The Weak Charge and Current Operators}
\label{sec:weakc}

The nuclear weak charge and current operators have scalar/polar-vector (V)
and pseudoscalar/axial-vector (A) components
\begin{eqnarray}
\rho_\pm({\bf q})&=&\rho_\pm({\bf q};{\rm V})+\rho_\pm({\bf q};{\rm A}) \ ,  \\
{\bf j}_\pm({\bf q})&=&{\bf j}_\pm({\bf q};{\rm V})+
{\bf j}_\pm({\bf q};{\rm A}) \ ,
\end{eqnarray}
where ${\bf q}$ is the momentum transfer, ${\bf q}={\bf p}_e+{\bf p}_\nu$,
and the subscripts $\pm$ denote charge raising (+) or lowering (--) isospin
indices.  Each component, in turn, consists of one-, two-, and many-body
terms that operate on the nucleon degrees of freedom:
\begin{eqnarray}
\rho({\bf q};{\rm a})&=& \sum_i \rho^{(1)}_i({\bf q};{\rm a})
             +\sum_{i<j} \rho^{(2)}_{ij}({\bf q};{\rm a})+ \dots \ , 
\label{eq1}\\
{\bf j}({\bf q};{\rm a})&=& \sum_i {\bf j}^{(1)}_i({\bf q};{\rm a})
             +\sum_{i<j} {\bf j}^{(2)}_{ij}({\bf q};{\rm a})+ \dots 
\label{eq2} \ ,
\end{eqnarray}
where ${\rm a}$=V, A and the isospin indices have been suppressed
to simplify the notation.
The one-body operators $\rho^{(1)}_i$ and ${\bf j}^{(1)}_i$ have the
standard expressions obtained from a non-relativistic reduction of the
covariant single-nucleon V and A currents,
and are listed below for convenience.  The V-charge operator is written as
\begin{equation}
\rho^{(1)}_i({\bf q};{\rm V})= \rho^{(1)}_{i,{\rm NR}}({\bf q};{\rm V})+
                       \rho^{(1)}_{i,{\rm RC}}({\bf q};{\rm V}) \>\>, 
\label{eq6}
\end{equation}
with
\begin{equation}
\rho^{(1)}_{i,{\rm NR}}({\bf q};{\rm V})= \tau_{i,\pm} \>
 {\rm e}^{{\rm i}{\bf q}\cdot {\bf r}_i} \label{eq7} \ ,
\end{equation}
\begin{equation}
\rho^{(1)}_{i,{\rm RC}}({\bf q};{\rm V})= 
- {\rm i}\frac{\left ( 2\, \mu^v-1 \right )}{4 m^2} \tau_{i,\pm} \>
{\bf q} \cdot (\bbox{\sigma}_i \times {\bf p}_i) \>
{\rm e}^{ {\rm i} {\bf q} \cdot {\bf r}_i } \ .
\label{eq8}
\end{equation}
The V-current operator is expressed as
\begin{equation} \label{eq:jIA}
     {\bf j}^{(1)}_i({\bf q};{\rm V})={\frac {1} {2m}} \tau_{i,\pm} \>
   \bigg[ {\bf p}_i\>,\>{\rm e}^{{\rm i} {\bf q} \cdot {\bf r}_i} \bigg]_+
   -{\rm i}  \frac{\mu^v}{2m} \tau_{i,\pm}\>
     {\bf q} \times \bbox{\sigma}_i \> {\rm e}^{{\rm i} {\bf q} \cdot
   {\bf r}_i}  \label{eq9}\>\>\>,
\end{equation}
where $[ \cdots \, ,\, \cdots ]_+$ denotes the anticommutator,
${\bf p}$, $\bbox{\sigma}$, and $\bbox{\tau}$ are
the nucleon's momentum, Pauli spin
and isospin operators, respectively, and $\mu^v$ is
the isovector nucleon magnetic moment ($\mu^v=4.709$ n.m.).  Finally,
the isospin raising and lowering operators are defined as
\begin{equation}
\tau_{i,\pm} \equiv ( \tau_{i,x} \pm {\rm i} \, \tau_{i,y} )/2  \ .
\label{taupm}
\end{equation}
The term proportional to $1/m^2$ in $\rho^{(1)}_{i,{\rm RC}}({\bf q};{\rm V})$ 
is the well known~\cite{Def66,Fri73} spin-orbit relativistic correction.  
The vector charge and current operators above are
simply obtained from the corresponding isovector electromagnetic
operators by the replacement $\tau_{i,z}/2 \rightarrow \tau_{i,\pm}$,
in accordance with the conserved-vector-current (CVC) hypothesis.
The $q$-dependence of the nucleon's vector form factors
(and, in fact, also that of the axial-vector form factors below) has
been ignored, since the weak transition under consideration
here involves very small momentum transfers, $q \le 20$ MeV/c.
For this same reason, the Darwin-Foldy relativistic correction 
proportional to $q^2/(8\, m^2)$
in $\rho^{(1)}_{i,{\rm RC}}({\bf q};{\rm V})$ has also been neglected.
The A-charge operator is given, to leading order, by 
\begin{equation}
\rho_{i}^{(1)}({\bf q};{\rm A})=
-\frac{g_{A}}{2\,m}\,\tau_{i,\pm}\,{\bbox{\sigma}}
_{i}\cdot\bigg[ {\bf{p}}_{i}\> ,\> 
{\rm e}^{{\rm{i}}{\bf{q}}\cdot{\bf{r}}_{i}}\bigg ]_+ \ ,
\label{eq:rho1}
\end{equation}
while the A-current operator considered in the present work includes
leading and next-to-leading order corrections in an expansion in
powers of $p/m$, i.e.
\begin{equation}
{\bf j}^{(1)}_i({\bf q};{\rm A})= {\bf j}^{(1)}_{i,{\rm NR}}({\bf q};{\rm A})+
                       {\bf j}^{(1)}_{i,{\rm RC}}({\bf q};{\rm A}) \ , 
\label{ja1}
\end{equation}
with
\begin{equation}
{\bf j}_{i,{\rm NR}}^{(1)}({\bf q};{\rm A})=
-g_{A}\, \tau_{i,\pm} \,{\bbox{\sigma}}_{i}\,
{\rm e}^{{\rm{i}}{\bf{q}}\cdot{\bf{r}}_{i}} \ ,
\label{eq:1baj}
\end{equation}
\begin{eqnarray}
{\bf j}_{i,{\rm RC}}^{(1)}({\bf q};{\rm A})=&&  
\frac{g_{A}}{4m^2}\,\tau_{i,\pm}\,\Bigg(
{\bbox{\sigma}}_i \bigg[ {\bf p}^2_i\> ,\> 
{\rm e}^{{\rm{i}}{\bf{q}}\cdot{\bf{r}}_{i}}\bigg ]_+
-\bigg[ {\bbox{\sigma}}_i \cdot {\bf p}_i \> {\bf{p}}_{i}\> ,\> 
{\rm e}^{{\rm{i}}{\bf{q}}\cdot{\bf{r}}_{i}}\bigg ]_+
-\frac{1}{2}{\bbox{\sigma}}_i \cdot {\bf q}  \>\bigg[ {\bf{p}}_{i}\> ,\> 
{\rm e}^{{\rm{i}}{\bf{q}}\cdot{\bf{r}}_{i}}\bigg ]_+ \nonumber \\
&& -\frac{1}{2} {\bf q} \> \bigg[ {\bbox{\sigma}}_i \cdot {\bf{p}}_{i}\> ,\> 
{\rm e}^{{\rm{i}}{\bf{q}}\cdot{\bf{r}}_{i}}\bigg ]_+ 
+{\rm i}\> {\bf q} \times {\bf p}_i \>
{\rm e}^{{\rm{i}}{\bf{q}}\cdot{\bf{r}}_{i}} \Bigg) 
-\frac{g_P}{2\, m\, m_\mu}\, \tau_{i,\pm} {\bf q} \, 
{\bbox \sigma}_i \cdot {\bf q}
\, {\rm e}^{ {\rm i} {\bf q} \cdot {\bf r}_i }\ .
\label{eq:1arc}
\end{eqnarray}
The axial coupling constant $g_A$ is taken to be~\cite{Ade99}
1.2654$\pm$0.0042, by averaging values obtained, respectively,
from the beta asymmetry in the decay of polarized neutrons
(1.2626$\pm$0.0033~\cite{PDG96,Abe97}) and the half-lives of
the neutron and superallowed $0^+ \rightarrow 0^+$ transitions, i.e. 
$[ 2 ft(0^+ \rightarrow 0^+)/ft(n)-1 ]$=1.2681$\pm$0.0033~\cite{Ade99}.
The last term in Eq.~(\ref{eq:1arc}) is the induced pseudoscalar contribution 
($m_\mu$ is the muon mass), for which the coupling constant $g_P$ is taken
as~\cite{GT87} $g_P$=--6.78 $g_A$.
As already mentioned in Sec.~\ref{sec:intro}, in $^3$S$_1$ capture
matrix elements of ${\bf j}^{(1)}_{i,{\rm NR}}$ are suppressed.
Consequently, the relativistic terms included in ${\bf j}^{(1)}_{i,{\rm RC}}$,
which would otherwise contribute at the percent level, give in fact
a 20 \% contribution relative to that of the leading
${\bf j}^{(1)}_{i,{\rm NR}}$ at $q$=0.
Among these, one would naively expect the induced pseudoscalar
term to be dominant, due to the relatively large value
of $g_P$.  This is not the case, however, since matrix elements
of the induced pseudoscalar term scale with 
$g_P q^2/( 2 g_A m m_\mu)$ ($\leq 0.014$ in the $q$-range of
interest) relative to those
$\hat{\bf q} \cdot {\bf j}_{i,{\rm NR}}^{(1)}({\bf q};{\rm A})$.
Note that in the limit $q$=0, the expressions
for $\rho^{(1)}_{i,{\rm NR}}({\bf q};{\rm V})$ and
${\bf j}^{(1)}_{i,{\rm NR}}({\bf q};{\rm A})$ reduce to the
familiar Fermi and Gamow-Teller operators.

In the next five subsections we describe: i) the two-body
V-current and V-charge operators, required by the CVC hypothesis;
ii) the two-body A-current and A-charge operators due
to $\pi$- and $\rho$-meson exchanges, and the $\rho\pi$ mechanism;
iii) the V and A current and charge operators associated with excitation
of $\Delta$-isobar resonances, treated in perturbation theory and within the
transition-correlation-operator method.  Since the
expressions for these operators are scattered in a number of
papers~\cite{Sch92,Car91,Sch89,Sch90}, we collect them here for completeness.
\subsection{Two-Body Weak Vector Current Operators}
\label{sec:vcnt}

The weak vector (V) current and charge operators
are derived from the corresponding electromagnetic operators by
making use of the CVC hypothesis, which for two-body terms implies
\begin{equation}
\left [ \frac{1}{2}(\tau_{i,a}+\tau_{j,a})\> ,
 \> {\bf j}^{(2)}_{ij,z}({\bf q};\gamma) \right ] = {\rm i} \> \epsilon_{azb}\>
{\bf j}^{(2)}_{ij,b}({\bf q};{\rm V}) \ ,
\end{equation}
where ${\bf j}^{(2)}_{ij,z}({\bf q};\gamma)$
are the isovector (charge-conserving)
two-body electromagnetic currents, and $a,b=x,y,z$
are isospin Cartesian components.  A similar
relation holds between the electromagnetic charge operators
and its weak vector counterparts.  The charge-raising or
lowering weak vector current (or charge) operators are then simply obtained
from the linear combinations
\begin{equation}
{\bf j}^{(2)}_{ij,\pm}({\bf q};{\rm V})=
{\bf j}^{(2)}_{ij,x}({\bf q};{\rm V}) \pm {\rm i}
\> {\bf j}^{(2)}_{ij,y}({\bf q};{\rm V}) \ .
\end{equation}

The two-body electromagnetic currents
have \lq\lq model-independent\rq\rq (MI) and 
\lq\lq model-dependent\rq\rq (MD) components,
in the classification scheme of Riska~\cite{Ris89}.
The MI terms are obtained from the two-nucleon interaction, and
by construction satisfy current conservation with it~\cite{Sch89}.
Studies of the electromagnetic structure of $A$=2--6 nuclei,
such as, for example, the threshold electrodisintegration of the deuteron
at backward angles~\cite{SR91}, the magnetic form factors
of the trinucleons~\cite{Mar98}, the magnetic dipole transition
form factors in $^6$Li~\cite{WS98}, and finally the neutron
and proton radiative captures on hydrogen and helium
isotopes~\cite{Car90,SR91,VSK96}--properties in which
the isovector two-body currents play a large role and
are, in fact, essential for the satisfactory description
of the experimental data--have shown that the leading
operator is the (isovector) \lq\lq $\pi$-like\rq\rq current obtained
from the isospin-dependent spin-spin and tensor interactions.
The latter also generate an isovector \lq\lq $\rho$-like\rq\rq current.
There are additional MI isovector currents, which arise from the
central and momentum-dependent interactions, but these
are short-ranged and have been found to be numerically
far less important than the $\pi$-like current~\cite{Sch89,SR91}.  Their
contributions are neglected in the present study.

Use of the CVC relation leads to the $\pi$-like and $\rho$-like
weak vector currents below:
\begin{eqnarray}
   {\bf j}^{(2)}_{ij}({\bf k}_i,{\bf k}_j;\pi{\rm V}) &= & {\rm i}\,
   (\bbox{\tau}_i \times \bbox{\tau}_j)_\pm\, \bigg[ v_{PS}(k_j) 
\bbox{\sigma}_i (\bbox{\sigma}_j \cdot
    {\bf k}_j) - v_{PS}(k_i) \bbox{\sigma}_j (\bbox{\sigma}_i \cdot {\bf k}_i) 
    \nonumber \\
\noalign{\medskip}
   && \qquad + { {\bf k}_i - {\bf k}_j \over k_i^2 -k_j^2 } \bigl [ 
       v_{PS}(k_i)- v_{PS}(k_j) \bigr ] (\bbox{\sigma}_i \cdot    {\bf k}_i)
      (\bbox{\sigma}_j \cdot {\bf k}_j) \bigg] \ , \label{eq34}
\end{eqnarray}
\begin{eqnarray}
   {\bf j}^{(2)}_{ij}({\bf k}_i,{\bf k}_j;\rho{\rm V}) &= & -{\rm i}\,
   (\bbox{\tau}_i \times \bbox{\tau}_j)_\pm\, \bigg[ v_V(k_j) 
\bbox{\sigma}_i \times
(\bbox{\sigma}_j \times {\bf k}_j) - v_V(k_i) \bbox{\sigma}_j \times 
(\bbox{\sigma}_i \times {\bf k}_i) \nonumber \\
\noalign{\medskip}
   && \qquad - { v_V (k_i) - v_V (k_j) \over k_i^2 -k_j^2 } \bigl [ 
       ({\bf k}_i-{\bf k}_j)(\bbox{\sigma}_i \times {\bf k}_i)\cdot
                            (\bbox{\sigma}_j \times {\bf k}_j) \nonumber \\
\noalign{\medskip}
&&\qquad +(\bbox{\sigma}_i \times {\bf k}_i)\> \bbox{\sigma}_j \cdot ({\bf 
k}_i\times {\bf k}_j)
 +(\bbox{\sigma}_j \times {\bf k}_j)\> \bbox{\sigma}_i \cdot ({\bf k}_i\times 
{\bf k}_j)
\bigr ] \nonumber \\
\noalign{\medskip}
&&\qquad +{ {\bf k}_i - {\bf k}_j \over k_i^2 -k_j^2 }
[ v_{VS}(k_i)- v_{VS}(k_j) ] \bigg] \ , \label{eq35}
\end{eqnarray}
where ${\bf k}_i$ and ${\bf k}_j$ are the 
momenta delivered to nucleons $i$ and $j$ with
${\bf q}= {\bf k}_i + {\bf k}_j$, the isospin
operators are defined as
\begin{equation}
(\bbox{\tau}_i \times \bbox{\tau}_j)_\pm \equiv
(\bbox{\tau}_i \times \bbox{\tau}_j)_x \pm {\rm i} \>
(\bbox{\tau}_i \times \bbox{\tau}_j)_y \ ,
\end{equation}
and $v_{PS}(k)$, $v_V(k)$, and $v_{VS}(k)$
are given by
\begin{eqnarray}
   v_{PS}(k)&=& v^{\sigma \tau}(k)-
2 \> v^{t\tau}(k) \>\>\>, \label{eq36} \\
   v_{V}(k)&=&  v^{\sigma \tau}(k) + 
v^{t\tau}(k)  \>\>\>,  \label{eq37} \\
   v_{VS}(k)&=& v^{\tau}(k) \>\>\>, \label{eq38}
\end{eqnarray}
with 
\begin{eqnarray}
v^\tau(k)&=&4\pi \int_0^\infty r^2 dr\, j_0(kr)v^\tau(r) \label{eq39}\>\>, \\
v^{\sigma \tau}(k)&=&\frac{4\pi}{k^2} \int_0^\infty r^2 dr \, 
\left [ j_0(kr)-1 \right ] v^{\sigma \tau}(r) \label{eq40}\>\>, \\
v^{t \tau}(k)&=&\frac{4\pi}{k^2} 
\int_0^\infty r^2 dr\, j_2(kr)v^{t\tau}(r) \label{eq41} \>\>.
\end{eqnarray}
Here $v^{\tau}(r)$, $v^{\sigma \tau}(r)$, $v^{t\tau}(r)$ are the
isospin-dependent central, spin-spin, and tensor components
of the two-nucleon interaction (either the AV14 or AV18
in the present study).  The factor $j_0(kr)-1$ in the
expression for $v^{\sigma \tau}(k)$ ensures that its volume integral 
vanishes.  Configuration-space expressions are obtained from
\begin{equation}
{\bf j}^{(2)}_{ij}({\bf q};{\rm a})=\int d{\bf x}\>
{\rm e}^{ {\rm i} {\bf q} \cdot {\bf x} }
\int \frac { d{\bf k}_i } {(2\pi)^3} \frac { d{\bf k}_j } {(2\pi)^3} 
{\rm e}^{ {\rm i} {\bf k}_i \cdot ({\bf r}_i-{\bf x}) }
{\rm e}^{ {\rm i} {\bf k}_j \cdot ({\bf r}_j-{\bf x}) } 
{\bf j}^{(2)}_{ij}({\bf k}_i,{\bf k}_j;{\rm a})\ , \label{jrs}
\end{equation}
where ${\rm a}$=$\pi$V or $\rho$V.  Techniques to carry out
the Fourier transforms above are discussed in Ref.~\cite{Sch89}.

In a one-boson-exchange (OBE) model, in which
the isospin-dependent central, spin-spin, and tensor interactions
are due to $\pi$- and $\rho$-meson exchanges, the functions
$v_{PS}(k)$, $v_V(k)$, and $v_{VS}(k)$ are simply given by 
\begin{eqnarray}
v_{PS}(k)&\rightarrow&v_\pi(k) \equiv -{\frac {f_\pi^2} {m_\pi^2} }
{\frac {f^2_\pi(k)} { k^2+m_\pi^2} } \ , \label{eq26} \\
v_V(k) &\rightarrow&v_\rho(k) \equiv  -{\frac {g^2_\rho
(1+\kappa_\rho)^2}{4 m^2} } {\frac {f^2_\rho(k)} { k^2+
m_\rho^2} } \label{eq27} \ , \\
v_{VS}(k) &\rightarrow&v_{\rho S} \equiv g^2_\rho \frac {f^2_\rho(k)} 
{ k^2+m_\rho^2}
\ ,\label{eq28}
\end{eqnarray}
where $m_\pi$ and $m_\rho$ are the meson masses, $f_\pi$, $g_\rho$
and $\kappa_\rho$ are the pseudovector $\pi$$N$$N$, vector
and tensor $\rho$$N$$N$ coupling constants, respectively,
$f_\pi(k)$ and $f_\rho(k)$ denote
$\pi$$N$$N$ and $\rho$$N$$N$ monopole form factors, i.e.
\begin{equation}
f_\alpha(k) = \frac{ \Lambda_\alpha^2 - m_\alpha^2}
{\Lambda_\alpha^2+k^2} \ ,
\label{eq:mono}
\end{equation}
with $\alpha$=$\pi$ or $\rho$.  For example, in the CD-Bonn
OBE model~\cite{Mac96} the values for the couplings and cutoff masses are: 
$f_\pi^2/4\pi=0.075$, $g_\rho^2/4 \pi=0.84$, $k_\rho=6.1$,
$\Lambda_\pi=1.7$ GeV/c, and $\Lambda_\rho=1.31$ GeV/c.  Even though
the AV14 and AV18 are not OBE models, the functions $v_{PS}(k)$
and, to a less extent, $v_V(k)$ and $v_{VS}(k)$ projected out from
their $v^\tau$, $v^{\sigma \tau}$, and $v^{t\tau}$ components are quite
similar to those of $\pi$- and $\rho$-meson exchanges in
Eqs.~(\ref{eq26})--(\ref{eq28}) (with cutoff masses of order 1 GeV/c), as shown
in Refs.~\cite{Sch89,VSK96}.

Among the MD (purely transverse) isovector currents, those
due to excitation of $\Delta$ isobars have been
found to be the most important, particularly at low momentum
transfers, in studies of electromagnetic structure~\cite{Mar98}
and reactions~\cite{Sch92} of few-nucleon systems.  Their
contribution, however, is still relatively small when compared
to that of the leading $\pi$-like current.  Discussion of the weak
vector currents associated with $\Delta$ degrees of freedom is
deferred to Sec.~\ref{sec:vadlt}.
\subsection{Two-Body Weak Vector Charge Operators}
\label{sec:vcrg}

While the main parts of the two-body electromagnetic or weak vector
current are linked to the form of the nucleon-nucleon
interaction through the continuity equation, the most important two-body
electromagnetic or weak vector charge operators are
model dependent, and should be viewed as relativistic
corrections.  Indeed, a consistent calculation
of two-body charge effects in nuclei would
require the inclusion of relativistic effects
in both the interaction models and nuclear wave functions.
Such a program is yet to be carried out, at least
for systems with $A\geq 3$.

There are nevertheless rather clear
indications for the relevance of two-body
electromagnetic charge operators from the failure
of the impulse approximation in
predicting the deuteron tensor polarization
observable~\cite{T2098}, and charge form factors of the three- and
four-nucleon systems~\cite{Mar98,Car98}.
The model commonly used~\cite{Sch90} includes the $\pi$-, $\rho$-, and
$\omega$-meson exchange charge operators with both isoscalar and isovector
components, as well as the (isoscalar) $\rho \pi \gamma$ and (isovector)
$\omega \pi \gamma$ charge transition couplings (in addition to the
single-nucleon Darwin-Foldy and spin-orbit relativistic corrections).
The $\pi$- and $\rho$-meson exchange charge operators are constructed from the
isospin-dependent spin-spin and tensor interactions, using the same
prescription adopted for the corresponding current operators~\cite{Sch90}.
At moderate values of momentum transfer ($q \!< \! 5$ fm$^{-1}$), the
contribution due to the \lq\lq $\pi$-like\rq\rq 
exchange charge operator has been found to
be typically an order of magnitude larger
than that of any of the remaining two-body mechanisms
and one-body relativistic corrections~\cite{Mar98}.  

In the present study we
retain, in addition to the one-body operator
of Eq.~(\ref{eq6}), only the \lq\lq $\pi$-like\rq\rq and 
\lq\lq $\rho$-like\rq\rq weak vector charge operators.  In
the notation of the previous subsection, these are given by
\begin{equation}
\rho_{ij}^{(2)}({\bf k}_i,{\bf k}_j;\pi{\rm V}) = -\frac{1}{m} 
\Bigg [ \tau_{j,\pm}\> v_{PS}(k_j)\> \bbox{\sigma}_i \cdot {\bf q}
\, \bbox{\sigma}_j \cdot {\bf k}_j  
+\tau_{i,\pm}\> v_{PS}(k_i)\> \bbox{\sigma}_i \cdot {\bf k}_i
\, \bbox{\sigma}_j \cdot {\bf q} 
\Bigg ] \ , \label{eq68}
\end{equation}
\begin{eqnarray}
\rho_{ij}^{(2)}({\bf k}_i,{\bf k}_j;\rho{\rm V}) = -\frac {1}{m} 
\Bigg [&& \tau_{j,\pm}\> v_V(k_j) ( \bbox{\sigma}_i \times {\bf q})
\cdot (\bbox{\sigma}_j \times {\bf k}_j) \nonumber \\
+&&\tau_{i,\pm}\> v_V(k_i) (\bbox{\sigma}_j \times {\bf q} )
\cdot (\bbox{\sigma}_i \times {\bf k}_i) \Bigg ] \ ,
\label{eq69}
\end{eqnarray}
where non-local terms from retardation effects in
the meson propagators or from direct couplings
to the exchanged mesons have been neglected~\cite{Fri77,Sch96}.  In the
$\rho_{ij}^{(2)}({\bf k}_i,{\bf k}_j;\rho{\rm V})$ operator terms
proportional to powers of $1/(1+\kappa_\rho)$, because of the large
$\rho$-meson tensor coupling ($\kappa_\rho \simeq$ 6--7),
have also been neglected.  Indeed, these terms have been
ignored also in most studies of nuclear charge form factors.
\subsection{Two-Body Weak Axial Current Operators}
\label{sec:acnt}

In contrast to the electromagnetic case, the axial current operator is not
conserved.  Its two-body components cannot be linked to the nucleon-nucleon
interaction and, in this sense, should be viewed as model dependent.
Among the two-body axial current operators, the
leading term is that associated with
excitation of $\Delta$-isobar resonances.  We again defer its 
discussion to Sec.~\ref{sec:vadlt}.  In the present section we list the
two-body axial current operators due to $\pi$- and $\rho$-meson
exchanges (the $\pi$A and $\rho$A currents, respectively),
and the $\rho\pi$-transition mechanism (the $\rho\pi$A current).  Their
individual contributions have been found numerically far less important
than those from $\Delta$-excitation currents in studies
of weak transitions involving light nuclei~\cite{Car91,Sch98,SW99}.
These studies~\cite{Car91,Sch98} have also found that the 
$\pi$A and $\rho$A current
contributions interfere destructively, making their combined contribution
almost entirely negligible.  These conclusions are confirmed
in the present work.

The $\pi$A, $\rho$A, and $\rho\pi$A current operators
were first described in a systematic way by
Chemtob and Rho~\cite{CR71}.  Their derivation has been given in a 
number of articles, including the original reference mentioned
above and the more recent review by Towner~\cite{Tow87}.  Their
momentum-space expressions are given by
\begin{eqnarray}
{\bf j}^{(2)}_{ij}({\bf k}_i,{\bf k}_j;\pi{\rm A}) \, =&-&{g_A \over 2\, m} \, 
 (\bbox{\tau}_i \times \bbox{\tau}_j)_\pm \, v_\pi(k_j) \,
 \bbox{\sigma}_i \times {\bf k}_j \, \bbox{\sigma}_j \cdot {\bf k}_j 
\nonumber \\ 
&+&{g_A \over m} \, 
\tau_{j,\pm} \, v_\pi(k_j) \, \left ( {\bf q}  + {\rm i}\, \bbox{\sigma}_i
 \times {\bf P}_i \right )\,  \bbox{\sigma}_j \cdot {\bf k}_j 
 + \, i \rightleftharpoons j \ ,
\label{eq:a2pi}
\end{eqnarray}
\begin{eqnarray}
{\bf j}^{(2)}_{ij}({\bf k}_i,{\bf k}_j;\rho{\rm A})=&&\frac{g_A}{2\, m}
(\bbox{\tau}_i \times \bbox{\tau}_j)_\pm \, v_\rho(k_j)\, 
\Big [\, {\bf q} \> \bbox{\sigma}_i \cdot ( \bbox{\sigma}_j
\times {\bf k}_j)+ {\rm i} (\bbox{\sigma}_j \times {\bf k}_j) 
\times {\bf P}_i \nonumber \\
&&\qquad \qquad \qquad \qquad \qquad \qquad \qquad \qquad
 - \lbrack \bbox{\sigma}_i \times (\bbox{\sigma}_j \times {\bf k}_j)
\rbrack \times {\bf k}_j \Big ] \nonumber \\
&&+ \frac{g_A}{m} \tau_{j,\pm}\, v_\rho(k_j) \, 
\Big [ (\bbox{\sigma}_j \times  {\bf k}_j) \times {\bf k}_j - {\rm i}
\lbrack \bbox{\sigma}_i \times (\bbox{\sigma}_j \times {\bf k}_j ) \rbrack 
\times {\bf P}_i \Big ] + i \rightleftharpoons j  \ ,
\label{eq:a2ro}
\end{eqnarray}
\begin{eqnarray}
{\bf j}^{(2)}_{ij}({\bf k}_i,{\bf k}_j;\rho\pi{\rm A}) =
 &&- {g_A\over m} \, g_\rho^2 \,  
(\bbox{\tau}_i \times \bbox{\tau}_j)_\pm \,  
{f_\rho(k_i) \over k_i^2 + m_\rho^2} { f_\pi(k_j) \over k_j^2 + m_\pi^2 } \,
 \bbox{\sigma}_j \cdot {\bf k}_j \nonumber \\ 
&&\times \Big [ (1 + \kappa_\rho )\, \bbox{\sigma}_i \times {\bf k}_i  
- {\rm i} {\bf P}_i \Big ] 
+ i \rightleftharpoons j \ ,
\label{eq:a2rp}
\end{eqnarray}
where ${\bf P}_i={\bf p}_i+{\bf p}_i^\prime$ is the sum of the 
initial and final momenta
of nucleon $i$, respectively ${\bf p}_i$ and ${\bf p}^\prime_i$, and
the functions $v_\pi(k)$ and $v_\rho(k)$ have already been defined
in Eqs.~(\ref{eq26})--(\ref{eq27}).  Configuration-space expressions are
obtained by carrying out the Fourier transforms in Eq.~(\ref{jrs}).
The values used for the $\pi$$N$$N$
and $\rho$$N$$N$ coupling constants and cutoff masses are the following:
$f_\pi^2/4 \pi=0.075$, $g^2_\rho/4 \pi= 0.5$,
$\kappa_\rho =6.6$, $\Lambda_\pi=4.8$ fm$^{-1}$, and
$\Lambda_\rho=6.8$ fm$^{-1}$.  The $\rho$-meson coupling
constants are taken from the older Bonn OBE model~\cite{Mac89}, rather
than from the more recent CD-Bonn interaction~\cite{Mac96}
($g_\rho^2/4 \pi =0.81$ and $\kappa_\rho=6.1$).  This uncertainty
has in fact essentially no impact on the results reported
in the present work for two reasons.  Firstly, the contribution
from ${\bf j}^{(2)}(\rho{\rm A})$, as already mentioned above,
is very small.  Secondly, the complete two-body
axial current model, including the currents due to $\Delta$-excitation 
discussed
below, is constrained to reproduce the Gamow-Teller matrix
element in tritium $\beta$-decay by appropriately tuning the value of the
$N$$\Delta$-transition axial coupling $g_A^*$.  Hence changes in $g_\rho$ and
$\kappa_\rho$ only require a slight readjustament of the $g_A^*$ value.

Finally, note that the replacements $v_\pi(k) \rightarrow v_{PS}(k)$
and $v_\rho(k) \rightarrow v_{V}(k)$ could have been made
in the expressions for ${\bf j}^{(2)}(\pi{\rm A})$ and
${\bf j}^{(2)}(\rho{\rm A})$ above, thus eliminating the need
for the inclusion of {\it ad hoc} form factors.  While this
procedure would have been more satisfactory, since it constrains
the short-range behavior of these currents in a way consistent
with that of the two-nucleon interaction, its impact
on the present calculations would still be marginal for
the same reasons given above.

\subsection{Two-Body Weak Axial Charge Operators}
\label{sec:acrg}

The model for the weak axial charge operator adopted here includes
a term of pion-range as well as short-range terms associated with scalar-
and vector-meson exchanges~\cite{Kir92}.  The experimental evidence for the
presence of these two-body axial charge mechanisms rests on
studies of $0^+ \rightleftharpoons 0^-$ weak transitions,
such as the processes $^{16}$N(0$^-$,120 keV)$\rightarrow$$^{16}$O(0$^+$)
and $^{16}$O(0$^+$)+$\mu^-$$\rightarrow$$^{16}$N(0$^-$,120 keV)+$\nu_\mu$,
and first-forbidden $\beta$-decays in the lead region~\cite{War91}.
Shell-model calculations of these transitions suggest that the effective
axial charge coupling of a bound nucleon may be enhanced by roughly
a factor of two over its free nucleon value.  There are rather
strong indications that such an enhancement can be explained by two-body
axial charge contributions~\cite{Kir92}. 

The pion-range operator is taken as
\begin{equation}
\rho_{ij}^{(2)}({\bf k}_i,{\bf k}_j;\pi{\rm A})=
-{\rm i} \frac{g_A}{4 \,\overline{f}_\pi^2}\,
({\bbox \tau}_i \times{\bbox \tau}_j)_\pm \,
\frac{f_\pi^2(k_i) }{ k_i^2+ m_\pi^2 }
\, {\bbox \sigma}_i \cdot{\bf k}_i + i \rightleftharpoons j\ ,
\label{eq:rho2pi}
\end{equation}
where $\overline{f}_\pi$ is the pion decay constant 
($\overline{f}_\pi$=93 MeV),
${\bf k}_i$ is the momentum transfer to nucleon
$i$, and $f_\pi(k)$ is the monopole form factor of Eq.~(\ref{eq:mono})
with $\Lambda_\pi$=4.8 fm$^{-1}$.  The structure and
overall strength of this operator are determined by soft pion
theorem and current algebra arguments~\cite{Kub78,Tow92}, and should therefore
be viewed as \lq\lq model independent\rq\rq.  It can also be
derived, however, by considering nucleon-antinucleon pair contributions
with pseudoscalar $\pi$$N$ coupling.

The short-range axial charge
operators can be obtained in a \lq\lq model-independent\rq\rq 
way, consistently with the two-nucleon
interaction model.  The procedure is described
in Ref.~\cite{Kir92}, and is similar to the one used to derive the 
\lq\lq model-independent\rq\rq electromagnetic or weak vector currents. 
Here we consider the charge operators associated only with the
central and spin-orbit components of the interaction, since these are
expected to give the largest contributions, after the
$\rho^{(2)}(\pi{\rm A})$ operator above.  This expectation is
in fact confirmed in the present study.  The momentum-space
expressions are given by
\begin{equation}
\rho_{ij}^{(2)}({\bf k}_i,{\bf k}_j;{\rm sA})= \frac{g_A}{2\, m^2}
\left[ \tau_{i,\pm} \, \overline{v}^{\,{\rm s}}(k_j)
      +\tau_{j,\pm} \, \overline{v}^{\,{\rm s}\tau}(k_j) \right ]
{\bbox \sigma}_i \cdot {\bf P}_i + i \rightleftharpoons j \ ,
\label{eq:rho12s}
\end{equation}
\begin{eqnarray}
\rho_{ij}^{(2)}({\bf k}_i,{\bf k}_j;{\rm vA})=&&\frac{g_A}{2\, m^2} \, 
\left [ \tau_{i,\pm}\, \overline{v}^{\,{\rm v}}(k_j)
       +\tau_{j,\pm}\, \overline{v}^{\, {\rm v}\tau}(k_j) \right ]
\left [ {\bbox \sigma}_i \cdot {\bf P}_j  
+{\rm i}\,({\bbox \sigma}_i \times {\bbox \sigma}_j) \cdot{\bf k}_j \right ]
\nonumber \\
&&-{\rm i}\,\frac{g_A}{4\, m^2} ({\bbox \tau}_i \times {\bbox \tau}_j)_\pm \,
\overline{v}^{\, {\rm v}\tau}(k_j) \,{\bbox \sigma}_i \cdot {\bf k}_i
+ i \rightleftharpoons j \ ,
\label{eq:rho12v}
\end{eqnarray}
where ${\bf P}_i={\bf p}_i+{\bf p}^\prime_i$, and
\begin{equation}
\overline{v}^{\, {\alpha}}(k)=4 \pi
\int_{0}^{\infty} dr\,r^2\,
j_{0}(kr)\,\overline{v}^{\,{\alpha}}(r) \ ,
\label{eq:vjrk}
\end{equation}
with $\alpha$=s, s$\tau$, v, and v$\tau$.  The following definitions
have been introduced 
\begin{eqnarray}
\overline{v}^{\,{\rm s}}(r)&=&\frac{3}{4}v^c(r)+\frac{m^2}{2}
\int_r^\infty dr^\prime\,r^\prime \left[ v^b(r^\prime)
-\frac{1}{2}v^{bb}(r^\prime) \right]  \nonumber \\
\overline{v}^{\,{\rm v}}(r)&=&\frac{1}{4}v^c(r)-\frac{m^2}{2}
\int_r^\infty dr^\prime\,r^\prime \left[ v^b(r^\prime)
-\frac{1}{2}v^{bb}(r^\prime) \right]  \ ,
\label{eq:vj}
\end{eqnarray}
where $v^c(r)$, $v^b(r)$ and $v^{bb}(r)$ are the isospin-independent
central, spin-orbit, and $({\bf L}\cdot{\bf S})^2$ components of
the AV14 or AV18 interactions, respectively.  The definitions
for $\overline{v}^{\,{\rm s}\tau}(r)$ and $\overline{v}^{\,{\rm v}\tau}(r)$
can be obtained from those above, by replacing the isospin-independent
$v^c(r)$, $v^b(r)$ and $v^{bb}(r)$ with the
isospin-dependent $v^{c\tau}(r)$, $v^{b\tau}(r)$ and $v^{bb\tau}(r)$.
\subsection{$\Delta$-Isobar Contributions}
\label{sec:vadlt}
In this section we review the treatment of the weak current and
charge operators associated with excitation of $\Delta$ isobars
in perturbation theory and within the context of the
transition-correlation-operator (TCO) method~\cite{Sch92}.
Among the two-body axial current operators, those 
associated with $\Delta$ degrees
of freedom have in fact been found to be the most important
ones~\cite{Sch92,Car91}.

In the TCO approach, the nuclear wave function is written as
\begin{equation}
\Psi_{N+\Delta}=\left[{\cal{S}}\prod_{i<j}\left(1\,+\,U^{\rm TR}_{ij}\right)
\right]\,\Psi \ ,
\label{eq:psiNDtco}
\end{equation}
where $\Psi$ is the purely nucleonic component, $\cal{S}$ is the 
symmetrizer, and the transition operators $U^{\rm TR}_{ij}$ convert $NN$ pairs 
into $N\Delta$ and $\Delta\Delta$ pairs.  The latter are defined as
\begin{equation}
U_{ij}^{\rm TR}\,=\,U_{ij}^{N\Delta}\,+\,U_{ij}^{\Delta N}\,+\,U_{ij}^
{\Delta\Delta} \ , 
\label{eq:tcoop}
\end{equation}
\begin{eqnarray}
U_{ij}^{N\Delta}&=&\left[u^{\sigma\tau II}(r_{ij})\bbox{\sigma}_{i}\cdot
{\bf S}_{j}\,+\,u^{t\tau II}(r_{ij})S_{ij}^{II}\right]\,\bbox{\tau}_{i}
\cdot{\bf T}_{j} \ , \label{eq:tcops1}\\
U_{ij}^{\Delta\Delta}&=&\left[u^{\sigma\tau III}(r_{ij}){\bf S}_{i}\cdot
{\bf S}_{j}\,+\,u^{t\tau III}(r_{ij})S_{ij}^{III}\right]\,{\bf T}_{i}\cdot
{\bf T}_{j} \ .
\label{eq:tcops2}
\end{eqnarray}
Here, ${\bf S}_{i}$ and ${\bf T}_{i}$ are spin- and isospin-transition 
operators which convert nucleon $i$ into a $\Delta$ isobar, $S_{ij}^{II}$ 
and $S_{ij}^{III}$ are tensor operators in which, respectively, the Pauli 
spin operators of either particle $i$ or $j$, and both particles $i$ and $j$ 
are replaced by corresponding spin-transition operators.
The $U_{ij}^{\rm TR}$ 
vanishes in the limit of large interparticle separations, since no 
$\Delta$-components can exist asymptotically.

In the present study the $\Psi$ 
is taken from CHH solutions of the AV14/UVIII or
AV18/UIX Hamiltonians with nucleons only 
interactions, while the $U^{\rm TR}_{ij}$ is obtained from two-body 
bound and low-energy scattering-state solutions of the full $N$-$\Delta$ 
coupled-channel problem with the Argonne
$v_{28Q}$~\cite{Wirpc} (AV28Q) interaction, containing
explicit $N$ and $\Delta$ degrees of freedom.
This aspect of the present calculations, including the validity
of the approximation inherent to Eq.~(\ref{eq:psiNDtco}),
were discussed at length in the original work~\cite{Sch92}, and
have been reviewed more recently in Ref.~\cite{Mar98}, making a further
review here unnecessary.  
The AV28Q interaction provided an excellent description
of the $N$$N$ database available in the early eighties.  No attempt
has been made to refit this model to the more recent and much
more extensive Nijmegen database~\cite{Sto93}.  

In the TCO scheme, the perturbation theory description of $\Delta$-admixtures 
is equivalent to the replacements:
\begin{eqnarray}
U_{ij}^{N\Delta,{\rm PT}}&=&\frac{v_{ij}(NN\rightarrow N\Delta)}
{m-m_{\Delta}} \ , \label{eq:UNDpt} \\
U_{ij}^{\Delta\Delta,{\rm PT}}&=&\frac{v_{ij}(NN\rightarrow \Delta\Delta)}
{2(m-m_{\Delta})} \ , \label{eq:UDDpt} 
\end{eqnarray}
where the kinetic energy contributions in the denominators of 
Eqs.~(\ref{eq:UNDpt}) and~(\ref{eq:UDDpt}) have been neglected 
(static $\Delta$ approximation).
The transition interactions $v_{ij}(NN\rightarrow N\Delta)$ 
and $v_{ij}(NN\rightarrow \Delta\Delta)$ have the same operator structure 
as $U_{ij}^{N\Delta}$ and $U_{ij}^{\Delta\Delta}$ of Eqs.~(\ref{eq:tcops1}) 
and~(\ref{eq:tcops2}), but with 
the $u^{\sigma\tau\alpha}(r)$ and $u^{t\tau\alpha}(r)$ functions 
replaced by, respectively,
\begin{eqnarray}
v^{\sigma\tau\alpha}(r)&=&
\frac{(ff)_{\alpha}}{4\pi}\frac{m_{\pi}}{3}\frac{{\rm{e}}^{-x}}{x}\,C(x) \ ,
\label{eq:uvst} \\
v^{t\tau\alpha}(r)&=&
\frac{(ff)_{\alpha}}{4\pi}\frac{m_{\pi}}{3}\left(1+\frac{3}{x}+\frac{3}{x^2}
\right)\frac{{\rm{e}}^{-x}}{x}\,C^{2}(x) \ .
\label{eq:uvtt} 
\end{eqnarray}
Here $\alpha$ = II, III, $x\equiv m_{\pi}r$, $(ff)_{\alpha}=f_\pi
f_\pi^*$, $f_\pi^* f_\pi^*$ for $\alpha$ = II, III, 
respectively, $f_\pi^*$ being the $\pi N\Delta$ coupling constant,  
and the cutoff function $C(x)\,=\,1-e^{-\lambda x^{2}}$.  In
the AV28Q interaction $f_\pi^*=(6\sqrt{2}/5) f_{\pi}$, as
obtained in the quark-model, and $\lambda$ = 4.09.  
When compared to $U^{\rm TR}_{ij}$, the perturbation theory
$U^{\rm TR,PT}_{ij}$ corresponding to Eqs.~(\ref{eq:UNDpt}) and
(\ref{eq:UDDpt}) produces $N$$\Delta$ and $\Delta$$\Delta$
admixtures that are too large at short distances, and
therefore leads to a substantial overprediction of the effects
associated with $\Delta$ isobars in electroweak observables~\cite{Sch92}. 
 
We now turn our attention to the discussion of
$N$$\Delta$ and $\Delta$$\Delta$ weak transition operators.
The axial current and charge 
operators associated with excitation of $\Delta$ isobars
are modeled as
\begin{eqnarray}
{\bf{j}}_{i}^{(1)}({\bf{q}}; N\rightarrow\Delta,{\rm A})
&=&-g_{A}^{*}\,T_{i,\pm} \, {\bf{S}}_{i}
\,{\rm e}^{{\rm{i}}{\bf{q}}\cdot{\bf{r}}_{i}} \ ,
\label{eq:j1bnd} \\
{\bf{j}}_{i}^{(1)}({\bf{q}}; \Delta\rightarrow\Delta,{\rm A})&=&
-{\overline{g}}_{A}\,\Theta_{i,\pm} \,{\bbox{\Sigma}}_{i}
\,{\rm e}^{{\rm{i}}{\bf{q}}\cdot{\bf{r}}_{i}} \ ,
\label{eq:j1bdd}
\end{eqnarray}
and
\begin{eqnarray}
{\rho}_{i}^{(1)}({\bf{q}}; N\rightarrow\Delta,{\rm A})&=&
-\frac{g_{A}^{*}}{m_{\Delta}} \, T_{i,\pm} \,
{\bf{S}}_{i}\cdot{\bf{p}}_{i}\,{\rm e}^{{\rm{i}}{\bf{q}}\cdot{\bf{r}}_{i}} 
\label{eq:rho1bnd} \\
{\rho}_{i}^{(1)}({\bf{q}}; \Delta\rightarrow\Delta,{\rm A})&=&
-\frac{{\overline{g}}_{A}}{2\,m_{\Delta}}\, \Theta_{i,\pm} \,  
{\bbox{\Sigma}}_{i}\cdot \bigg[ {\bf{p}}_{i}\,
 , \,{\rm e}^{{\rm{i}}{\bf{q}}\cdot{\bf{r}}_{i}} \bigg]_+ \ ,
\label{eq:rho1bdd}
\end{eqnarray}
where $m_{\Delta}$ is the $\Delta$-isobar mass,
${\bbox{\Sigma}}$ (${\bbox{\Theta}}$) is the Pauli operator
for the $\Delta$ spin 3/2 (isospin 3/2), and $T_{i,\pm}$ and
$\Theta_{i,\pm}$ are defined in analogy to Eq.~(\ref{taupm}).
The expression for 
${\bf{j}}_{i}^{(1)}({\bf{q}}; \Delta\rightarrow  N,{\rm A})$ 
(${\rho}_{i}^{(1)}({\bf{q}}; \Delta\rightarrow  N,{\rm A})$) is 
obtained from that for
${\bf{j}}_{i}^{(1)}({\bf{q}}; N\rightarrow\Delta,{\rm A})$ 
(${\rho}_{i}^{(1)}({\bf{q}}; N\rightarrow\Delta,{\rm A})$)
by replacing ${\bf S}_i$ and ${\bf T}_i$ by their hermitian
conjugates.  The coupling constants
$g_{A}^{*}$ and ${\overline{g}}_{A}$ are not well known.
In the quark-model, they are related to the axial coupling constant
of the nucleon by the relations $g_{A}^{*}=(6\sqrt{2}/5) g_A$ and
${\overline{g}}_{A}=(1/5) g_A$.  These values have often been used in the
literature in the calculation of $\Delta$-induced axial current contributions
to weak transitions.  However, given the uncertainties inherent 
to quark-model predictions, a more reliable estimate for $g_A^*$
is obtained by determining its value phenomenologically 
in the following way.  It is well established by now~\cite{Sch98}
that the one-body axial current of Eq.~(\ref{eq:1baj})
leads to a $\simeq$ 4 \% underprediction of the measured
Gamow-Teller matrix element in tritium $\beta$-decay, see Table~\ref{tb:h3gt}.
Since the contributions of
$\Delta\rightarrow\Delta$ axial currents (as well as those due to 
the two-body operators of Sec.~\ref{sec:acnt})
are found to be numerically very small, as
can be seen again from Table~\ref{tb:h3gt}, this
4 \% discrepancy can then be used to determine $g_{A}^{*}$~\cite{Note}.
Obviously, this procedure produces 
different values for $g_{A}^{*}$ depending on 
how the $\Delta$-isobar degrees of freedom are treated.  These values 
are listed in Table~\ref{tb:gas} for comparison.  The $g_{A}^{*}$ value that
is determined in the context of a TCO calculation based on
the AV28Q interaction, is about 40 \% larger than the naive
quark-model estimate.  However, when perturbation theory is used for the 
treatment of the $\Delta$ isobars, the
$g_{A}^{*}$ value required to reproduce
the Gamow-Teller matrix element of tritium $\beta$-decay
is much smaller than the TCO estimate, as expected.  
Finally, the $N \rightarrow \Delta$
axial current derived in perturbation theory from Eqs.~(\ref{eq:UNDpt})
and~(\ref{eq:j1bnd}) is, of course, identical to the
expression given in Refs.~\cite{Car91,Sch98}.

The $N \rightarrow \Delta$ and $\Delta \rightarrow \Delta$
weak vector currents are modeled, consistently with the CVC hypothesis, as 
\begin{eqnarray}
{\bf j}^{(1)}_{i}({\bf q};N\rightarrow \Delta,{\rm V})&=&-{\rm i}\,
\frac{\mu^*}{m}\, T_{i,\pm} \,
{\bf q}\times{\bf S}_{i} \,
 {\rm e}^{{\rm i}{\bf q}\cdot{\bf r}_{i}} \ , \label{eq:j1bND} \\
{\bf j}^{(1)}_{i}({\bf q};\Delta\rightarrow \Delta,{\rm V})&=&-{\rm i}
\frac{\overline{\mu}}{12\, m} \, \Theta_{i,\pm} \,
{\bf q}\times\bbox{\Sigma}_{i} \,
{\rm e}^{{\rm i}{\bf q}\cdot{\bf r}_{i}} \ , \label{eq:j1bDD}
\end{eqnarray}
where the $N\Delta$-transition magnetic moment $\mu^*$ is 
taken equal to 3 n.m., as obtained from an analysis of $\gamma N$ 
data in the $\Delta$-resonance region~\cite{CAR86}, while the
value used for the $\Delta$ magnetic moment $\overline{\mu}$ 
is 4.35 n.m.\ by averaging results of a soft-photon analysis of 
pion-proton bremsstrahlung data near the $\Delta^{++}$ resonance~\cite{LIN91}.
The contributions due to the weak vector currents above
have been in fact found to be very small in the $p$$\,^3$He capture
process.  Finally, $\Delta$ to $\Delta$ weak vector charge operators
are ignored in the present study, since their associated
contributions are expected to be negligible. 
\section{Calculation}
\label{sec:calc}

The calculation of the $p$$\,^3$He weak capture cross section
proceeds in two steps: firstly, the Monte Carlo evaluation
of the weak charge and current operator
matrix elements, and the subsequent decomposition of these in terms
of reduced matrix elements; secondly, the evaluation of the cross section
by carrying out the integrations in Eq.~(\ref{eq:xsc1}).
\subsection{Monte Carlo Calculation of Matrix Elements}
\label{sec:calmc}

In a frame where the direction of the momentum transfer $\hat{\bf q}$ also
defines the quantization axis of the nuclear spins, the matrix element
of, as an example, the weak axial (or vector) current has the multipole
expansion
\begin{equation}
\langle\Psi_{4}\,|\,{\hat{\bf{e}}}^{*}_{q\lambda}\cdot
{\bf{j}}^{\dag}({\bf{q}})\,|\,
{\overline{\Psi}}_{1+3}^{LSJ,J_{z}=\lambda}\rangle
=\sqrt{2\pi}\, {\rm i}^J\, \left[ \lambda M_{J}^{LSJ}(q) + 
E_{J}^{LSJ}(q)\right] \ , 
\label{eq:meq}
\end{equation}
with $\lambda = \pm 1$.  The expansion above is easily obtained from
that in Eq.~(\ref{eq:me}), in which the quantization axis for the
nuclear spins was taken along the direction of the relative momentum
$\hat{\bf p}$, by setting $\theta$=$\phi$=0 and using
$D^J_{J_z^\prime,J_z}(0,0,0)=\delta_{J_z^\prime,J_z}$.  Then, again as
an example, the reduced matrix element of the axial electric dipole operator
involving a transition from the $p$$\,^3$He $^3$S$_1$ state is simply given by
\begin{equation}
E_1^{011}(q;{\rm A})=-\frac{\rm i}{\sqrt{2 \pi}}
\langle\Psi_{4}\,|\,{\hat{\bf{e}}}^{*}_{q\lambda}\cdot
{\bf{j}}^{\dag}({\bf q};{\rm A})\,|\,
{\overline{\Psi}}_{1+3}^{011,J_{z}=\lambda}\rangle \ .
\label{e1cal}
\end{equation}

\begin{figure}[bth]
\let\picnaturalsize=N
\def\picsize{5in}
\def\picfilenamea{deltaaj.eps}
\ifx\nopictures Y\else{\ifx\epsfloaded Y\else\input epsf \fi
\let\epsfloaded=Y
\centerline{
\ifx\picnaturalsize N\epsfxsize \picsize\fi \epsfbox{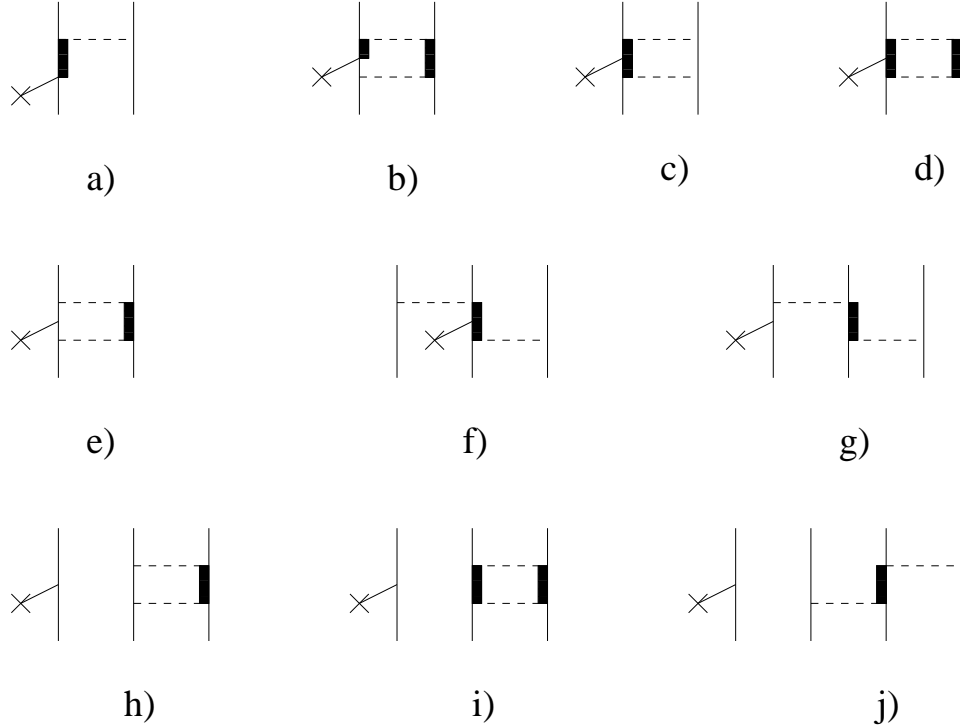}
 }}\fi
\caption{Diagrammatic representation of the operators 
included in $O(\Delta)$, due to the one-body current 
and charge operators
$O^{(1)}(N\rightarrow\Delta)$, $O^{(1)}(\Delta\rightarrow N)$, and
$O^{(1)}(\Delta\rightarrow\Delta)$, given in
Eqs.~(\ref{eq:j1bnd})--(\ref{eq:j1bDD}), and to the transition
correlations $U^{\Delta N}$, $U^{N \Delta}$, $U^{\Delta \Delta}$,
and corresponding hermitian conjugates.  Thin, thick, and dashed lines
denote, respectively, nucleons, $\Delta$ isobars, and transition
correlations $U^{B B^{'}}$ or ${U^{B B^{'}}}^{\dagger}$,
with $B,B^{'}\equiv N,\Delta$.}
\label{fig:jd}
\end{figure}

The problem is now reduced to the evaluation of matrix elements
of the same type as on the right-hand-side of Eq.~(\ref{e1cal}).  These
can schematically be written as
\begin{equation}
\frac{\langle\Psi_{4,N+\Delta}\,|\,O\,|\,\Psi_{1+3,N+\Delta}\rangle}
{\left[ \langle\Psi_{4,N+\Delta}\,|\Psi_{4,N+\Delta}\rangle
 \langle\Psi_{1+3,N+\Delta}\,|\Psi_{1+3,N+\Delta}\rangle\right]^{1/2} } \ ,
\label{eq:schem}
\end{equation}
where the initial and final states have the form
of Eq.~(\ref{eq:psiNDtco}).  It is convenient to expand the latter as
\begin{equation}
\Psi_{N+\Delta} = \Psi + \sum_{i<j}U_{ij}^{\rm TR}\Psi + 
\cdots \ ,
\label{eq:psind}
\end{equation}
so that the numerator of Eq.~(\ref{eq:schem}) can be expressed as
\begin{equation}
\langle\Psi_{4,N+\Delta}\,|\,O\,|\,\Psi_{1+3,N+\Delta}\rangle\,=\,\langle
\Psi_{4}\,|\,O(N\,{\rm only})\,|\,\Psi_{1+3}\rangle 
\,+\,\langle\Psi_{4}\,|\,O(\Delta)\,|\,\Psi_{1+3}\rangle \ ,
\label{eq:schemj}
\end{equation}
where the operator $O(N \,{\rm only})$ 
denotes all one- and two-body contributions to the weak
charge or current operator $O$, involving only nucleon degrees of freedom,
i.e.\ $O(N\,{\rm only})=O^{(1)}(N\rightarrow N)+O^{(2)}(NN\rightarrow NN)$, 
while $O(\Delta)$ includes terms that involve the $\Delta$-isobar 
degrees of freedom, associated with the explicit $\Delta$ transitions
$O^{(1)}(N\rightarrow\Delta)$, $O^{(1)}(\Delta\rightarrow N)$, $O^{(1)}(\Delta
\rightarrow\Delta)$, and with the transition operators $U_{ij}^{\rm TR}$.
A diagrammatical illustration of the terms contributing to $O(\Delta)$ 
is given in Fig.~\ref{fig:jd}: 
the terms (a)--(e), (f)--(i), and (j) represent, respectively,
two-, three-, and four-body operators.  The terms (e) and
(g)--(j) are to be viewed as renormalization corrections to the 
\lq\lq nucleonic\rq\rq matrix element of $O^{(1)}(N\rightarrow N)$, due 
to the presence of $\Delta$-admixtures in the wave functions. 
Connected three-body terms containing more than a single $\Delta$ isobar
have been ignored, since their contributions are expected to be negligible.
Indeed, the contribution from diagram (d) has already been found numerically
very small.

The two-body terms of Fig.~\ref{fig:jd} are expanded as operators 
acting on the nucleons' coordinates. For example, the terms (a) and (c) in 
Fig.~\ref{fig:jd} have the structure, respectively,
\begin{eqnarray}
({\rm a})&=&{U_{ij}^{\Delta N}}^{\dagger}\,
O_{i}^{(1)}(N\rightarrow\Delta) \ , 
\label{eq:aterm} \\
({\rm c})&=&{U_{ij}^{\Delta N}}^{\dagger}\,
O^{(1)}_{i}(\Delta\rightarrow\Delta)\,U_{ij}^{\Delta N} \ , 
\label{eq:cterm}
\end{eqnarray}
which can be reduced to operators involving only Pauli spin and isospin 
matrices by using the identities:
\begin{eqnarray}
{\bf S}^{\dagger}\cdot{\bf A}\,{\bf S}\cdot{\bf B}&=&\frac{2}{3}{\bf A}
\cdot{\bf B}-\frac{\rm i}{3}\bbox{\sigma}\cdot ({\bf A}\times{\bf B}) \ ,
\label{eq:SdagS} \\
{\bf S}^{\dagger}\cdot{\bf A}\,\bbox{\Sigma}\cdot{\bf B}\,{\bf S}
\cdot{\bf C}&=&\frac{5}{3}\,{\rm i}\,{\bf A}\cdot({\bf B}\times{\bf C})-
\frac{1}{3}\bbox{\sigma}\cdot{\bf A}\,{\bf B}\cdot{\bf C} \nonumber \\
& & -\frac{1}{3}{\bf A}\cdot {\bf B}\,{\bf C}\cdot\bbox{\sigma}+\frac{4}{3}
{\bf A}\cdot({\bf B}\cdot\bbox{\sigma}){\bf C} \ ,
\label{eq:SdagSigmaS}
\end{eqnarray}
where ${\bf A}$, ${\bf B}$ and ${\bf C}$ are vector operators that commute 
with $\bbox{\sigma}$, but not necessarily among themselves.
While the three- and four-body 
terms in Fig.~\ref{fig:jd} could have been reduced in precisely 
the same way, the resulting expressions in terms of $\bbox{\sigma}$ and 
$\bbox{\tau}$ matrices become too cumbersome.  Thus, for these it was 
found to be more convenient to retain the explicit representation of 
${\bf S}$ $({\bf S}^{\dagger})$ as a $4 \times 2\,(2\times 4)$ matrix

$ \hspace{5cm}
  {\bf S}= \left[ \begin{array}{cc}
           -\hat{{\bf e}}_{-} & 0  \\
           \sqrt{\frac{2}{3}}\hat{{\bf e}}_{0} & -\frac{1}{\sqrt{3}}
	                     \hat{{\bf e}}_{-} \\
           -\frac{1}{\sqrt{3}}\hat{{\bf e}}_{+}& \sqrt{\frac{2}{3}}
                              \hat{{\bf e}}_{0} \\
	   0  & -\hat{{\bf e}}_{+}
           \end{array} \right] \ , 
$

\noindent
and of ${\bbox{\Sigma}}$ as a $4 \times 4$ matrix

\noindent
$ \hspace{5cm}
  {\bbox \Sigma}= \left[ \begin{array}{cccc}
           3\hat{{\bf e}}_{0} & \sqrt{6}\hat{{\bf e}}_{-} & 0 & 0  \\
           -\sqrt{6}\hat{{\bf e}}_{+} & \hat{{\bf e}}_{0} & 
	   \sqrt{8}\hat{{\bf e}}_{-} & 0 \\
           0 & -\sqrt{8}\hat{{\bf e}}_{+} & -\hat{{\bf e}}_{0} & 
	   \sqrt{6}\hat{{\bf e}}_{-} \\
	   0 & 0 &  -\sqrt{6}\hat{{\bf e}}_{+} & -3\hat{{\bf e}}_{0}
           \end{array} \right] \ , 
$

\noindent
where $\hat{{\bf e}}_{\pm}=\mp(\hat{\bf x}\pm {\rm i}\hat{\bf y})/{\sqrt{2}}$, 
$\hat{\bf e}_{0}=\hat{\bf z}$, and $\hat{\bf e}_{\mu}^{*}=
(-)^{\mu}\hat{\bf e}_{-\mu}$, and derive the result of terms such as 
(f)=${U_{ij}^{N\Delta}}^{\dagger}\,O^{(1)}_{j}
(\Delta\rightarrow\Delta)\,U_{jk}^{\Delta N}$ 
on state $|\Psi\rangle$ by first operating 
with $U_{jk}^{\Delta N}$, then with $O^{(1)}_{j}
(\Delta\rightarrow\Delta)$, and finally 
with ${U_{ij}^{N\Delta}}^{\dagger}$.  These terms
(as well as three-body contributions to the wave
function normalizations, see below) were neglected
in the calculations reported in Ref.~\cite{Sch92}.

Of course, the presence of $\Delta$-admixtures also
influences the normalization of the wave functions, as is obvious
from Eq.~(\ref{eq:schem}):
\begin{eqnarray}
\langle \Psi_{N+\Delta}\,|\,\Psi_{N+\Delta}\rangle&=&
\langle \Psi\,|\,1\,+\,\sum_{i<j}[\,2\,{U_{ij}^{\Delta N}}^{\dagger}
U_{ij}^{\Delta N}\,+\,{U_{ij}^{\Delta \Delta}}^{\dagger}U_{ij}^{\Delta 
\Delta}] \nonumber \\
&+&\sum_{i<j\, , \, k\neq i,j} [{U_{ij}^{\Delta N}}^{\dagger}
U_{ik}^{\Delta N}\,+\,{U_{ij}^{N\Delta}}^{\dagger}U_{kj}^{N\Delta}]
\,|\,\Psi\rangle  + \dots \ .
\label{eq:normal}
\end{eqnarray}
The wave function normalization ratios 
$\langle \Psi_{N+\Delta}\,|\,\Psi_{N+\Delta}\rangle/
\langle \Psi\,|\,\Psi\rangle$, obtained
for the bound three- and four-nucleon systems, are listed
in Table~\ref{tb:normr}.  Note that the normalization of
the $p$$\,^3$He continuum state is the same as that
of $^3$He, up to corrections of order (volume)$^{-1}$.

The matrix elements in Eqs.~(\ref{eq:schemj}) and (\ref{eq:normal})
are computed, without any approximation, by Monte Carlo
integrations.  The wave functions 
are written as vectors in the spin-isospin space
of the four nucleons for any given spatial configuration
${\bf R}=({\bf r}_1,\dots,{\bf r}_4)$.  For the given
${\bf R}$ we calculate the state vector $\left[ O({\bf R},N\>{\rm only})
+O({\bf R},\Delta)\right] \Psi({\bf R})$ with the techniques developed in
Refs.~\cite{Mar98,Sch89}.  The spatial integrations are
carried out with the Monte Carlo method by sampling ${\bf R}$ configurations
according to the Metropolis {\it et al.} algorithm~\cite{Met53}, using
a probability density $W({\bf R})$ proportional to
\begin{equation}
W({\bf R}) \propto 
\sqrt{\langle \Psi^\dagger_4({\bf R}) \Psi_4({\bf R})\rangle} \ ,
\end{equation}
where the notation $\langle \cdots \rangle$ implies sums over the spin-isospin
states of the $^4$He wave function.
Typically 200,000 configurations are enough to achieve a 
relative error $\leq$ 5 \% on the total $S$-factor.
\subsection{Calculation of Cross Section}
\label{sec:calxs}

Once the reduced matrix elements (RMEs) have been
obtained, the calculation of the cross section $\sigma(E)$
is reduced to performing the integrations over the
electron and neutrino momenta in Eq.~(\ref{eq:xsc1})
numerically.  We write
\begin{equation}
\sigma(E) = \frac{1}{(2\, \pi)^2}\,\frac{G_V^2}{v_{\rm rel}} 
\int_0^{p_e^*} dp_e\, p_e^2 \int_{-1}^1 dx_e \int_{-1}^1 dx_\nu \int_0^{2\pi}
d\phi\, p_\nu^2 \, f^{-1} \, L_{\sigma \tau} N^{\sigma \tau} \ ,
\label{xscint}
\end{equation}
where one of the azimuthal integrations has been carried out, since
the integrand only depends on the difference $\phi=\phi_e-\phi_\nu$.
The $\delta$-function occurring in Eq.~(\ref{eq:xsc1}) has also
been integrated out resulting in the factor $f^{-1}$, with
\begin{equation}
f=\left | 1 + \frac{p_e\, x_{e \nu}}{m_4} + \frac{p_\nu}{m_4} \right | \ .
\end{equation}
The magnitude of the neutrino momentum is fixed by
energy conservation to be
\begin{equation}
p_\nu = \frac{2 \, \overline{\Delta} }
{ 1+p_e\, x_{e \nu}/m_4 + \sqrt{ (1+p_e \, x_{e \nu}/m_4 )^2+2\,
\overline{\Delta}/m_4 } } \ ,
\end{equation}
where $\overline{\Delta} = \Delta m+E-E_e-p^2_e/2 m_4$.  The variable
$x_{e \nu}$ is defined as
\begin{equation}
x_{e \nu} = \hat{\bf p}_e \cdot \hat{\bf p}_\nu
= x_e\, x_\nu +\sqrt{1-x_e^2} \sqrt{ 1-x_\nu^2}\,{\rm cos}\, \phi \ ,
\end{equation}
where $x_e = {\rm cos}\,\theta_e$ and
$x_\nu = {\rm cos}\,\theta_\nu$.  Finally, the integration over the
magnitude of the electron momentum extends from zero up to
\begin{equation}
p_e^* = \sqrt{ \left[ \sqrt{m_4^2+m_e^2+2\,
 m_4\,(\Delta m+E) }\,-\,m_4^2 \right ]^2-m_e^2} \ .
\end{equation}
The lepton tensor is explicitly given by Eq.~(\ref{eq:lalb}), while
the nuclear tensor is constructed using Eqs.~(\ref{nuclt})--(\ref{nuclx}).
Computer codes have been developed to calculate the required rotation
matrices corresponding to the $\hat{\bf q}$-direction ($\theta,\phi$)
with
\begin{eqnarray}
{\rm cos}\, \theta= \hat{\bf z} \cdot \hat{\bf q}&=& 
\frac{\hat{\bf z} \cdot \left( {\bf p}_e+{\bf p}_\nu \right ) }
{ \left| {\bf p}_e+{\bf p}_\nu \right|} \nonumber \\
&=& \frac{ p_e \, x_e + p_\nu \, x_\nu}
{\sqrt{p_e^2 +p_\nu^2 + 2\, p_e\, p_\nu \, x_{e \nu} } } \ .
\label{cost}
\end{eqnarray}
Finally, note that the nuclear tensor requires the values of the RMEs
at the momentum transfer $q$ (the denominator in the second line of
Eq.~(\ref{cost})).  It has been found convenient to make
the dependence upon $q$ of the RMEs explicit by expanding
\begin{equation}
T^{LSJ}_J(q) = q^m\, \sum_{n \ge 0} t^{LSJ}_{2n} \, q^{2n} \ ,
\end{equation}
consistently with Eqs.~(\ref{eq:cdef})--(\ref{eq:mdef}).  Here 
$m=J,J\pm 1$, depending on the RME 
considered.  For example,
$m=1$ for the $L^{110}_0({\rm A})$ RME.  Given the low momentum
transfers involved, $q \leq 20$ MeV/c, the leading and
next-to-leading order terms $t_0$ and $t_2$ are
sufficient to reproduce accurately $T(q)$.  Note that the
long-wavelength-approximation corresponds, typically, to retaining only
the $t_0$ term. 

A moderate number of Gauss points (of the order of 10) for each
of the integrations in Eq.~(\ref{xscint}) 
is sufficient to achieve convergence within better
than one part in $10^3$.  The computer program has been successfully
tested by reproducing the result obtained analytically by retaining only
the $^3$S$_1$ $E_1({\rm A})$ and $L_1({\rm A})$ and
$^3$P$_0$ $C_0({\rm A})$ RMEs.
\section{Results and Discussion}
\label{sec:res}

The $S$-factor calculated values are
listed in Table~\ref{tb:sfact}, and their implications 
to the recoil electron spectrum measured in the SK experiment, 
see Fig.~\ref{fig:ratio}, have already been discussed
in the introduction.  In Tables~\ref{tb:rme1s0},~\ref{tb:rme3s1},
and~\ref{tb:rme3p0}-\ref{tb:rme3p2}, we present
our results, obtained with the AV18/UIX Hamiltonian model,
for the reduced matrix elements (RMEs) connecting any of the $p\,^3$He
S- and P-wave channels to the $^4$He bound state.  The values for these
RMEs are given at zero energy and a lepton momentum transfer $q$=$19.2$ MeV/c.
Note that the RMEs listed in all tables are related to those defined in
Eqs.~(\ref{eq:c})--(\ref{eq:me}) via 
\begin{equation}
    \overline{T_J}^{LSJ}=
      \sqrt{ {v_{\rm rel} \over 4\pi\alpha }
      [{\rm exp}(4\pi\alpha/v_{\rm rel})-1]} \,
      T_J^{LSJ} \label{newRME2} \ ,
\end{equation}
which can be shown to remain finite in the limit $v_{\rm rel} \rightarrow 0$,
corresponding to zero energy.  

In Table~\ref{tb:sfc_contr} we list the individual contributions
of the S- and P-wave capture channels to the total $S$-factor
at zero c.m.\ energy, obtained 
with the AV18/UIX, the AV18 only (to study the effects
of the three-nucleon interaction), and the older AV14/UVIII
(to study the model dependence and to make contact with 
the earlier calculations of Refs.~\cite{Sch92,Car91}).
The model dependence is discussed in Sec.~\ref{subsec:modd}.

In Tables~\ref{tb:sfact},~\ref{tb:rme1s0},~\ref{tb:rme3s1}, 
and~\ref{tb:rme3s1_c}-\ref{tb:rme3p2},
the cumulative nucleonic contributions are normalized as
\begin{equation}
\left [ {\rm one\!-\!body}\!+\!{\rm mesonic} \right ]=
{ \langle \Psi_4 | O(N \> {\rm only})
| \Psi_{1+3} \rangle \over 
\left [\langle \Psi_4 | \Psi_4 \rangle \langle \Psi_{1+3} | 
\Psi_{1+3} \rangle \right ]^{1/2} } \ . \label{res1}
\end{equation}
However, when the $\Delta$-isobar contributions are added to the 
cumulative sum, the normalization changes to
\begin{equation}
\left [{\rm one\!-\!body}\!+\!{\rm mesonic}\! +\!
\Delta \right ]={ \langle \Psi_{4,N+\Delta} | O(N \> {\rm only})
+ O(\Delta) | \Psi_{1+3,N+\Delta} \rangle \over 
\left [ \langle \Psi_{4,N+\Delta} | \Psi_{4,N+\Delta} \rangle 
\langle \Psi_{1+3,N+\Delta} | 
\Psi_{1+3,N+\Delta} \rangle \right ]^{1/2} } \>\>\>. \label{res2}
\end{equation}
As already pointed out earlier in Sec.~\ref{sec:calmc},
the normalization of the initial scattering state
is the same as that of $^3$He, up to corrections of order (volume)$^{-1}$.
In Table~\ref{tb:rme3s1_c} we also report results in which the 
$\Delta$-components in the nuclear
wave functions are treated in perturbation theory, as discussed in 
Secs.~\ref{sec:vadlt}
and~\ref{sec:calmc}, and the $O(\Delta)$ only includes
the operators in panel (a) of Fig.~\ref{fig:jd}.
In this case, the cumulative contributions [one-body+mesonic+$\Delta_{\rm
PT}$] are normalized as in Eq.~(\ref{res1}).

\subsection{$^1$S$_0$ Capture}
\label{subsec:1s0}

The $^1$S$_0$ capture is induced by the weak vector charge and longitudinal
component of the weak vector current via the $C_0({\rm V})$ and 
$L_0({\rm V})$ multipoles,
respectively.  The associated RMEs, while small, are not negligible--they
are about 20 \% of the \lq\lq large\rq\rq $E_1({\rm A})$ RME in 
$^3$S$_1$ capture,
see Table~\ref{tb:rme3s1}.  These $^1$S$_0$ transitions are inhibited 
by an isospin
selection rule, indeed they vanish at $q$=$0$, since in this limit
\begin{equation} 
C_0(q;{\rm V}) \rightarrow \frac{1}{\sqrt{4 \pi}} \, \sum_i \tau_{i,\pm}
\equiv \frac{1}{\sqrt{4\pi}} \, T_{\pm} \ ,
\end{equation}
and
\begin{equation}
L_0(q;{\rm V}) = -\frac{1}{q} \, \left[ H \, , \,
 \int d{\bf x}\>j_0(qx) \,Y_{00}(\hat{\bf x})
\>\rho ({\bf x};{\rm V}) \right]\, \rightarrow -\frac{1}{q}
 \, \left[ H \, , \, \frac{1}{\sqrt{4 \pi}}\, T_{\pm} \right] \ ,
\label{eq:lw1}
\end{equation}
where the expression for $L_0({\rm V})$ has been obtained
by integrating Eq.~(\ref{eq:ldef}) by parts, and then using the
continuity equation to relate $\nabla \cdot {\bf j}({\bf x};{\rm V})$
to the commutator $-{\rm i} [ H \, ,\, \rho({\bf x};{\rm V})]$.
The $^4$He and $p\,^3$He states have total isospins $T,T_z$=0,0 and
1,1, respectively, ignoring additional, but very small, isospin admixtures
induced by isospin-symmetry-breaking components of the interaction.
Therefore matrix elements of the (total) isospin raising
or lowering operators $T_{\pm}$ between these $T,T_z$ states vanish. 

Equation~(\ref{eq:lw1}) shows that, if the initial and final CHH
wave functions were to be exact eigenfunctions of the AV18/UIX Hamiltonian,
then one would expect, neglecting the kinetic energy of the recoiling $^4$He:
\begin{equation}
L_0(q;{\rm V}) = \frac{E_3-E_4}{q}\, C_0(q;{\rm V}) \ ,
\label{eq:c0l0}
\end{equation}
where $E_3$ and $E_4$ are the three- and four-nucleon ground-state
energies.  Note that the relation above is in fact valid for any 
$C_J(q;{\rm V})$ and
$L_J(q;{\rm V})$ multipoles.  For $q$=$19.2$ MeV/c the ratio $L_0/C_0$ is
expected to be $1.07$, which is in perfect agreement with that obtained
in the calculation, when the two-body current contributions are taken
into account, see Table~\ref{tb:rme1s0}.  As already discussed
in Sec.~\ref{sec:vcnt}, the present model for the weak vector current satisfies
current conservation with the $v_6$ part of the nucleon-nucleon interaction
(either AV14 or AV18).  The spin-orbit and quadratic momentum-dependent
components of the interaction, however, require additional short-range currents
that have been neglected in this work.  If their contributions were to be
completely negligible, then the degree of agreement between the expected
and calculated values for the ratio $L_0/C_0$ would simply reflect the
extent to which the present variational wave functions are truly exact
eigenfunctions of the AV18/UIX Hamiltonian.  However, the CHH wave
function used here gives a ground-state energy of $-27.9$ MeV for $^4$He,
which is about $400$ keV higher than predicted for the
AV18/UIX model in GFMC calculations~\cite{Pud97}.  In view of
these considerations, the perfect agreement referred to above may
be accidental.

Finally, the $C_1({\rm V})$ and $L_1({\rm V})$ RMEs interfere destructively 
in the cross section (see discussion at the end of Sec.~\ref{sec:xs}),
substantially reducing the $^1$S$_0$ channel contribution to the $S$-factor, 
see Table~\ref{tb:sfc_contr}.

\subsection{$^3$S$_1$ Capture}
\label{subsec:3s1}

The $^3$S$_1$ capture is induced by the weak axial charge and current,
and weak vector current operators via the multipoles 
$C_1({\rm A})$, $L_1({\rm A})$,
$E_1({\rm A})$, and $M_1({\rm V})$.  All earlier studies only retained
the dominant $L_1({\rm A})$ and $E_1({\rm A})$ transitions.  However,
as is evident from Table~\ref{tb:rme3s1}, the $M_1({\rm V})$ and especially
$C_1({\rm A})$ RMEs are not negligible.  Furthermore, the $C_1({\rm A})$ and
$L_1({\rm A})$ RMEs interfere constructively in the cross section, 
since their signs are opposite.  For example, neglecting the
$C_1({\rm A})$ contribution would produce an $S$-factor value
$4.94\times 10^{-20}$ keV~b, 30 \% smaller than the $^3$S$_1$ total 
result $6.38\times 10^{-20}$ keV~b (see Table~\ref{tb:sfc_contr}).

The destructive interference between the one- and many-body axial
current contributions in the $L_1({\rm A})$ and $E_1({\rm A})$
RMEs, first obtained in Refs.~\cite{Sch92,Car91}, is confirmed
in the present work.
The axial currents associated with $\Delta$-excitation play
a crucial role.  The (suppressed) one-body contribution
comes mostly from transitions involving the D-state components
of the $^3$He and $^4$He wave functions, while the many-body
contributions are predominantly due to transitions connecting
the S-state of $^3$He to the D-state of $^4$He, or viceversa.
To clarify this point, it is useful to define the one- and two-body
densities
\begin{equation}
\rho^{(1)}(x) = \langle ^4{\rm He}|\sum_i
\delta(x-|{\bf r}_i-{\bf R}_{jkl}|) O^{(1)}_i |p\,^3{\rm He}\rangle \ ,
\label{eq:den1}
\end{equation}
\begin{equation}
\rho^{(2)}(x) = \langle ^4{\rm He}|\sum_{i<j}
\delta(x-r_{ij}) O^{(2)}_{ij} |p\,^3{\rm He}\rangle \ ,
\label{eq:den2}
\end{equation}
where $O^{(1)}_i$ is the (lowest order) Gamow-Teller operator of
Eq.~(\ref{eq:1baj}) at $q$=$0$, and $O^{(2)}_{ij}$ is the most
important $\Delta$-excitation current associated with diagrams of
type (a) in Fig.~\ref{fig:jd}.  These densities are normalized
such that
\begin{equation}
\int_0^\infty \, dx \, \rho^{(\alpha)} (x) = 
O^{(\alpha)}\!-\!{\rm contribution} \ .
\end{equation}

\begin{figure}[bth]
\let\picnaturalsize=N
\def\picsize{5in}
\def\picfilenamea{dens1.eps}
\ifx\nopictures Y\else{\ifx\epsfloaded Y\else\input epsf \fi
\let\epsfloaded=Y
\centerline{
\ifx\picnaturalsize N\epsfxsize \picsize\fi \epsfbox{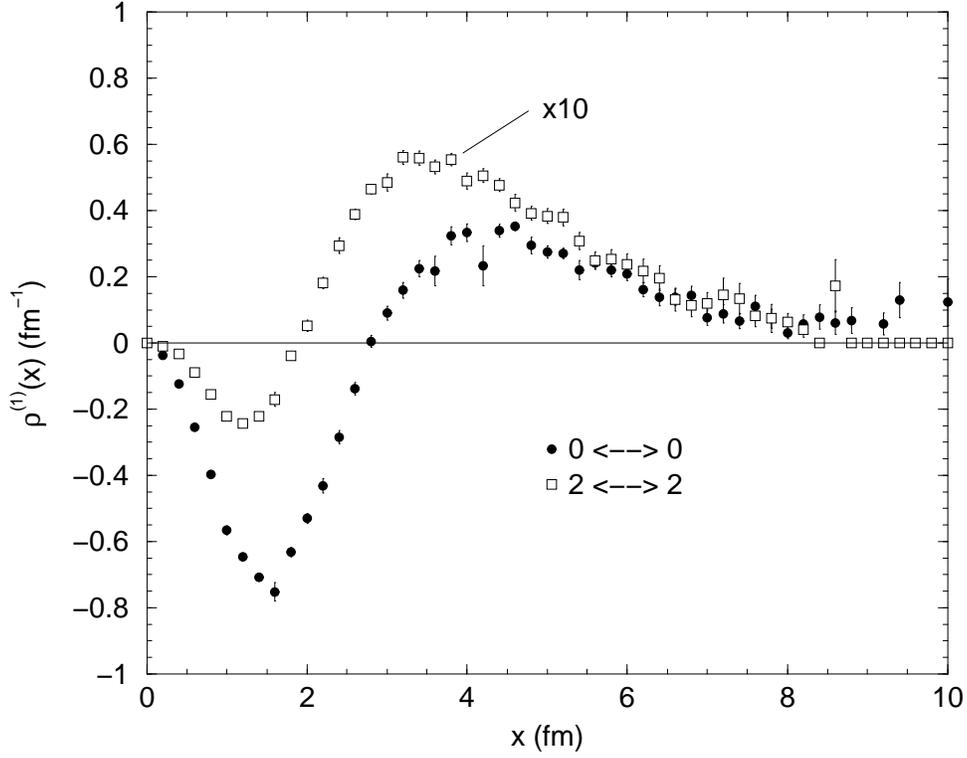}
 }}\fi
\caption{Contributions to the density function $\rho^{(1)}(x)$, defined in 
Eq.~(\ref{eq:den1}), due to transitions involving 
the $L\!=\!0 \rightarrow L\!=\!0$ (filled circles) 
and $L\!=\!2 \rightarrow L\!=\!2$ (opaque squares) components in the 
$^{3}$He and $^{4}$He wave functions. Note that the $2\rightarrow 2$ 
density function 
has been multiplied by a factor of 10, for ease of presentation.}
\label{fig:den1}
\end{figure}

In Fig.~\ref{fig:den1} we display separately the contributions
to $\rho^{(1)}(x)$ due to transitions involving the $L$=$0\rightarrow $$L$=$0$
and $L$=$2\rightarrow $$L$=$2$ components of the $^3$He and $^4$He wave
functions.  Note that the $L$=$0 \rightleftharpoons $$L$=$2$ transitions vanish, since the
Gamow-Teller operator has no dependence on the spatial coordinates
in the $q$=$0$ limit.   The $0 \rightarrow 0$ density, while much
larger than the $2 \rightarrow 2$ density, consists of positive
and negative pieces, which nearly cancel out in the integral.
Indeed, out of a total integral of $0.19$, the
$0 \rightarrow 0$ and $2 \rightarrow 2$ contributions are, respectively,
$0.02$ and $0.17$.  
It is important to reemphasize that in the $0 \rightarrow 0$ integral the whole 
contribution comes from the mixed symmetry
S$^\prime$-states of the $^{3}$He and $^{4}$He 
wave functions, since the Gamow-Teller operator, in the 
$q=0$ limit, cannot connect their dominant (symmetric) S-states,
as already pointed out in Sec.~\ref{subsec:mot}. 
This fact has been analitically verified using a simplified 
form for the nuclear wave functions, given by (for $^4$He, as an example):

\begin{equation}
\Psi_4 \simeq \Bigg[ 1 + \sum_{i<j} u^{\sigma,4}(r_{ij}) \,
\bbox{\sigma}_i \cdot \bbox{\sigma}_j + u^{t\tau,4}(r_{ij})\, S_{ij}\,
\bbox{\tau}_i \cdot \bbox{\tau}_j \Bigg]
\Bigg[ \prod_{i<j} f^{c,4}(r_{ij}) \Bigg]\Phi_4 \ , 
\label{eq:simplepsi}
\end{equation}
where $\Phi_4 = {\rm det} [ p\!\uparrow_1, p\!\downarrow_2,
n\!\uparrow_3, n\!\downarrow_4]$ 
is the spin-isospin Slater determinant, $f^{c,4}(r)$,
$u^{\sigma,4}(r)$, and $u^{t\tau,4}(r)$
are central, spin-spin, and tensor correlation functions, respectively. 
The non-central terms in Eq.~(\ref{eq:simplepsi})
generate the S$^\prime$- and D-state components.

\begin{figure}[bth]
\let\picnaturalsize=N
\def\picsize{5in}
\def\picfilenamea{dens2.eps}
\ifx\nopictures Y\else{\ifx\epsfloaded Y\else\input epsf \fi
\let\epsfloaded=Y
\centerline{
\ifx\picnaturalsize N\epsfxsize \picsize\fi \epsfbox{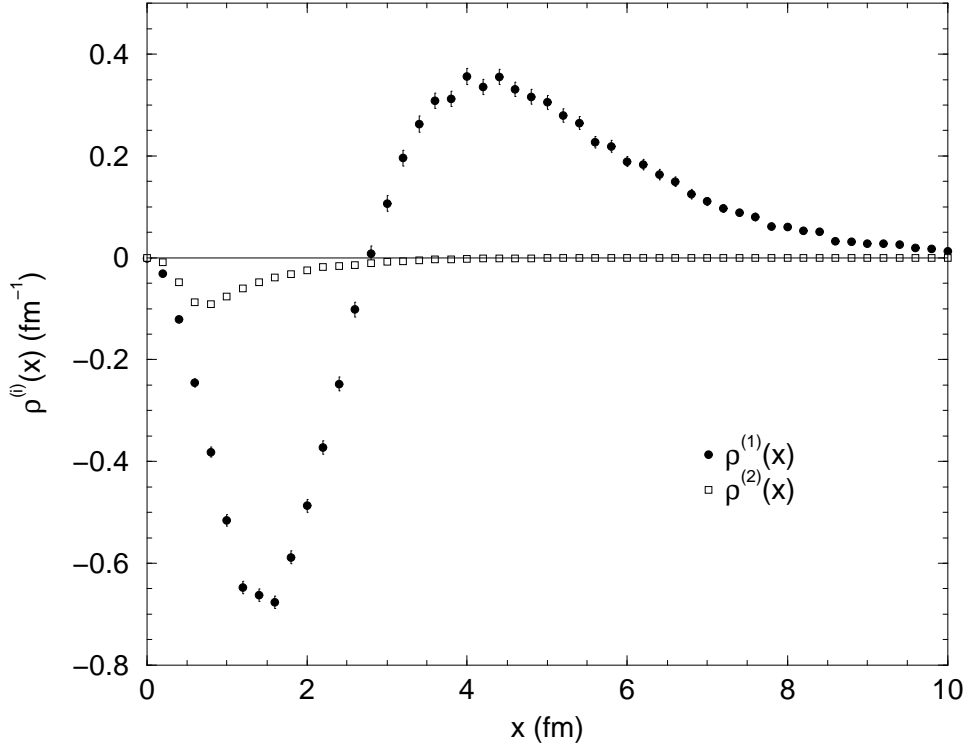}
 }}\fi
\caption{Density functions $\rho^{(1)}(x)$ (filled circles) 
and $\rho^{(2)}(x)$ (opaque squares),  defined in 
Eqs.~(\ref{eq:den1})--(\ref{eq:den2}).}
\label{fig:den2}
\end{figure}

Finally, in Fig.~\ref{fig:den2}, we display both the 
density functions $\rho^{(1)}(x)$ and 
$\rho^{(2)}(x)$.  The density function $\rho^{(2)}(x)$, 
although much smaller than $\rho^{(1)}(x)$, 
has no zeros, and consequently its integral is comparable
to that of $\rho^{(1)}(x)$.

It is interesting to examine the \lq\lq small\rq\rq $M_1$ RME induced
by the weak vector current.  It is dominated by the contributions
due to two-body currents, which interfere destructively with
(and, in fact, are much larger in magnitude than) those from
one-body currents.  This matrix element can be approximately
related to that occurring in the $n\,^3$He radiative
capture at thermal neutron energies~\cite{Sch92}.
Ignoring isospin-symmetry breaking, one has

\begin{equation}
\mid p\,^3{\rm He}\rangle \simeq
\frac{C_0}{\sqrt{2}}\, T_{+}\,\mid T\!=\!1, M_T\!=\!0 \rangle \ ,
\label{eq:n3he1}
\end{equation}
and hence, in a schematic notation, 
\begin{eqnarray}
\langle ^4{\rm He}\mid \hat{{\bf e}}_{\lambda}^* \cdot
{\bf j}_z^\dagger(\gamma) \mid n\,^3{\rm He}\rangle 
&\simeq& \frac{1}{\sqrt{2}} 
\langle ^4{\rm He}\mid \hat{{\bf e}}_{\lambda}^* \cdot
{\bf j}_z^\dagger(\gamma) \mid T\!=\!1, M_T\!=\!0 \rangle \nonumber \\
&\simeq&-\frac{1}{2\,C_0} 
\langle ^4{\rm He}\mid \hat{{\bf e}}_{\lambda}^* \cdot
{\bf j}_+^\dagger({\rm V}) \mid p\,^3{\rm He}\rangle 
\label{eq:n3he}
\end{eqnarray}
where $C_0$ is the Gamow penetration factor,
${\bf j}_z(\gamma)$ is the electromagnetic current, and use
has been made of the CVC relation to relate the commutator
$\left[ T_{+}\, , \, {\bf j}_{+}^\dagger({\rm V}) \right]$
to ${\bf j}^\dagger_z(\gamma)$.  Note that in the first line
of Eq.~(\ref{eq:n3he}) the contribution from the $T,M_T=0,0$
(1+3)-state has been neglected, since the isoscalar magnetic moment
of the nucleon is a factor $\simeq 5$ smaller than the isovector
one, and the dominant two-body electromagnetic currents are isovector.
On the basis of Eq.~(\ref{eq:n3he}), one would
predict $n\,^3$He radiative capture cross sections, at
zero energy, of 227 $\mu$b, 142 $\mu$b, and 480 $\mu$b with
one-body, one- plus two-body, and full currents--the latter
include the $\Delta$-excitation currents treated in perturbation
theory (PT), which severely overestimates their contribution~\cite{Sch92}. 
The value 480 $\mu$b is almost an order of magnitude larger than
the measured cross section, $(55\pm3) \mu$b~\cite{Wer91}.
Ignoring the $\Delta$ contribution, for which the PT estimate
is known to be unrealistic, the result obtained with
one- and two-body currents (the model-independent
ones of Sec.~\ref{sec:vcnt}), 142 $\mu$b, is still too large by
a factor $\simeq 2.6$.  However, the approximations made
in Eqs.~(\ref{eq:n3he1})--(\ref{eq:n3he}) are presumably too rough
for a reaction as delicate as the $n\,^3$He capture (see discussion
in Sec.~\ref{subsec:mot}).  Indeed, this process
provides a sensitive testing ground for models of interactions and
currents.  A calculation of its cross section with CHH wave functions
is currently underway.

In Table~\ref{tb:rme3s1_1b} we list the 
one-body axial current contributions at two values of $q$,
0 and 19.2 MeV/c, corresponding to the lowest and
highest momentum transfers allowed by the $p\,^3$He kinematics. 
A number of comments are in order.  Firstly, the RME associated
with the Gamow-Teller operator, labelled NR in the table, has a rather
strong dependence on $q$.  At $q$=0 this RME is suppressed (see
discussion above).  When $q > 0$, however, the next term
in the expansion of the plane wave in Eq.~(\ref{eq:1baj}) leads
to an operator having the structure
${\tau}_{i,\pm} \,{\bbox{\sigma}}_{i} \,r_{i,z}^{2}$, which
can connect the \lq\lq large\rq\rq S- and D-state components
of the bound-state wave functions.  Its contribution, although
of order $(qR)^2 \simeq 0.02$ ($R \simeq 1.4$ fm is the $\alpha$-particle
radius), is not negligible.
Secondly, the suppression mechanism referred to above also makes
the relativistic corrections to the Gamow-Teller operator of 
Eq.~(\ref{eq:1arc}) relatively important.  Thirdly, the induced
pseudoscalar term, last term in Eq.~(\ref{eq:1arc}), is purely
longitudinal, and itself suppressed, since it is proportional
to the NR operator.

In Table~\ref{tb:rme3s1_c} we report the cumulative contributions 
to the $L_1({\rm A})$ and $E_1({\rm A})$ RMEs at $q$=0 and
19.2 MeV/c.  The momentum transfer dependence of the results
originates from that of the one-body currents,
the mesonic and $\Delta$-excitation current contributions are, in
fact, very weakly dependent on $q$.  Note that the results
obtained by treating the $\Delta$-currents either with the TCO
method or in perturbation theory (PT) differ by 1--2 \%.  This is because
the $N$$\Delta$ axial coupling constant $g_A^*$ is determined by
fitting, independently in the TCO and PT schemes, the Gamow-Teller
matrix element of tritium $\beta$-decay.  This procedure severely reduces
the model dependence of the weak axial current.
Finally, we note that, if the quark model value were to be used for $g_A^*$
($g_A^*=6\sqrt{2}/5 g_A$), the $L_1({\rm A})$ ($E_1({\rm A})$) RME
at $q$=19.2 MeV/c would have been $-0.489\times 10^{-1}$
($-0.716\times 10^{-1}$) using the TCO method and $-0.150\times 10^{-1}$
($-0.234\times 10^{-1}$) in the PT treatment, respectively.

\subsection{P-wave Capture}
\label{subsec:pwave}

There are four P-wave capture channels: 
$^3$P$_0$,  $^1$P$_1$,  $^3$P$_1$, and $^3$P$_2$. Note that $^1$P$_1$ and $^3$P$_1$ 
are coupled channels (see Sec.~\ref{sec:chhs}).
The $^3$P$_0$ capture is induced by the weak axial charge and 
the longitudinal component of the weak axial current 
via the $C_{0}({\rm A})$ and $L_{0}({\rm A})$ multipoles, 
respectively.  The associated RMEs, as defined in Eq.~(\ref{newRME2}), are
listed in Table~\ref{tb:rme3p0}.  The two-body axial charge operators
of Sec.~\ref{sec:acrg}, among which the pion-exchange term is dominant,
give a $\simeq 20$ \% correction to the one-body
contribution in the $C_0({\rm A})$ RME.  The $L_0({\rm A})$ RME
is about 40 \% of, and has the same sign as, the $C_0({\rm A})$
RME.  This positive relative sign produces a destructive
interference between these RMEs in the cross section, substantially
reducing the $^3$P$_0$ overall contribution to the $S$-factor, as discussed
in Sec.~\ref{sec:xs}.  The $C_0({\rm A})$ and $L_0({\rm A})$
RMEs are expected to have the same sign, as justified by
the following argument.  The $C_0(q;{\rm A})$ multipole operator can be written, in the
$q \rightarrow 0$ limit, as

\begin{eqnarray}
C_0(q;{\rm A})&\rightarrow&- \frac{1}{\sqrt{4\pi}} \frac{g_A}{2\, m}
\sum_i \Big[ \tau_{i,\pm} \, \bbox{\sigma}_i\cdot \> ,\> {\bf p}_i \Big]_+ \nonumber \\
&\simeq& \frac{\rm i}{\sqrt{4\pi}} \frac{g_A}{2} \sum_i 
\Big[ \tau_{i,\pm} \, \bbox{\sigma}_i\cdot \>{\bf r}_i\> ,\> H \,\Big] \ ,
\label{eq:simplec0} 
\end{eqnarray}
where we have used the approximate relation ${\bf p}_i \simeq
-{\rm i}\, m \left[ {\bf r}_i \>,\> H \,\right]$ (violated by
the momentum-dependent components of the two-nucleon interaction), and in
the second line of Eq.~(\ref{eq:simplec0}) have ignored, in a rather
cavalier fashion, terms like
$\tau_{i,\pm} \, \bbox{\sigma}_{i}\cdot H {\bf r}_{i} -
{\bf r}_{i} H \cdot \bbox{\sigma}_{i} \tau_{i,\pm}$.  For the $L_0(q;{\rm A})$
multipole we find in the same limit

\begin{equation}
L_0(q;{\rm A})\rightarrow 
\frac{\rm i}{\sqrt{4\pi}}\frac{ g_A}{3} \, q \,
\sum_{i}\tau_{i,\pm} {\bbox{\sigma}}_{i}\cdot{\bf{r}}_{i} \ ,
\label{eq:simplel0} 
\end{equation}
and therefore we would expect the $C_0({\rm A})$ and $L_0({\rm A})$
RMEs to be approximately in the ratio

\begin{equation}
\frac{C_0({\rm A})}{L_0({\rm A})} \simeq \frac{3}{2} \frac{E_3-E_4}{q} \ ,
\end{equation}
which, given the rather severe
approximations made in deriving Eq.~(\ref{eq:simplec0}),
is reasonably close to the (one-body) value
obtained in the calculation (1.6 versus 2.0).

The $^1$P$_1$ and $^3$P$_1$ captures are induced by the weak vector charge
and current, and weak axial current via the multipoles
$C_1({\rm V})$, $L_1({\rm V})$, $E_1({\rm V})$, 
and $M_1({\rm A})$.  The calculated values for the associated RMEs
are listed in Tables~\ref{tb:rme1p1} and~\ref{tb:rme3p1}.
The RME magnitudes of the weak vector transitions in $^3$P$_1$
capture are much smaller than those in $^1$P$_1$ capture.  In the long-wavelength
approximation, the one-body $C_1({\rm V})$, $L_1({\rm V})$, and
$E_1({\rm V})$ multipoles are independent of spin, and therefore
cannot connect the dominant part of the $^3$P$_1$ wave function,
which has total spin $S$=1, to the S-wave component of $^4$He, which
has $S$=0.  This is not the case for the $^1$P$_1$ channel, in which
the total spin $S$=0 term is in fact largest.  Indeed, because of
this suppression, the two-body weak vector charge and current
contributions are found to be dominant in $^3$P$_1$ capture.
The situation is reversed for the axial transition, since there
the spin-flip nature of the $M_1({\rm A})$ multipole
makes the associated RME in $^3$P$_1$ larger than that in
$^1$P$_1$ (in absolute value).

The $E_1({\rm V})$ operator can be shown to have the
long-wavelength form~\cite{Viv00}

\begin{equation}
E_1(q;{\rm V }) =
-{\sqrt{2} \over q} \left[ H \, , \, C_1 (q;{\rm V}) \right]  \ ,
\end{equation}
and so the $E_1({\rm V})$ and $C_1({\rm V})$
RMEs would be expected to be in the ratio

\begin{equation}
\frac{E_1(q;{\rm V})}{C_1(q;{\rm V})} \simeq \sqrt{2} \,
\frac{E_3-E_4}{q} \simeq 1.51 \ ,
\end{equation}
assuming the validity of the long-wavelength approximation, and
that the CHH wave functions are truly exact eigenfunctions of the
Hamiltonian.  We reiterate here that the currents used in the present work
satisfy the continuity equation only with the $v_6$ part of the
AV14 and AV18 interactions, namely in momentum space
${\bf q} \cdot {\bf j}({\bf q};{\rm V}) =
\left[ T+v_6 \, , \, \rho^{(1)}_{\rm NR}({\bf q};{\rm V})\right]$.
The currents induced by the momentum dependent components of the
interactions, such as the spin-orbit term, have been neglected.  Thus
the ratio obtained in the calculation is 1.34 for the $^1$P$_1$ channel,
somewhat smaller than the expected value presumably because of the
\lq\lq missing\rq\rq currents and the approximate eigenstate property
satisfied by the present CHH (variational) wave functions.
These same cautionary remarks also apply to the comparison between
the $C_1({\rm V})$ and $L_1({\rm V})$ RMEs, which should be related to
each other via Eq.~(\ref{eq:c0l0}). 

The situation is more delicate in $^3$P$_1$ capture, since this transition
is suppressed.  Here the long-wavelength approximation of the $E_1({\rm V})$
multipole is inadequate, and higher order terms in the power expansion
in $q$ need to be retained, so called retardation terms.  In fact the situation
is closely related to that of electric dipole transitions in
$pd$ radiative capture at very low energies (0--100 keV).  We refer
the reader to Ref.~\cite{Viv00} for a thorough discussion of these
issues.

\begin{figure}[bth]
\let\picnaturalsize=N
\def\picsize{5in}
\def\picfilenamea{dens_ia_3p2.eps}
\ifx\nopictures Y\else{\ifx\epsfloaded Y\else\input epsf \fi
\let\epsfloaded=Y
\centerline{
\ifx\picnaturalsize N\epsfxsize \picsize\fi \epsfbox{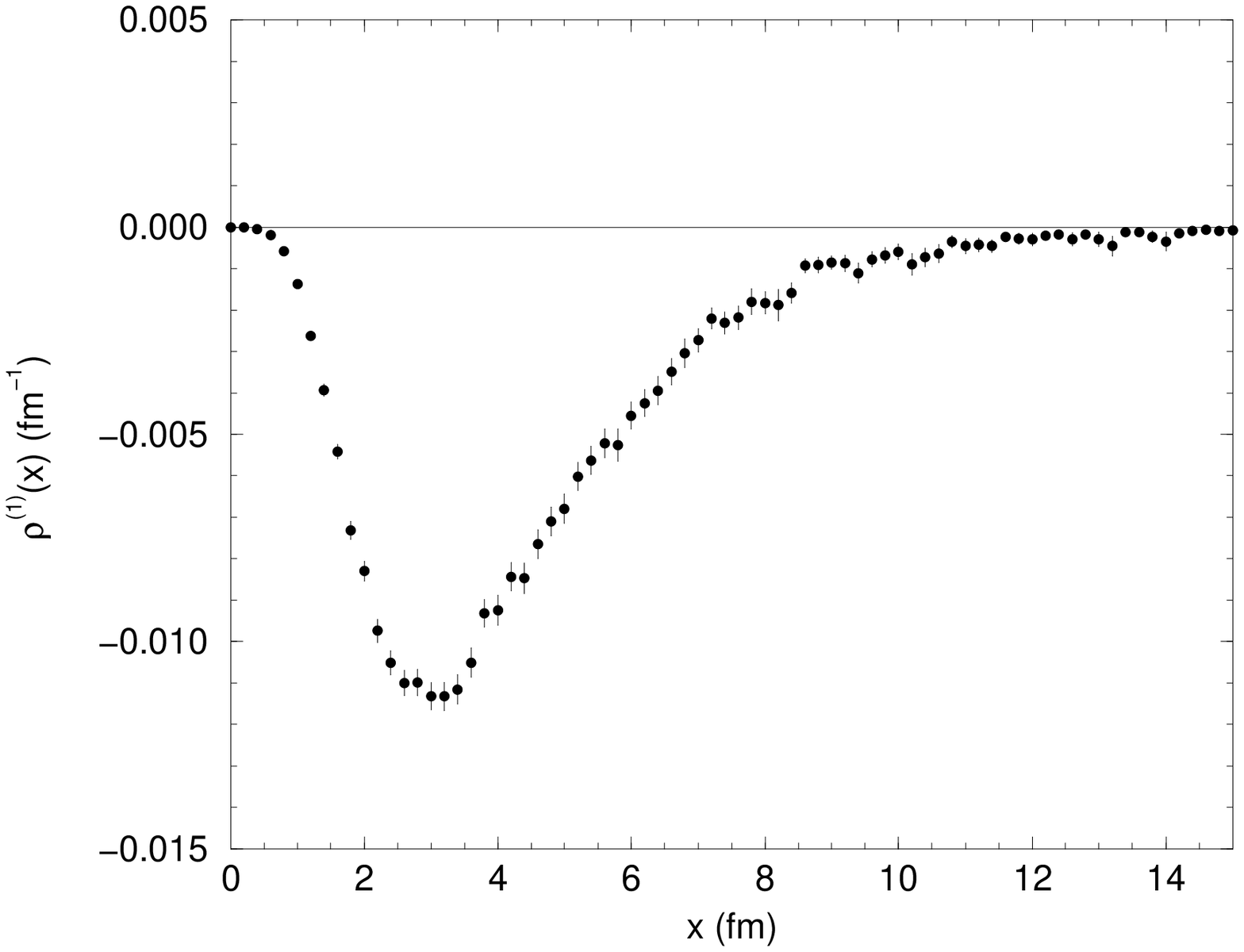}
 }}\fi
\caption{Density function $\rho^{(1)}(x)$  defined in 
Eq.~(\ref{eq:den1}) in the $^3$P$_2$ capture channel.}
\label{fig:den3p2}
\end{figure}

The $^3$P$_2$ capture is induced by the weak axial charge and current, 
and weak vector current operators via the multipoles 
$C_2({\rm A})$, $L_2({\rm A})$, $E_2({\rm A})$, and $M_2({\rm V})$.
The associated RMEs are listed in Table~\ref{tb:rme3p2}.
The $L_2({\rm A})$ and $E_2({\rm A})$ RMEs are comparable
to the $L_1({\rm A})$ and $E_1({\rm A})$ RMEs in $^3$S$_1$
capture, and are dominated by the contributions of one-body currents.
In fact, the latter can now connect the large
S-wave components of the three- and four-nucleon bound states.
The density function $\rho^{(1)}(x)$, defined in analogy to
Eq.~(\ref{eq:den1}) (but for the $^3$P$_2$ channel), is
displayed in Fig.~\ref{fig:den3p2}, and should be compared
to that in Fig.~\ref{fig:den2} for $^3$S$_1$ capture.
While smaller in magnitude than the latter--after all, the $^3$P$_2$
transition is inhibited with respect to the $^3$S$_1$ transition
by a factor $\simeq qR$ and the presence of the centrifugal barrier--the $^3$P$_2$
density has the same sign, and therefore its integral turns out to be
comparable to that of the $^3$S$_1$ density. 

\subsection{Model Dependence}
\label{subsec:modd}

In Table~\ref{tb:sfc_contr} we list, for all S- and P-wave channels,
the $S$-factor values obtained with the AV18/UIX, AV18, and AV14/UVIII
interactions.  Note that the sum of the channel contributions is a few
\% smaller than the total result reported at the bottom of the
table (see end of Sec.~\ref{sec:xs}).  The $N$$\Delta$ axial
coupling constant is determined by fitting the Gamow-Teller matrix element
in tritium $\beta$-decay, within each given Hamiltonian model.  As a result
of this procedure the model dependence of the $S$-factor predictions
is substantially reduced.

Inspection of Table~\ref{tb:sfc_contr} shows that inclusion
of the three-nucleon interaction reduces the total $S$-factor by
about 20 \% (compare the AV18 and AV18/UIX results).  This decrease
is mostly in the $^3$S$_1$ contribution, and can be traced back
to a corresponding reduction in the magnitude of the one-body
axial current matrix elements.  The latter are sensitive to the
triplet scattering length, for which the AV18 and AV18/UIX
models predict, respectively, 10.0 fm and 9.13 fm (see Table~\ref{tb:scl}).

The comparison between the AV18/UIX and AV14/UVIII models, which
both reproduce the measured bound-state properties and low-energy
scattering parameters of the three- and four-nucleon systems, suggests
a rather weak model dependence.  It is important to reiterate
that this is accomplished by virtue of the procedure used to constrain
the axial current.  Indeed, the AV18/UIX and AV14/UVIII
$^3$S$_1$ contributions to the $S$-factor obtained with
one-body currents only are, respectively, $26.4\times 10^{-20}$
keV~b and $35.8\times 10^{-20}$ keV~b.  This difference is presumably due
to the stronger tensor component of AV14 as compared to that of AV18. 

Finally, the $^3$S$_1$ contribution to the $S$-factor
obtained with the AV14/UVIII model in the present work, $6.60\times 10^{-20}$
keV~b, is to be compared with the older prediction of Ref.~\cite{Sch92},
$1.3\times 10^{-20}$ keV~b.  It is important to point out that the
older calculation (i) used the long-wavelength form of the
$E_1({\rm A})$ and $L_1({\rm A})$ operators, (ii) ignored the contributions
of transitions induced by the axial charge and vector current, (iii) retained
only the leading non-relativistic (Gamow-Teller) term of the
single-nucleon axial current, and (iv) employed
bound and continuum wave functions, obtained with the Variational
Monte Carlo (VMC) method.  In regard to this last point, we note that, for
example, the $\overline{E}_{1}(q\!=\!0; {\rm A})$ RME calculated 
in Ref.~\cite{Sch92} with the Gamow-Teller operator is
$0.613 \times 10^{-1}$ fm$^{3/2}$ versus a value of
0.119 fm$^{3/2}$ obtained here.  The factor $\simeq 2$ increase is
only due to differences in the wave functions.  The present CHH
wave functions are expected to be more accurate than
the VMC wave functions of Ref.~\cite{Sch92}.
\section*{Acknowledgments}
The authors wish to thank V.R.\ Pandharipande,
D.O.\ Riska, P.\ Vogel, and R.B.\ Wiringa for useful discussions,
and J.\ Carlson for a critical reading of the manuscript.
M.V.\ and R.S.\ acknowledge partial financial support of NATO
through the Collaborative Research Grant No. 930741.  The support
of the U.S. Department of Energy under contract number DE-AC05-84ER40150
is gratefully acknowledged by L.E.M.\ and R.S.  Finally, some
of the calculations were made possible by grants of computing time from
the National Energy Research Supercomputer Center in Livermore.
\begin{table}
\caption{The $hep$ $S$-factor, in units of $10^{-20}$ keV~b, calculated
with CHH wave functions corresponding to the AV18/UIX Hamiltonian model,
at $p\,^3$He c.m.\ energies $E$=0, 5, and 10 keV.  The rows
labelled \lq\lq one-body\rq\rq and \lq\lq full\rq\rq list the
contributions obtained by retaining the one-body only and both
one- and many-body terms in the nuclear weak current.  The contributions due 
the $^3$S$_1$ channel only and all S- and P-wave channels are
listed separately.}
\begin{tabular}{ccccccc}
& \multicolumn{2}{c} {$E$=$0$ keV} &
  \multicolumn{2}{c} {$E$=$5$ keV} &
  \multicolumn{2}{c} {$E$=$10$ keV} \\
& $^3$S$_1$ & S+P & $^3$S$_1$ & S+P & $^3$S$_1$ & S+P\\
\tableline
one-body  &26.4  & 29.0 & 25.9 & 28.7 & 26.2 & 29.3 \\
full      &6.38  & 9.64 & 6.20 & 9.70 & 6.36 & 10.1
\end{tabular}
\label{tb:sfact}
\end{table}
\begin{table}
\caption{Binding energies in MeV of $^{4}$He calculated with the CHH method
using the AV18 and AV18/UIX, and the older AV14 and AV14/UVIII, 
Hamiltonian models.  Also listed are the corresponding 
\lq\lq exact\rq\rq GFMC results~\protect\cite{Pud97,Car90a} 
and the experimental value.}
\begin{tabular}{lcc}
    Model    &  CHH & GFMC  \\
\tableline
AV18     &  24.01  & 24.1(1)  \\
AV18/UIX   & 27.89 & 28.3(1)  \\
AV14     &  23.98  & 24.2(2)  \\
AV14/UVIII & 27.50 & 28.3(2)  \\
\tableline
EXP&\multicolumn{2}{c}{28.3}
\end{tabular}
\label{tb:be}
\end{table}
\begin{table}
\caption{Binding energies, $B_3$, of $^3$He, and $p\,^3$He singlet
and triplet S-wave scattering lengths, $a_{\rm s}$ and
$a_{\rm t}$, calculated with the CHH method
using using the AV18 and AV18/UIX, and the older AV14 and AV14/UVIII, 
Hamiltonian models.  The corresponding experimental values are also listed.}
\begin{tabular}{lccc}
Model      & $B_3$(MeV)  & $a_{\rm s}$(fm) & $a_{\rm t}$(fm) \\
\tableline
AV14       & 7.03 &         &        \\
AV18       & 6.93 &  12.9   & 10.0   \\
AV14/UVIII & 7.73 &         & 9.24   \\
AV18/UIX   & 7.74 &  11.5   & 9.13   \\
\tableline
EXP        & 7.72 &  10.8$\pm$2.6~\protect\cite{AK93} 
                  & 8.1$\pm$0.5~\protect\cite{AK93} \\
           &      &         & 10.2$\pm$1.5~\protect\cite{TEG83}
\end{tabular}
\label{tb:scl}
\end{table}
\begin{table}
\caption{Phase-shift and mixing-angle parameters (in deg) for $p\,^3$He
elastic scattering at c.m.\ energy of $1.2$ MeV, calculated with the CHH
method using the AV18 and AV18/UIX Hamiltonian models.  The corresponding
experimental values obtained in the phase-shift analysis of
Ref.~\protect\cite{AK93} are also listed.}  
\begin{tabular}{lddd}
parameter   & AV18  & AV18/UIX & PSA \\
\tableline
$^1$S$_0$  & --33.3 &  --31.3   &  --27.4$\pm$3.5      \\
$^3$S$_1$  & --28.8 &  --27.1   &  --26.5$\pm$0.6      \\
$^3$P$_0$  &    4.1 &     3.2   &     2.6$\pm$0.6      \\
$^3$P$_1$  &    8.1 &     7.4   &    10.1$\pm$0.5      \\
$^3$P$_2$  &    7.7 &     6.9   &     8.9$\pm$0.5      \\
$^1$P$_1$  &    6.5 &     5.5   &     4.2$\pm$1.5      \\
$\epsilon(1^-)$  & --14.7   &   --13.2   &   --7.8$\pm$0.6      \\
\end{tabular}
\label{tb:ps}
\end{table}
\begin{table}
\caption{Contributions to the Gamow-Teller (GT) matrix element of tritium 
$\beta$-decay, obtained with the CHH trinucleon wave functions corresponding
to the AV18/UIX Hamiltonian model.  The rows labelled 
\lq\lq one-body NR\rq\rq and
\lq\lq one-body RC\rq\rq list the contributions associated 
with the single-nucleon axial current operators of Eq.~(\ref{eq:1baj})
and Eq.~(\ref{eq:1arc}), respectively, while the row labelled 
\lq\lq mesonic\rq\rq 
lists the sum of the contributions due to the $\pi$-, $\rho$-,
and $\rho \pi$-exchange axial current operators of
Eqs.~(\ref{eq:a2pi})--(\ref{eq:a2rp}) with cutoff masses
$\Lambda_\pi=4.8$ fm$^{-1}$ and $\Lambda_\rho=6.8$ fm$^{-1}$.  Finally,
the rows labelled \lq\lq $\Delta$-$g_{A}^{*}$\rq\rq,
\lq\lq $\Delta$-${\overline{g}}_{A}$\rq\rq,
and \lq\lq $\Delta$-renormalization\rq\rq list, respectively, the contributions
associated with panels (a)-(b), (c)-(d) and (f), and (e) and (g)-(j),
of Fig.~\ref{fig:jd}.  The cumulative result reproduces the
\lq\lq experimental value\rq\rq  0.957 for the GT matrix
element~\protect\cite{Sch98}, once the change in normalization
of the wave functions due to the presence of $\Delta$-components
is taken into account.}
\begin{tabular}{ld}
               &   GT matrix element   \\
\tableline
one-body NR       &   0.9218  \\
one-body RC       &  --0.0084  \\
mesonic           &   0.0050  \\
$\Delta$-$g_{A}^{*}$ & 0.0509  \\
$\Delta$-${\overline{g}}_{A}$ & 0.0028  \\
$\Delta$-renormalization &  0.0074 
\end{tabular}
\label{tb:h3gt}
\end{table}
\begin{table}
\caption{The values of the $N\!\rightarrow\!\Delta$ axial coupling 
constant $g_{A}^{*}$ in units of $g_{A}$, when the $\Delta$-isobar 
degrees of freedom are treated in perturbation theory (PT), or in the 
context of a TCO calculation based on the AV28Q interaction.
The purely nucleonic CHH wave functions correspond to the AV18/UIX
Hamiltonian model.}
\begin{tabular}{cc}
$\Delta$-isobar treatment & $g_{A}^{*}/g_{A}$ \\
\tableline
PT & 1.224 \\
TCO & 2.868 
\end{tabular}
\label{tb:gas}
\end{table}
\begin{table}
\caption{The wave function normalization ratios 
$\langle \Psi_{N+\Delta}\,|\,\Psi_{N+\Delta}\rangle/
\langle \Psi\,|\,\Psi\rangle$ obtained for the bound three- and 
four-nucleon systems, when the TCO calculation is based
on the AV28Q interaction.  The purely nucleonic 
CHH wave functions $|\,\Psi\rangle$ 
correspond to the AV18/UIX Hamiltonian model.}
\begin{tabular}{cccc}
Model & $^{3}$H & $^{3}$He & $^{4}$He \\  
\tableline
AV28Q & 1.0238 & 1.0234 & 1.0650 
\end{tabular}
\label{tb:normr}
\end{table}
\begin{table}
\caption{Cumulative contributions to the reduced matrix elements (RMEs)
$\overline{C_0}(q;{\rm{V}})$ and
$\overline{L_0}(q;{\rm{V}})$ in $^1$S$_0$ capture
at zero $p\,^3$He c.m.\ energy.  The momentum transfer $q$ is
19.2 MeV/c, and the results correspond to the AV18/UIX Hamiltonian
model.  The row labelled \lq\lq one-body\rq\rq lists the contributions
associated with the operators in Eq.~(4.5) for the weak vector charge 
$\rho({\rm V})$
and Eq.~(4.8) for the weak vector current ${\bf j}({\rm V})$; the row
labelled \lq\lq mesonic\rq\rq lists the results obtained by including, 
in addition,
the contributions associated with the operators in Eqs.~(4.30)--(4.31)
for $\rho({\rm V})$, and Eqs.~(4.16)--(4.17) for ${\bf j}({\rm V})$. 
The $\Delta$ terms in $\rho({\rm V})$ are neglected, while those in 
${\bf j}({\rm V})$
are purely transverse and therefore do not contribute to
the $\overline{L_0}$ RME.  Note that the RMEs are purely real and 
in fm$^{3/2}$ units.}
\begin{tabular}{ccc}
         & $\overline{C_0} (q;{\rm V})$  & $\overline{L_0} (q;{\rm V})$ \\
\tableline
one-body & $-0.857\times 10^{-2}$ & $-0.864 \times 10^{-2}$ \\
mesonic  & $-0.856\times 10^{-2}$ & $-0.919 \times 10^{-2}$    
\end{tabular}
\label{tb:rme1s0}
\end{table}
\begin{table}
\caption{Cumulative contributions to the reduced matrix elements (RMEs)
$\overline{C_1}(q;{\rm A})$, $\overline{L_1}(q;{\rm A})$,
$\overline{E_1}(q;{\rm A})$ and $\overline{M_1}(q;{\rm V})$
in $^3$S$_1$ capture at zero $p\,^3$He c.m.\ energy.  
The momentum transfer $q$ is
19.2 MeV/c, and the results correspond to the AV18/UIX Hamiltonian
model.  The row labelled \lq\lq one-body\rq\rq lists the contributions
associated with the operators in Eq.~(4.10) for the weak axial charge 
$\rho({\rm A})$,
Eq.~(4.11) for the weak axial current ${\bf j}({\rm A})$, and Eq.~(4.8)
for the weak vector current ${\bf j}({\rm V})$; the row
labelled \lq\lq mesonic\rq\rq lists the results obtained by including, 
in addition,
the contributions associated with the operators in Eqs.~(4.35)--(4.37)
for $\rho({\rm A})$, Eqs.~(4.32)--(4.34) for ${\bf j}({\rm A})$, and
Eqs.~(4.16)--(4.17) for ${\bf j}({\rm V})$; finally,
the row labelled \lq\lq $\Delta$\rq\rq lists the results obtained
by also including the contributions of the operators in Eqs.~(4.50)--(4.51)
for $\rho({\rm A})$, Eqs.~(4.48)--(4.49) for ${\bf j}({\rm A})$, and
Eqs.~(4.52)--(4.53) for ${\bf j}({\rm V})$.
The $\Delta$ contributions in both $\rho({\rm A})$ and ${\bf j}({\rm A})$
are calculated with the TCO method, and take into account
the change in normalization of the wave functions due to the presence
of $\Delta$-components.  Those in ${\bf j}({\rm V})$
are calculated in perturbation theory.  Note that the RMEs are
purely imaginary and in fm$^{3/2}$ units.}
\begin{tabular}{ccccc}
& $\overline{C_1}(q;{\rm A})$ & $\overline{L_1}(q;{\rm A})$ 
& $\overline{E_1}(q;{\rm A})$ & $\overline{M_1}(q;{\rm V})$ \\
\tableline
one-body & $0.147\times 10^{-1}$ & $-0.730\times 10^{-1}$ & $-0.106$
         & $ 0.333\times 10^{-2}$ \\
mesonic  & $0.156\times 10^{-1}$ & $-0.679\times 10^{-1}$ & 
						$-0.984\times 10^{-1}$ 
         & $-0.263\times 10^{-2}$ \\
$\Delta$ & $0.155\times 10^{-1}$ & $-0.293\times 10^{-1}$ & 
						$-0.440\times 10^{-1}$ 
         & $-0.484\times 10^{-2}$ 
\end{tabular}
\label{tb:rme3s1}
\end{table}
\begin{table}
\caption{One-body contributions, at momentum transfers
$q$=0 and 19.2 MeV/c, to the reduced matrix elements (RMEs)
$\overline{L_1}(q;{\rm A})$ and $\overline{E_1}(q;{\rm A})$ 
in $^3$S$_1$ capture at zero $p\,^3$He c.m.\ energy.  The results
correspond to the AV18/UIX Hamiltonian model.  The rows
labelled \lq\lq NR\rq\rq and \lq\lq RC\rq\rq
list the contributions obtained with the operators 
of Eq.~(4.12) and Eq.~(4.13),
respectively; the row labelled \lq\lq IPS\rq\rq lists the contribution
of the induced pseudoscalar current only (last term in Eq.~(4.13)).
Note that the RMEs are purely imaginary and in fm$^{3/2}$ units.}
\begin{tabular}{ccccc}
& \multicolumn{2}{c} {$\overline{L_1}(q;{\rm A})$} & 
  \multicolumn{2}{c} {$\overline{E_1}(q;{\rm A})$} \\
& $q$=0 MeV/c & $q$=19.2 MeV/c 
& $q$=0 MeV/c & $q$=19.2 MeV/c \\
\tableline
NR  &  $-0.726\times 10^{-1}$ & $-0.586\times 10^{-1}$ & $-0.103$ 
    &  $-0.838\times 10^{-1}$ \\
RC  &  $-0.154\times 10^{-1}$ & $-0.145\times 10^{-1}$ & $-0.220\times 10^{-1}$
    &  $-0.219\times 10^{-1}$ \\
IPS &                         & $ 0.741\times 10^{-3}$   &   
                              &   
\end{tabular}
\label{tb:rme3s1_1b}
\end{table}
\begin{table}
\caption{Cumulative contributions, at momentum transfers
$q$=0 and 19.2 MeV/c, to the reduced matrix elements (RMEs)
$\overline{L_1}(q;{\rm A})$ and $\overline{E_1}(q;{\rm A})$ of the weak
axial currrent in $^3$S$_1$ capture at zero $p\,^3$He c.m.\ energy.  
The results correspond to the AV18/UIX Hamiltonian model.
The row labelled \lq\lq one-body\rq\rq lists the contributions
associated with the operator in Eq.~(4.11); the row
labelled \lq\lq mesonic\rq\rq lists the results obtained by including, 
in addition,
the contributions associated with the operators in Eqs.~(4.32)--(4.34); 
finally,
the rows labelled \lq\lq $\Delta$-TCO\rq\rq and $\Delta$-PT list 
the results obtained
by also including the contributions of the operators
in Eqs.~(4.48)--(4.49), calculated either in the TCO scheme or in 
perturbation theory (PT).
The $\Delta$-TCO results also take into account the change in 
normalization of the
wave functions due to the presence of $\Delta$-components.
Note that the RMEs are purely imaginary and in fm$^{3/2}$ units.}
\begin{tabular}{ccccc}
& \multicolumn{2}{c} {$\overline{L_1}(q;{\rm A})$} & 
  \multicolumn{2}{c} {$\overline{E_1}(q;{\rm A})$} \\
& $q$=0 MeV/c & $q$=19.2 MeV/c 
& $q$=0 MeV/c & $q$=19.2 MeV/c \\
\tableline
one-body     & $-0.880\times 10^{-1}$ & $-0.730\times 10^{-1}$ & --0.125 
                                                               & --0.106 \\
mesonic      & $-0.829\times 10^{-1}$ & $-0.679\times 10^{-1}$ & --0.117
                                                    & $-0.984\times 10^{-1}$ \\
$\Delta$-TCO & $-0.440\times 10^{-1}$ & $-0.293\times 10^{-1}$ 
                         &  $-0.625\times 10^{-1}$ & $-0.440\times 10^{-1}$ \\ 
\tableline
$\Delta$-PT  & $-0.447\times 10^{-1}$ & $-0.298\times 10^{-1}$ 
                             & $-0.631\times 10^{-1}$ & $-0.443\times 10^{-1}$ 
\end{tabular}
\label{tb:rme3s1_c}
\end{table}
\begin{table}
\caption{Cumulative contributions to the reduced matrix elements (RMEs)
$\overline{C_0}(q;{\rm A})$ and
$\overline{L_0}(q;{\rm A})$ in $^3$P$_0$ capture
at zero $p\,^3$He c.m.\ energy.  The momentum transfer $q$ is
19.2 MeV/c, and the results correspond to the AV18/UIX Hamiltonian
model.  The row labelled \lq\lq one-body\rq\rq lists the contributions
associated with the operators in Eq.~(4.10) 
for the weak axial charge $\rho({\rm A})$
and Eq.~(4.11) for the weak axial current ${\bf j}({\rm A})$; the row
labelled \lq\lq mesonic\rq\rq lists the results obtained by including, 
in addition,
the contributions associated with the operators in Eqs.~(4.35)--(4.37)
for $\rho({\rm A})$, and Eqs.~(4.32)--(4.34) for ${\bf j}({\rm A})$; finally,
the row labelled \lq\lq $\Delta$\rq\rq lists the results obtained
by also including the contributions of the operators in Eqs.~(4.50)--(4.51)
for $\rho({\rm A})$, and Eqs.~(4.48)--(4.49) for ${\bf j}({\rm A})$.
The $\Delta$ contributions in both $\rho({\rm A})$ and ${\bf j}({\rm A})$
are calculated with the TCO method, and take into account
the change in normalization of the wave functions due to the presence
of $\Delta$-components.  Note that the RMEs are purely imaginary 
and in fm$^{3/2}$ units.}
\begin{tabular}{ccc}
         & $\overline{C_0}(q;{\rm A})$  & $\overline{L_0}(q;{\rm A})$ \\
\tableline
one-body &  $0.371\times 10^{-1}$ & $0.182\times 10^{-1}$ \\
mesonic  &  $0.444\times 10^{-1}$ & $0.183\times 10^{-1}$ \\
$\Delta$ &  $0.459\times 10^{-1}$ & $0.188\times 10^{-1}$ 
\end{tabular}
\label{tb:rme3p0}
\end{table}
\begin{table}
\caption{Cumulative contributions to the reduced matrix elements (RMEs) 
$\overline{C_1}(q;{\rm V})$, $\overline{L_1}(q;{\rm V})$,
$\overline{E_1}(q;{\rm V})$ and $\overline{M_1}(q;{\rm A})$ 
in $^1$P$_1$ capture at zero $p\,^3$He c.m.\ energy.  The momentum
transfer $q$ is 19.2 MeV/c, and the results correspond to the AV18/UIX 
Hamiltonian
model.  The row labelled \lq\lq one-body\rq\rq lists the contributions
associated with the operators in Eq.~(4.5) for the weak vector charge 
$\rho({\rm V})$,
Eq.~(4.8) for the weak vector current ${\bf j}({\rm V})$, and Eq.~(4.11)
for the weak axial current ${\bf j}({\rm A})$; the row
labelled \lq\lq mesonic\rq\rq lists the results obtained by including, 
in addition,
the contributions associated with the operators in Eqs.~(4.30)--(4.31)
for $\rho({\rm V})$, Eqs.~(4.16)--(4.17) for ${\bf j}({\rm V})$, and
Eqs.~(4.32)--(4.34) for ${\bf j}({\rm A})$; finally,
the row labelled \lq\lq $\Delta$\rq\rq lists the results obtained
by also including the contributions of the operators in
Eqs.~(4.52)--(4.53) for ${\bf j}({\rm V})$, and
Eqs.~(4.48)--(4.49) for ${\bf j}({\rm A})$.
The $\Delta$ contributions in ${\bf j}({\rm A})$
are calculated with the TCO method, and take into account
the change in normalization of the wave functions due to the presence
of $\Delta$-components.  Those in ${\bf j}({\rm V})$
are calculated in perturbation theory.  The $\Delta$ terms
in $\rho({\rm V})$ are neglected, while those in ${\bf j}({\rm V})$
are purely transverse and therefore do not contribute to
the $\overline{L_1}$ RME.  Note that the RMEs are purely
real and in fm$^{3/2}$ units.}
\begin{tabular}{ccccc}
& $\overline{C_1}(q;{\rm V})$ & $\overline{L_1}(q;{\rm V})$ 
& $\overline{E_1}(q;{\rm V})$ & $\overline{M_1}(q;{\rm A})$ \\
\tableline
one-body & $-0.222\times 10^{-1}$ & $-0.162\times 10^{-1}$ 
						& $-0.231\times 10^{-1}$ 
         & $-0.100\times 10^{-2}$ \\
mesonic  & $-0.222\times 10^{-1}$ & $-0.209\times 10^{-1}$ 
						& $-0.298\times 10^{-1}$ 
         & $-0.779\times 10^{-3}$ \\
$\Delta$ & & & $-0.298\times 10^{-1}$ & $-0.809\times 10^{-3}$ 
\end{tabular}
\label{tb:rme1p1}
\end{table}
\begin{table}
\caption{Cumulative contributions to the reduced matrix elements (RMEs) 
$\overline{C_1}(q;{\rm V})$, $\overline{L_1}(q;{\rm V})$,
$\overline{E_1}(q;{\rm V})$ and $\overline{M_1}(q;{\rm A})$ 
in the $^3$P$_1$ capture at zero $p\,^3$He c.m.\ energy.  The momentum
transfer $q$ is 19.2 MeV/c, and the results correspond 
to the AV18/UIX Hamiltonian
model.  Notation as in Table~\ref{tb:rme1p1}.  Note that the RMEs 
are purely real and in fm$^{3/2}$ units.}
\begin{tabular}{ccccc}
& $\overline{C_1}(q;{\rm V})$ & $\overline{L_1}(q;{\rm V})$ 
& $\overline{E_1}(q;{\rm V})$ & $\overline{M_1}(q;{\rm A})$ \\
\tableline
one-body & $0.953\times 10^{-3}$ & $0.118\times 10^{-2}$ 
						& $0.521\times 10^{-3}$ 
         & $0.304\times 10^{-1}$ \\
mesonic  & $0.217\times 10^{-2}$ & $0.174\times 10^{-2}$ 
						& $0.128\times 10^{-2}$ 
         & $0.304\times 10^{-1}$ \\
$\Delta$ & & & $0.127\times 10^{-2}$ & $0.303\times 10^{-1}$ 
\end{tabular}
\label{tb:rme3p1}
\end{table}
\begin{table}
\caption{Cumulative contributions to the reduced matrix elements (RMEs) 
$\overline{C_2}(q;{\rm A})$, $\overline{L_2}(q;{\rm A})$,
$\overline{E_2}(q;{\rm A})$ and $\overline{M_2}(q;{\rm V})$ 
in the $^3$P$_2$ capture at zero $p\,^3$He c.m.\ energy.  The momentum
transfer $q$ is 19.2 MeV/c, and the results correspond to the AV18/UIX 
Hamiltonian model.  Notation as in Table~\ref{tb:rme3s1}.  Note that the RMEs 
are purely imaginary and in fm$^{3/2}$ units.}
\begin{tabular}{ccccc}
& $\overline{C_2}(q;{\rm A})$ & $\overline{L_2}(q;{\rm A})$ 
& $\overline{E_2}(q;{\rm A})$ & $\overline{M_2}(q;{\rm V})$ \\
\tableline
one-body & $-0.146\times 10^{-3}$ & $0.236\times 10^{-1}$ 
						& $0.292\times 10^{-1}$ 
         & $-0.110\times 10^{-2}$ \\
mesonic  & $-0.114\times 10^{-3}$ & $0.236\times 10^{-1}$ 
						& $0.293\times 10^{-1}$ 
         & $-0.116\times 10^{-2}$ \\
$\Delta$ & $-0.988\times 10^{-4}$ & $0.238\times 10^{-1}$ 
						& $0.295\times 10^{-1}$ 
         & $-0.118\times 10^{-2}$ 
\end{tabular}
\label{tb:rme3p2}
\end{table}
\begin{table}
\caption{Contributions of the S- and P-wave capture channels to 
the $hep$ $S$-factor at zero $p\,^3$He c.m.\ energy in 10$^{-20}$ keV~b. 
The results correspond to the AV18/UIX, AV18 and   
AV14/UVIII Hamiltonian models.}
\begin{tabular}{cccc}
& AV18/UIX & AV18 & AV14/UVIII \\
\tableline
$^{1}$S$_{0}$ & 0.02 & 0.01 & 0.01\\
$^{3}$S$_{1}$ & 6.38 & 7.69 & 6.60\\
$^{3}$P$_{0}$ & 0.82 & 0.89 & 0.79\\
$^{1}$P$_{1}$ & 1.00 & 1.14 & 1.05\\
$^{3}$P$_{1}$ & 0.30 & 0.52 & 0.38\\
$^{3}$P$_{2}$ & 0.97 & 1.78 & 1.24\\
\tableline
TOTAL         & 9.64 & 12.1 & 10.1
\end{tabular}
\label{tb:sfc_contr}
\end{table}

\begin{references}
\bibitem{BK98} J.N.\ Bahcall and P.I.\ Krastev,
               Phys.\ Lett.\ B {\bf 436}, 243 (1998).
%
\bibitem{Fio98} G.\ Fiorentini, V.\ Berezinsky, S.\ Degl'Innocenti, 
		and B.\ Ricci,
                Phys.\ Lett.\ B {\bf 444}, 387 (1998).
%
\bibitem{Esc98} R.\ Escribano, J.M.\ Fr\`{e}re, A.\ Gevaert, 
                and P.\ Monderen, 
		Phys.\ Lett.\ B {\bf 444}, 397 (1998). 
%
\bibitem{Hor99} C.J.\ Horowitz,
                Phys.\ Rev.\ C {\bf 60}, 022801 (1999).
%
\bibitem{Alb00} W.M.\ Alberico, S.M.\ Bilenky, and W.\ Grimus,
                hep-ph/0001245.
%
\bibitem{Alb002}W.M.\ Alberico, J.\ Barnab\'{e}u, S.M.\ Bilenky, 
		and W.\ Grimus, Phys.\ Lett.\ B {\bf 478}, 208 (2000).
%
\bibitem{Fuk99} Y.\ Fukuda {\it et al.},
                Phys.\ Rev.\ Lett.\ {\bf 82}, 2430 (1999).
%
\bibitem{Smy99} M.B.\ Smy,
                hep-ex/9903034.
%
\bibitem{Suz00} Y.\ Suzuki, contribution to Lepton-Photon Symposium 99 (1999), 
		http://www-sk.icrr.u-tokyo.ac.jp/doc/sk/pub/index.html.
%
\bibitem{BBP98} J.N.\ Bahcall, S.\ Basu, and M.H.\ Pinsonneault,
                Phys.\ Lett.\ B {\bf 433}, 1 (1998).
%
\bibitem{Sch92} R.\ Schiavilla, R.B.\ Wiringa, V.R.\ Pandharipande, 
		and J.\ Carlson,
                Phys.\ Rev.\ C {\bf 45}, 2628 (1992).
%
\bibitem{MS86} S.P.\ Mikheev and A.Y.\ Smirnov,
               Nuovo Cim.\ {\bf 9C}, 17 (1986).
%
\bibitem{CWC65} A.E.\ Cox, S.A.R.\ Wynchank, and C.H.\ Collie,
                Nucl.\ Phys.\ {\bf 74}, 497 (1965).
%
\bibitem{JBB82} E.T.\ Jurney, P.J.\ Bendt, and J.C.\ Browne,
                Phys.\ Rev.\ C {\bf 25}, 2810 (1982).
%
\bibitem{Wol89} F.L.H.\ Wolfs, S.J.\ Freedman, J.E.\ Nelson, M.S.\ Dewey, 
		and G.L.\ Greene,
                Phys.\ Rev.\ Lett.\ {\bf 63}, 2721 (1989).
%
\bibitem{Wer91} R.\ Wervelman, K.\ Abrahams, H.\ Postma, J.G.L.\ Booten, 
		and A.G.M.\ Van Hees,
                Nucl.\ Phys.\ {\bf A526}, 265 (1991).   
%
\bibitem{Sch37} L.I.\ Schiff,
                Phys.\ Rev.\ {\bf 52}, 242 (1937).
%
\bibitem{Viv00} M.\ Viviani, A.\ Kievsky, L.E.\ Marcucci, S.\ Rosati, 
		and R.\ Schiavilla,
                Phys.\ Rev.\ C {\bf 61}, 064001 (2000).
%
\bibitem{Car90} J.\ Carlson, D.O.\ Riska, R.\ Schiavilla, and R.B.\ Wiringa,
                Phys.\ Rev.\ C {\bf 42}, 830 (1990).
%
\bibitem{Car91} J.\ Carlson, D.O.\ Riska, R.\ Schiavilla, and R.B.\ Wiringa,
                Phys.\ Rev.\ C {\bf 44}, 619 (1991).
%
\bibitem{CR71} M.\ Chemtob and M.\ Rho,
               Nucl.\ Phys.\ {\bf A163}, 1 (1971).
%
\bibitem{Tow87} I.S.\ Towner,
                Phys.\ Rep.\ {\bf 155}, 263 (1987).
%
\bibitem{WB73} C.\ Werntz and J.G.\ Brennan,
               Phys.\ Rev.\ C {\bf 8}, 1545 (1973).
%
\bibitem{Sal57} E.E.\ Salpeter,
                Phys.\ Rev.\ {\bf 88}, 547 (1952).
%
\bibitem{Kuz65} V.A.\ Kuzmin,
                Phys.\ Lett.\ {\bf 17}, 27 (1965).
%
\bibitem{WB67} C.\ Werntz and J.G.\ Brennan,
               Phys.\ Rev.\ {\bf 157}, 759 (1967).
%
\bibitem{TEG83} P.E.\ Tegn\'er and C.\ Bargholtz,
                Astrophys. J. {\bf 272}, 311 (1983).
%
\bibitem{WSS95} R.B.\ Wiringa, V.G.J.\ Stoks, and R.\ Schiavilla,
                Phys.\ Rev.\ C {\bf 51}, 38 (1995).
%
\bibitem{Pud95} B.S.\ Pudliner, V.R.\ Pandharipande, J.\ Carlson, 
		and R.B.\ Wiringa,
                Phys.\ Rev.\ Lett.\ {\bf 74}, 4396 (1995).
%
\bibitem{WSA84} R.B.\ Wiringa, R.A.\ Smith, and T.L.\ Ainsworth,
                Phys.\ Rev.\ C {\bf 29}, 1207 (1984).
%
\bibitem{Wir91} R.B.\ Wiringa,
                Phys.\ Rev.\ C {\bf 43}, 1585 (1991).
%
\bibitem{Pud97} B.S.\ Pudliner, V.R.\ Pandharipande, J.\ Carlson, 
		S.C.\ Pieper, and R.B.\ Wiringa,
                Phys.\ Rev.\ C {\bf 56}, 1720 (1997).
%
\bibitem{Car90a} J.\ Carlson,
                 private communication.
%
\bibitem{VKR95} M.\ Viviani, A.\ Kievsky, and S.\ Rosati,
                Few-Body Syst.\ {\bf 18}, 25 (1995).
%
\bibitem{VRK98} M.\ Viviani, S.\ Rosati, and A.\ Kievsky,
                Phys.\ Rev.\ Lett.\ {\bf 81}, 1580 (1998). 
%
\bibitem{Viv99b} M.\ Viviani,
                 private communication.
%
\bibitem{Kie98} A.\ Kievsky {\it et al.},
		Phys.\ Rev.\ C {\bf 58}, 3085 (1998).
%
\bibitem{Glo96} W.\ Gl\"ockle {\it et al.},
                Phys.\ Rep.\ {\bf 274}, 107 (1996), and references therein; 
                A.\ Kievsky, Phys.\ Rev.\ C {\bf 60}, 034001 (1999).
%
\bibitem{Viv98} M.\ Viviani, Nucl.\ Phys.\ {\bf A631}, 111c (1998).
%
\bibitem{Phi80} T.W.\ Phillips, B.L.\ Berman, and J.D.\ Seagrave,
                Phys.\ Rev.\ C {\bf 22}, 384 (1980).
%
\bibitem{AK93} M.T.\ Alley and L.D.\ Knutson,
               Phys.\ Rev.\ C {\bf 48}, 1901 (1993).
%
\bibitem{Mar98} L.E.\ Marcucci, D.O.\ Riska, and R.\ Schiavilla,
                Phys.\ Rev.\ C {\bf 58}, 3069 (1998).
%
\bibitem{Kub78} K.\ Kubodera, J.\ Delorme, and M.\ Rho, 
                Phys.\ Rev.\ Lett.\ {\bf 40}, 755 (1978). 
%
\bibitem{Kir92} M.\ Kirchbach, D.O.\ Riska, and K.\ Tsushima, 
                Nucl.\ Phys.\ {\bf A542}, 616 (1992).
%
\bibitem{Sch98} R.\ Schiavilla {\it et al.},
                Phys.\ Rev.\ C {\bf 58}, 1263 (1998).
%
\bibitem{Mac96} R.\ Machleidt, F.\ Sammarruca, and Y.\ Song,
                Phys.\ Rev.\ C {\bf 53}, 1483 (1996).
%
\bibitem{8Bshape} J.\ N.\ Bahcall, E.\ Lisi, D.E.\ Alburger, 
		  L.\ De Braeckeleer, S.J.\ Freedman, 
	         and J.\ Napolitano, Phys.\ Rev.\ C {\bf 54}, 411 (1996).
%
\bibitem{hepshape} {\tt http://www.sns.ias.edu/\verb+~+jnb}.
%
\bibitem{Nak99} M.\ Nakahata {\it{et al.}}, Nucl.\ Instrum.\ Methods 
		Phys. Res., Sect. A {\bf 421}, 113 (1999).
%
\bibitem{Bea99} J.F.\ Beacom and P.\ Vogel, 
		Phys.\ Rev.\ Lett.\ {\bf 83}, 5222 (1999).
%
\bibitem{Ort00} C.E.\ Ortiz {\it{et al.}}, 
		nucl-ex/0003006.
%
\bibitem{WAL95} J.D.\ Walecka, 
                {\it{Theoretical Nuclear and Subnuclear Physics}}
                (Oxford University Press, New York, 1995).
%
\bibitem{Har90} J.C.\ Hardy, I.S.\ Towner, V.T.\ Koslowsky, E.\ Hagberg, 
		and H.\ Schmeing,
                Nucl.\ Phys.\ {\bf A509}, 429 (1990).
%
\bibitem{BJO64} J.D.\ Bjorken and S.D.\ Drell,
                {\it{Relativistic Quantum Mechanics}} 
		(McGraw-Hill, New York, 1964). 
%
\bibitem{EDM57} A.R.\ Edmonds,
                {\it{Angular Momentum in Quantum Mechanics}}
                (Princeton University Press, Princeton, 1957).
%
\bibitem{Kie93} A.\ Kievsky, M.\ Viviani, and S.\ Rosati,
  		Nucl.\ Phys.\ {\bf A551}, 241 (1993).
%
\bibitem{Kie94} A.\ Kievsky, M.\ Viviani, and S.\ Rosati,
  		Nucl.\ Phys.\ {\bf A577}, 511 (1994).
%
\bibitem{KVR95} A.\ Kievsky, M.\ Viviani, and S.\ Rosati,
                Phys.\ Rev.\ C {\bf 52}, R15 (1995).
%
\bibitem{K99} H.J.\ Karwowski, private communication.
%
\bibitem{GK99} E.\ George and L.D.\ Knutson, private communication.
%
\bibitem{A4} D.R.\ Tilley, H.R.\ Weller, and G.M.\ Hale,
             Nucl.\ Phys.\ {\bf A541}, 1 (1992).
%
\bibitem{Fea54} K.F.\ Famularo {\it et al.},
                Phys.\ Rev.\ {\bf 93}, 928 (1954).
%
\bibitem{KRV00} M.\ Viviani, A.\ Kievsky,  and S.\ Rosati,
                in preparation.
%
\bibitem{Def66} T.\ deForest and J.D.\ Walecka,
                Adv.\ Phys.\ {\bf 15}, 1 (1966).
%
\bibitem{Fri73} J.L.\ Friar,
                Ann.\ Phys.\ (N.Y.) {\bf 81}, 332 (1973).
%
\bibitem{Ade99} E.\ Adelberger {\it et al.},
                Rev.\ Mod.\ Phys.\ {\bf 70}, 1265 (1998).
%
\bibitem{PDG96} Particle Data Group, R.M.\ Barnett {\it et al.},
                Phys.\ Rev.\ D {\bf 54}, 1 (1996).
%
\bibitem{Abe97} H.\ Abele {\it et al.},
                Phys.\ Lett.\ B {\bf 407}, 212 (1997).
%
\bibitem{GT87} M.\ Gmitro and P.\ Tru\"ol,
               Adv.\ Nucl.\ Phys.\ {\bf 18}, 241 (1987).
%
\bibitem{Sch89} R.\ Schiavilla, D.O.\ Riska, and V.R.\ Pandharipande,
                Phys.\ Rev.\ C {\bf 40}, 2294 (1989).
%
\bibitem{Sch90} R.\ Schiavilla, D.O.\ Riska, and V.R.\ Pandharipande,
                Phys.\ Rev.\ C {\bf 41}, 309 (1990).
%
\bibitem{Ris89} D.O.\ Riska,
                Phys.\ Rep.\ {\bf 181}, 207 (1989).
%
\bibitem{SR91} R.\ Schiavilla and D.O.\ Riska,
               Phys.\ Rev.\ C {\bf 43}, 437 (1991).
%
\bibitem{WS98} R.B.\ Wiringa and R.\ Schiavilla,
               Phys.\ Rev.\ Lett.\ {\bf 81}, 4317 (1998).
%
\bibitem{VSK96} M.\ Viviani, R.\ Schiavilla, and A.\ Kievsky,
                Phys.\ Rev.\ C {\bf 54}, 534 (1996).
%
\bibitem{T2098} M.\ Gar\c con, private communication; D.\ Abbott $et$ $al.$, 
		nucl-ex/0001006 and nucl-ex/0002003.
%
\bibitem{Car98} J.\ Carlson and R.\ Schiavilla,
                Rev.\ Mod.\ Phys.\ {\bf 70}, 743 (1998).
%
\bibitem{Fri77} J.L.\ Friar,
                Ann.\ Phys.\ (N.Y.) {\bf 104}, 380 (1977).
%
\bibitem{Sch96} R.\ Schiavilla,
                {\it Perspectives in Nuclear Physics at Intermediate Energies},
                edited by S.\ Boffi, C.\ Ciofi degli Atti, and M.\ Giannini
                (World Scientific, Singapore, 1996), p.\ 490.
%
\bibitem{SW99} R.\ Schiavilla and R.B.\ Wiringa,
               to be published.
%
\bibitem{Mac89} R.\ Machleidt,
                Adv.\ Nucl.\ Phys.\ {\bf 19}, 189 (1989).
%
\bibitem{War91} E.K.\ Warburton,
                Phys.\ Rev.\ Lett.\ {\bf 66}, 1823 (1991);
                Phys.\ Rev.\ C {\bf 44}, 233 (1991).
%
\bibitem{Tow92} I.S.\ Towner,
                Nucl.\ Phys.\ {\bf A542}, 631 (1992).
%
\bibitem{Wirpc} R.B.\ Wiringa,
                private communication.
%
\bibitem{Sto93} V.G.J.\ Stoks, R.A.M.\ Klomp, M.C.M.\ Rentmeester, 
		and J.J.\ de Swart,
                Phys.\ Rev.\ C {\bf 48}, 792 (1993).
%
\bibitem{Note} In Ref.~\cite{Sch98} the relativistic corrections
to the single-nucleon axial current operator in Eq.~(\ref{eq:1arc}) were
neglected.  Furthermore, in the calculation of the contribution of
the $\rho$-exchange axial current operator, the non-local term (proportional
to ${\bf P}_i$) in the second line of Eq.~(\ref{eq:a2ro}) was inadvertently
omitted.  However, this has no impact on the conclusions of that
work, since the non-local term referred to above was also
omitted in the calculation of the $pp$ capture cross section.
%
\bibitem{CAR86} C.E.\ Carlson,
                Phys.\ Rev.\ D {\bf 34}, 2704 (1986).
%
\bibitem{LIN91} D.\ Lin and M.K.\ Liou,
                Phys.\ Rev.\ C {\bf 43}, R930 (1991).
%
\bibitem{Met53} N.\ Metropolis, A.W.\ Rosenbluth, M.N.\ Rosenbluth, 
		A.H.\ Teller, and E.\ Teller,
                J.\ Chem.\ Phys.\ {\bf 21}, 1087 (1953).
\end{references}
\end{document}